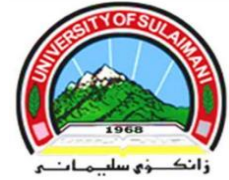

# RESPONSE OF DIFFERENT TOMATO ACCESSIONS TO BIOTIC AND ABIOTIC STRESSES

**A Dissertation**

**Submitted to the Council of the College of Agricultural Engineering Sciences at the University of Sulaimani in Partial Fulfillment of the Requirements for the Degree of Doctor of Philosophy**

**in**

**Horticulture**

**Plant Biotechnology**

By

**Kamaran Salh Rasul**

B.Sc., Horticulture (2008), College of Agriculture, University of Sulaimani

M.Sc., Plant Genetics Manipulation (2013), School of BioScience,

University of Nottingham

Supervisors

| | |
|---|---|
| **Dr. Nawroz Abdul-razzak Tahir** | **Dr. Florian M. W. Grundler** |
| Professor | Professor |

**2723 K.**          **2023 A.D.**

# Summary

Accessions are prospective sources of genetic variability, as well as valuable genetic resources to deal with present and future crop breeding difficulties. The assessment of population structure and genetic diversity of tomatoes (*Solanum lycopersicum* L.) that have been distributed in Iraqi Kurdistan region critical in breeding programs for the production of high-yielding cultivars as well as widening the genetic base of tomato. Using fruit quality indices and molecular markers, a panel of 64 tomato accessions taken from six provinces of Iraqi Kurdistan Region, were analyzed for genetic diversity and population structure. In the analysis of variance, the fruit phenotypic data revealed a high level of significant variability (P ≤ 0.01) among tomato accessions. The most important characteristics for explaining fruit morphological variability, according to principal component analysis (PCA), were fruit weight, fruit size, fruit diameter, total soluble solids, and moisture content. Seven clades with different fruit characteristics were revealed in the cluster analysis. Genetic diversity and relationships among accessions were analyzed using thirteen inter simple sequence repeat (ISSR), twenty-six start codon-targeted (SCoT) polymorphisms, and fifteen conserved DNA-derived polymorphisms (CDDP). The ISSR, SCoT, and CDDP markers generated 121, 294, and 170 polymorphic bands, respectively, showing a high prevalence of polymorphism. The average polymorphism information content (PIC) values for ISSR, SCoT, and CDDP were 0.81, 0.84, and 0.84, respectively. The accessions were divided into two groups based on the cluster and STRUCTURE analysis results. The Mantel test revealed that three sets of markers had positive and significant relationships. The increased genetic variation within the populations was found by the analysis of molecular variance (AMOVA), indicating considerable gene exchange between populations.

Drought stress is one of the most significant abiotic stresses on the sustainability of global agriculture. To evaluate and select drought tolerant and susceptible accessions, 64 tomato accessions were tested for drought at the seedling stage under *in vitro* conditions using polyethylene glycol (PEG-600). Two concentrations of PEG, 7.50% and 15%, compared with the control of 0% PEG, were performed, and significant changes in the morphological and biochemical profiles of tomato accession seedlings were observed among the tested materials. Based on the results, three accessions were selected as highly tolerant to drought stress, namely Raza Pashayi (AC61), Wrdi Be Tow (AC9), and Sandra (AC63), while Braw (AC13), Yadgar (AC30), and Israili (AC8) were highlighted as sensitive to drought stress. To confirm these results under field conditions, four tomato accessions, two sensitive (AC13 and AC30) and two tolerant (AC61 and AC63), were evaluated for drought stress under greenhouse conditions, with the use of oak leaf extract, biofertilizer, and oak leaf powder, to reduce the effect of drought on these accessions. In this study,

i

a factorial experimental design was used to investigate the effects of these treatments on the growth and biochemical parameters of four tomato accessions under water stress throughout the pre-flowering and pre-fruiting stages of plant development. The experiment had two factors. The first factor represented the accessions, while the second factor represented the treatment group, which included irrigated plants (SW), untreated and stressed plants (SS), treated plants with oak leaf powder and stressed (SOS), treated plants with oak leaf powder and oak leaf extract and stressed (SOES), and treated plants with oak leaf powder and biofertilizers and stressed (SOBS). The data analysis showed that drought stress under the treatments of SS, SOS, SOES, and SOBS conditions at two stages and their combination significantly lowered shoot length (12.95%), total fruit weight per plant (33.97%), relative water content (14.05%), and total chlorophyll content (26.30%). The reduction values for shoot length (17.58%), shoot fresh weight (22.08%), and total fruit weight per plant (42.61%) were significantly larger in two sensitive accessions compared to tolerant accessions, which recorded decreasing percentages of 8.36, 8.88, and 25.32% for shoot length, shoot fresh weight, and total fruit weight per plant, respectively. On the other hand, root fresh weight and root dry weight of accessions treated with SS, SOS, SOES, and SOBS, were increased in comparison to control plants. Tomato fruits from stressed plants treated with SS, SOS, SOES, and SOBS had considerably higher levels of titratable acidity, ascorbic acid, and total phenolic compounds than irrigated plants during all stress stages. Under water stress conditions, the application of oak leaf powder to soil, oak leaf extract, and biofertilizer improved the biochemical contents of leaves in all accessions. Furthermore, leaf lipid peroxidation was lower in plants treated with SOES and SOBS, as well as lower in the two tolerant accessions than in the two susceptible accessions.

In the case of the effects of heavy metal stress on the tomato accessions, cadmium (Cd) was used to assess 64 tomato accessions under *in vitro* conditions at the seedling stage. Three dosages of cadmium (150 μM, 300 μM, and 450 μM), compared to the control (0 μM) were exposed to all tomato accessions. There were significant changes among the phenotypic and phytochemical traits. The results revealed the best tolerate accessions and the susceptible accessions to heavy metal stress. Sirin (AC7), Karazi (AC5), and Balami (AC31) were indicated as having the highest tolerance to Cd stress, while Sewi Qaladze (AC56), Super (AC32), and Braw (AC13) showed more sensitivity to this stress. To approve these results, four tomato accessions, two sensitive (AC56 and AC32), and two tolerant (AC7 and AC5) to Cd stress, were evaluated under greenhouse conditions. The experiment had two factors, accessions were represented as the first factor, and the second factor represented the treatment group, which included untreated plants (control), treated plants with Cd (Cd+Soil), and treated plants with Cd and oak leaf residue together (Cd+Soil+Oak). Regarding the application of Cd, 35 mg per kg of soil was used, and oak leaf residue was used to decrease the absorption of Cd by the tomato plants. Compared to the control treatment, most morphological traits were significantly decreased in the Cd+Soil treatment. Root length, shoot length, shoot fresh weight, shoot dry weight and total fruit weight per plant, were decreased by 14.35, 18.00, 6.53, 5.88, and 13.85%, respectively. Under (Cd+Soil+Oak) treatment, significant reductions were observed in root length (10.68%), shoot length (10.22%), total fruit weight per plant (10.10%), and shoot fresh weight (0.46%), compared to the reductions observed under Cd+Soil treatment. On the other hand, Cd+Soil and Cd+Soil+Oak treatments significantly increased the values of both root fresh weight (20.77 and 27.73%) and root dry weight (17.47 and 24.21%), respectively. Furthermore, the biochemical values of proline content, soluble sugar content, total phenolic content, antioxidant capacity, guaiacol peroxidase activity, and catalase activity were significantly higher in both treatment groups than in the control group. These biochemical values were also higher in the two resistant accessions, Sirin (AC7) and Karazi (AC5), than in the two sensitive accessions. In the case of Cd accumulation in the parts of the tomato plant, the results showed that most of the Cd absorbed by the plant was accumulated in the root, while only a small amount was accumulated in other parts like the stem, leaf, and fruit. Using oak leaf residue reduced the amount of Cd accumulation in all plant parts.

Root knot-nematode (RKN) infection has severely harmed the tomato plant. The eggs of the nematode *Meloidogyne* spp. were used to evaluate the tolerance of four tomato accessions, Amad (AC14), Pamayi Kurdi (AC43), Kurdi Gawray Swr (AC53), and Sandra (AC63) using three treatments; control treatment, only use the soil without nematode eggs and oak leaf powder, while second treatment, approximately 15000 eggs per pot without oak leaf were used and third treatment, approximately 15000 eggs per pot with 80 g of oak leaf were used to reduce nematode infection. The results revealed that the accessions responded differently to the nematode infection and the infection severity revealed that AC14 was more affected (96.67%), however AC63 less affected was (60.00%). The values of all morphological traits (shoot length, shoot dry weight, root length, root dry weight) decreased under both treatments, and there was no significant effect from using the oak leaf powder.

# CHAPTER ONE
# INTRODUCTION

The Solanaceae family, which also called Nightshade family, is a large and diverse plant family that includes more than 3000 species. The *Lycopersicon* clade contains the domesticated tomato and its twelve closest wild relatives (Bauchet and Causse, 2012). Tomato (*Solanum lycopersicum* L.) is one of the most important, popular, and versatile vegetables in the world. Tomato is a self-pollinated annual crop with chromosome number 2n = 2x = 24 (Peralta *et al.*, 2008; Salim *et al.*, 2020). Evidence from the diversity of cultivated tomato, all demonstrated that tomato was originally domesticated in Mexico (Jenkins, 1948; Rick, 1974). Tomato was brought to Europe and gradually spread all over the world (Lin *et al.*, 2014). Tomatoes are currently one of the most important vegetable crops in the horticulture industry, and cultivated all over the world, either for fresh consumption or processing. It is the world's leading product of vegetables, accounting for approximately 14% of global production (Bauchet and Causse, 2012). According to FAOSTAT (2021), tomato production in the world was about 189 million tonnes. Li *et al.*, (2023) stated that the tomato fruit contains a variety of healthy nutrients, including ascorbic acid (ASC), amino acids, minerals, and carotenoid pigments (Viuda-Martos *et al.*, 2014). Carotenoids, in particular, act as antioxidants that protect against certain degenerative conditions like macular degeneration of the eye, a major cause of age-related blindness, as well as cardiovascular and cerebrovascular conditions and cancers like prostate cancer (Li *et al.*, 2023). Even though, a considerable amount of tomato fruits is processed to make the paste, juices, peeled tomatoes, and diced products. The demand for tomato has recently increased due to the nutritional value related to the carotenoids, fibers, and vitamins, and it is eaten freshly or added to other food items (Foolad, 2007; Collard and Mackill, 2009a). The tomato is a common model plant in several scientific fields, including genetics, developmental biology, and molecular biology, particularly for research on the control of ethylene signaling transduction in climacteric ripening of fruits (Li *et al.*, 2023).

Systematic study and evaluation of germplasm are important for current and future agronomic and genetic improvement of the crops (Reddy *et al.*, 2013). To identify and estimate the genetic diversity of plants, various methods can be used, including morphological, biochemical, and molecular markers. Genetic variation in tomato plants by morphological characters and molecular markers has been the subject of many studies around the world (Henareh *et al.*, 2015). For differentiating genotypes, morphological characteristics are crucial diagnostic factors (Osei, 2014). Traditional strategies for identifying genetic variants in local tomato collections have previously relied on morphological features. These agro-morphological features, however, have significant limitations due to their reliance on plant phenological stage and proclivity to be highly influenced by environmental influences. To date, several durable and effective molecular markers have been produced. DNA-based molecular markers have been developed and used to study genetic diversity in plants, however, only a limited level of polymorphism in tomato plant has been identified (Abdein *et al.*, 2018). Moreover, the evolution of molecular markers, such as amplified fragment length polymorphism (AFLP), restriction fragment length polymorphism (RFLP), randomly amplified polymorphic DNA (RAPD), single nucleotide polymorphism (SNP), simple sequence repeats (SSR), and sequence-related amplified polymorphism (SRAP) allow for greater precision in evaluating the genetic variation of tomato germplasm (García-Martínez *et al.*, 2006; Mata-Nicolás *et al.*, 2020; Caramante *et al.*, 2021; El-Mansy *et al.*, 2021). Initiating a trend away from random DNA markers called inter simple sequence repeat (ISSR) (García-Martínez *et al.*, 2006) towards gene-targeted markers, a novel marker system, such as conserved DNA-derived polymorphism (CDDP) and start codon targeted polymorphism (SCoT) (Collard and Mackill 2009a; Abdeldym *et al.*, 2020), which were developed based on the conserved regions within genes. These markers were widely used for variation analysis among tomato accessions (Sharifova *et al.*, 2017; Abdein *et al.*, 2018; Kiani *et al.*, 2018). The analysis of genetic diversity based on both fruit features and molecular markers (i.e., CDDP, SCoT, and ISSR) is noteworthy for tomato breeding and production in order to acquire high-quality data. Nowadays, there has been only a little research on evaluating genetic variation utilizing combined CDDP, SCoT, and ISSR markers, as well as morphometric features.

The tomato plant is seriously affected by abiotic stresses such as drought, heavy metals, and salinity as well as biotic stresses caused by fungi, nematodes, and bacteria (Toumi *et al.*, 2014).

Water resources around the world have decreased as a result of climate change and global warming. Agriculture productivity is significantly impacted by water constraints around the world (Abbasi *et al.*, 2020). The plant's internal water content is affected by low soil water availability, which inhibits its physiological and biochemical functions. Despite the tomato's economic importance, it is susceptible to drought stress, especially during its blooming and fruit enlargement phases (Srinivasa Rao *et al.*, 2000; Jangid *et al.*, 2016), which prevents seed germination, slows down plant development, and lowers fruit yields (Liu *et al.*, 2017). Additionally, little is known about the crucial role of stress-responsive genes, the processes behind their response to abiotic pressures, and the mechanisms underlying their response to biotic challenges (Tamburino *et al.*, 2017). The plant's response to drought stress is highly dependent on the duration and severity of the stress but is also influenced by the plant's genotype and its developmental stage (Janni *et al.*, 2019). The plants change their cellular activities by producing different defense mechanisms in response to water stress. Drought causes osmotic stress, which can result in turgor loss, membrane deterioration, protein degradation, and often high amounts of reactive oxygen species (ROS), which cause tissue oxidative damage (Stajner *et al.*, 2016).

The addition of plant tissue to soil improves soil quality by reducing the risk of soil erosion and increasing crop yields (Carvalho *et al.*, 2017). Plant tissue application also plays a crucial role in sustaining and improving the chemical, physical, and biological properties of the soil by providing mineral nutrients and protecting the soil's water content (Cherubin *et al.*, 2018) and may have an effect on plant water uptake (Armada *et al.*, 2014). Natural biofertilizer is a product made from living microorganisms that are extracted from root or cultivated soil. It is safe for the environment and soil health, and it is essential for atmospheric nitrogen fixation and phosphorus solubilization, which leads to increased nutrient uptake and tolerance to drought and moisture stress (Meena *et al.*, 2017).

In recent years, heavy metals dramatically increased due to the development of urban industrial and agricultural production, and environmental pollution (Chen *et al.*, 2019). Growth retardation, modifications in the activity of a number of enzymes, and problems with photosynthesis within the plant system are some of the negative impacts of heavy metals. Based on their genetic characteristics, several species of crops absorb heavy metals dramatically in different ways (Li *et al.*, 2010). One of the most stressful heavy metals for plants is cadmium (Cd). The physiological processes and morphological characteristics of the majority of crop species are impacted by excessive Cd. Numerous studies have shown that in Cd-contaminated conditions, photosynthetic efficiency, nutrient intake, biomass production, and crop yield all significantly decline. Additionally, excessive Cd levels have a deleterious impact on root architecture and growth, which affects the uptake of nutrients and the mobility of nutrients in plant tissues (Asati *et al.*, 2016; Jiang *et al.*, 2017; Chtouki *et al.*, 2021). The production of a large amount of tomatoes in greenhouses typically involves the reuse of water, fertilizers, and pesticides these increase the risk of heavy metals contamination including Cd (Borges *et al.*, 2018).

Root-knot nematodes (RKNs), *Meloidogyne* spp., are a serious pest of tomatoes in fields and greenhouses all over the world. Tomato plants infected with RKNs revealed yellow leaves and stunted growth, and in cases of severe infection, the yield loss could reach 85% (Karssen *et al.*, 2013). *Meloidogyne* spp. includes four major species, *M. arenaria*, *M. incognita*, *M. javanica* and *M. hapla* (Moens *et al.*, 2009). The annual losses caused by these species are estimated at about $10 billion (Toumi *et al.*, 2014).

The objectives of this study are to evaluate:

1- Genetic diversity among 64 tomato accessions using morphological characteristics and molecular markers (ISSR, SCoT, and CDDP), which were collected from six provinces in the Iraqi Kurdistan Region (Sulaimani, Erbil, Duhok, Halabja, Garmian, and Raparin).

2- The morphological and phytochemical responses of all tomato accessions to drought stress using polyethylene glycol (PEG-6000) under *in vitro* conditions at seedling stage.

3- The effect of oak leaf extract, biofertilizer, and oak leaf powder on tomato growth and biochemical characteristics under water stress conditions in a greenhouse.

4- The morphological and biochemical responses of all tomato accessions to heavy metal stress, using cadmium (Cd) under *in vitro* conditions at seedling stage.

5- The morphological and biochemical effects of oak leaf residue on tomato accessions exposed to cadmium stress conditions in a greenhouse.

6- The tomato accessions' response to root-knot nematode infection using oak leaf powder.

Based on the above objectives, it can be said that the aim of this project is to select or find the best tomato accessions to resist of biotic and abiotic stresses.

# CHAPTER TWO
# LITERATURE REVIEW

## 2.1 Origin and Botanical Description of Tomato

The origin of the tomato is the Andean region of Colombia, Chile, Peru, and Bolivia. However, there are indications that domestication occurred in Mexico (Saavedra *et al.*, 2016) and from there the cultivated tomato (*Solanum lycopersicum)* was produced and disseminated (Jenkins, 1948). Based on morphological and molecular evidence, *Solanum lycopersicum* var. cerasiforme or at least a specific group of accessions of *Solanum lycopersicum* var. cerasiforme considered to be the direct ancestor of cultivated tomatoes. The domestication process is expected as two-step process, which started with the pre-domestication of *Solanum lycopersicum* var. cerasiforme from the wild tomato *Solanum pimpinellifolium*, an indeterminate weedy plant bearing small round fruits, in Ecuador and Northern Peru, after that by migration of *Solanum lycopersicum* var. cerasiforme to Mesoamerica, where the true domestication occurred, and finally led to the production of the cultivated tomatoes bearing big fruits (Liang *et al.*, 2017; Li *et al.*, 2022). The conquistadors carried tomatoes to Europe in the sixteenth century, and as a result of subsequent migration and selection, this crop's genetic diversity decreased (Lin *et al.*, 2014).

From botanical point of view, the tomato is a fruit. Nevertheless, it contains a much lower sugar content compared to other fruits. It is a diploid plant with $2n = 24$ chromosomes. The tomato belongs to the Solanaceae family, which contains more than 3,000 species, including plants of economic importance such as potatoes, eggplants, peppers, tobacco and petunias (Gerszberg *et al.*, 2015). In 1753, Linnaeus placed the tomato in the *Solanum* genus (along with potato) under the specific name *Solanum lycopersicum* and defined three species of what are now known as tomatoes as members of the genus *Solanum* (*S. lycopersicum, S. peruvianum* L. *and S. pimpinellifolium*). In 1754, Philip Miller moved tomato to its own genus, naming it *Lycopersicum esculentum* (Foolad, 2007; Peralta *et al.*, 2008), separated the new genus *Lycopersicon* to accommodate *Solanum* species with multi-locular fruits, including the tomatoes, the potato (*S. tuberosum* L.) and several other species (Darwin *et al.*, 2003). However, the designation of the tomato was for a long time a subject of consideration and discussion by many scientists. The use of molecular data (genome mapping) and morphological information allowed for the verification of the Solanaceae classification when the genus *Lycopersicon* was re-introduced in the *Solanum* genus in the *Lycopersicon* section (Foolad, 2007; Gerszberg *et al.*, 2015).

Tomato fruit is a fleshy berry with a range sizes and colors. Tomato fruit has a pericarp, which is made up of an outer layer of exocarp and inner layers of mesocarp and endocarp, the tomato fruit exocarp (epidermis) comprises of a thin cuticle with no stomata. Tomato cuticle is generally composed of a lipid polymer known as cutin, and waxes, which are complex and variable (Rančić *et al.*, 2010). The mesocarp encompasses fruit vascular tissue linked to pedicel vascular tissue. Vascular tissue is positioned in the center of tomato fruit, providing seeds with required water and minerals, and is also parallel to the fruit surface. Within the unicellular endocarp boundary are seed-containing cavities derived from carpels, which called locules (Yeats *et al.*, 2012). The number of locules within a fruit can vary, changing the size and shape of the fruit. Locules are separated by a septum, with seeds bound to an elongated axial placenta. Furthermore, tomato seeds are identified to contain steroidal saponins called lycoperosides (Takeda *et al.*, 2021; Collins *et al.*, 2022).

## 2.2 Economic Importance of Tomato

Tomato plays an important role in food and commercial utilization all over the world due to its taste, flavor and nutritional value, it is unattached component of food. The tomato fruit is rich of minerals, vitamins and antibiotic characteristics (Jangid *et al.*, 2016). Tomato fruit is not only used fresh but also processed and marketed in a variety of forms, such as soup, juice, paste, sauce, powder, concentrated or whole. Tomato is one of the most consumed vegetables in the world, after potatoes and probably the most favorite garden crop, accounting for approximately 14% of global production (Bauchet and Causse, 2012). According to FAOSTATI (2021), global production of tomatoes expected to reach around 189 million tonnes and the production in Iraq (including Kurdistan Region) was about 744166 tonnes. Tomato is the tenth most significant agricultural crops after sugar cane, maize, rice, wheat, palm fruit, potatoes, soybeans, cassava and sugar beet (FAOSTAT, 2021). During the last 20 years, tomato production, as well as the area devoted to its culture, has folded. It is noteworthy that most countries cultivated the majority of tomatoes under controlled greenhouse conditions (Bergougnoux, 2014; Figueiredo *et al.*, 2016).

## 2.3 Tomato Nutritional Value

The tomato fruit is composed mostly of water (87-95%) with proteins, lipids (fat) and sugars (carbohydrates) content around 3% (glucose and fructose) in a low-level value (Vaughan and Geissler, 2009). Sass (2022) reported that a 100 g of ripe red tomato contains, 18 calories, <1 g of fat, 0 mg of cholesterol, 5 milligrams of sodium, 3.89 g of carbohydrates, 1.20 g of fiber, and <1 g of protein. Nevertheless, tomatoes denote an important source of phytochemicals and nutrients which are important for human health such as antioxidants, represented by the content in lycopene, *β*-carotene (vitamin A), ascorbic acid (vitamin C), potassium, iron and folate (Kumar *et al.*, 2012a). Thus, tomatoes represent the main source of lycopene, which has antioxidant properties and is considered to protect against cancer or cardiovascular diseases. Besides lycopene and vitamin C, tomatoes provide other antioxidants, such as *β*-carotene, and phenolic compounds, such as flavonoids, hydroxycinnamic acid, chlorogenic acid, homovanillic acid, and ferulic acid (Collins *et al.*, 2022).

Tomatoes are also a significant and noteworthy source of ascorbic acid. The L-galactose Wheeler-Smirnoff pathway (Smirnoff, 2000), is the major method of ascorbic acid biosynthesis, in which ascorbic acid is generated from mannose-6-phosphate via guanosine diphosphate GDP-mannose and GDP-L-galactose. More routes have been described, including one containing an L-galactonic acid intermediary derived from cell wall polymers (Di Matteo *et al.*, 2010). When compared to modern cultivated tomatoes, wild tomato varieties contain up to 5 times more ascorbic acid (Bergougnoux, 2014). Di Matteo *et al.* (2010) confirmed that the accumulation of ascorbic acid is attained by increasing pectin degradation and may be triggered by ethylene. Some cultivars with increased nutritional value were therefore effectively created; nevertheless, yield reduction in these new cultivars hampered their economic viability (Causse *et al.*, 2007). Ascorbic acid content in fresh tomatoes rises to a maximum and then declines during the ripening process. The salad tomatoes grown in field conditions contained 15–21 mg/100 g fresh weight (FW) of ascorbic acid compared to a range of industrial grades of tomatoes with an average vitamin C value of 19 mg/100 g FW (Collins *et al.*, 2022).

Total soluble solids (TSS) are crucial characteristics for processing tomatoes and help to define the concentrated tomato product. Soluble solids are sugars and organic acids, and their ratio, along with volatile aroma composition, describe the flavor of the fruit.  Non-soluble solids (NSS), characterized by components of the cell wall and proteins, determine the firmness of the fruit as well as the viscosity of the final products, such as tomato juice, ketchup, soups and paste (Bergougnoux, 2014). This horticultural crop is the main source of carotenoids. Only 25 of the approximately 40 carotenoids included in the human diet are detectable in human blood due to selective uptake by the digestive tract, which 9-20 of them are obtained from fresh and processed tomatoes, with the primary ones being lycopene, *α*- and, *β*-carotene, lutein, zeaxanthin, and cryptoxanthin (Dorais *et al.*, 2008). Lycopene accounts for roughly 80-90 % of the total carotenoid content in red ripe tomatoes. Lycopene is the most effective antioxidant among carotenoids due to its ability to quench singlet oxygen and scavenge peroxyl radicals (Erba *et al.*, 2013). On the other hand, *β*-carotene, a powerful dietary precursor of vitamin A, contributes for approximately 7% of tomato carotenoid content (Collins *et al.*, 2022). While ascorbic acid is a powerful antioxidant in plants, it is also a significant phytochemical in tomato fruit. Tomato fruits are not typically reported to comprise anthocyanin. The limited caloric supply, relatively high fiber content, and delivery of minerals, vitamins, and phenols such as flavonoids make the tomato fruit an excellent 'functional food' providing additional physiological profits as well as meeting basic nutritional requirements (Dorais *et al.*, 2008).

## 2.4 Genetic Diversity of Tomatoes

Many years ago, people started growing a variety of wild plants for a multiple of uses, such as food, drink, spices, herbs, oils, waxes, medicine, dyes, and ornaments. People first started domesticating plants when they selected out suitable plants from the wild and then grown them in the field (Kaneko, 2011). For plant breeding activities that aim to create varieties with high quality, yield, resistance to biotic and abiotic factors, among other qualities of economic value, information of the genetic diversity of a germplasm is very important (Vargas *et al.*, 2020).

Tomato accessions have distinctive organoleptic traits (flavor and aroma) and nutritional value. These genotypes have received intensive attention especially in fresh market tomato. Many landraces were continuously replaced by modern tomato cultivars in these regions in recent years; therefore this germplasm has experienced an overall reduction of its genetic basis (Terzopoulos *et al.*, 2008). In spite of the potential as a source of variability, the lack of information about agronomic traits and genetic constitution of landraces has limited their use in breeding programs. To identify and estimate the genetic diversity of plants, various methods can be used including morphological, biochemical and molecular markers (Henareh *et al.*, 2015). Morphological traits have limitations since they are influenced by environmental factors and the developmental stage of the plant. Molecular markers have proved valuable in crop breeding, especially in study of genetic diversity (Henareh *et al.*, 2016).

## 2.4.1 Morphological markers

Morphological-based estimation of genetic diversity in tomato has been the subject of many researchers in different regions of the world (Henareh, 2015). Morphological trait measurements can provide a simple technique of quantifying genetic variation while simultaneously assessing genotype performance under relevant growing environments. However, assessment of morphological traits is time consuming and phenotypic characters are generally influenced by environments and plant developmental stages (Meena *et al.*, 2015). There are numerous traits of morphology can be considered in genetic diversity of tomato such as, plan height, root length, shoot fresh and dry weights, root fresh and dry weights, branch number, leaf area, leaf shape, fruit color, fruit size, fruit number per plant, and etc.

Hu *et al.* (2012) used seventeen morphological traits to evaluate the variation of 67 tomato varieties. The results indicated that fruit shape had the largest variation with seven types (flat, oblate, round, high round, prelate round, ovate, and pear-shaped). No obvious differences for nine traits including leaf vein color, leaf shape, leaf state, stem and leaf hairiness, corolla color, abscission layer, fruit shoulder, inflorescence type, and plant posture were observed. Most taxonomic distances were between 0.5001 and 1.1000. Lines collected from different regions at different years were randomly clustered into different groups. These 67 tomato varieties formed 3 clusters at the average taxonomic distance of 0.88.

High heritability coupled with high genetic advance was observed for plant fruits, single fruit weight, fruit yield per plant, fruit clusters, plant and fruit yield indicating the presence of additive gene effects which may be utilized for improvement through phenotypic selection for yield improvement. High heritability with moderate to low genetic advance was observed for days to 50% flowering and fruiting, first and last picking, plant height and fruit diameter (Meitei *et al.*, 2014).

Osei (2014) used tomato morphological traits such as stem and fruit pubescence, leaf attitude, style, stamen length, color of immature fruit, fruit skin color, ease of fruit wall to peel and plant habit, to know the variation among 216 tomato germplasm, as a result, two main groups were generated through agglomerative hierarchical clustering based on the similarity matrix.

## 2.4.2 Biochemical markers

Tomato fruit contains 87–95% water and 5–12% organic compounds (solids), of which about 1% is skin and seeds. The percentage of solids in tomato varies which depends on variety, character of soil and particularly the irrigation amount at the growing and harvesting season. Total soluble solids (TSS) content is the most important quality measurement for tomato processing. Consequently, soluble solids content or acidity (pH), can be considered for tomato selection (Ahmet and Seniz, 2009).

Phytochemical markers are characteristics derived from the study of plant biochemical compounds such as primary and secondary metabolites. The components determine sweet-sour taste of tomato are: reducing sugar (fructose and glucose), free acids (citric acid), some volatile substances, minerals (potassium and phosphate) and free amino acids (glutamic acid, glutamine, gamma-aminobutyric acid, and aspartic acid). Tomato flavors are derived by volatile substances which are mainly detected by fatty acids and amino acids, and approximately 400 volatile compounds were identified in tomato fresh fruit such as cis-3-hexenal, trans-2-hexenal, 2-isobutylthiazole, hexanal, cis-3-hexen-1-ol, 2E,4E-decadienal, and 6-methyl-5-hepten-2-one (Petro-Turza, 1986). Tomato is a vital source of ascorbic acid, carotenoids (lycopene), potassium, and folic acid. During tomato fruit ripening, the pigments of carotenoids synthesized which are responsible for the final red color of the tomato. Consumption of tomato and tomato-based products contribute to the absorption of carotenoids and lycopenes in human serum. Tomato also comprises numerous active constituents, namely, neoxanthin, lutein, α-cryptoxanthin, α-carotene, β-carotene, cyclolycopene, and β-carotene 5, 6-epoxide (Perveen *et al.*, 2015).

Chemical analysis reveals that sugar and organic acids make a major contribution to the total dry solid, sugar content is positively associated with total soluble solids content in tomato fruit. While soluble solid content considered as an indicator of the sugar level in tomato fruit. The sugars present in tomato are glucose and fructose which constitute around 65% of total soluble solids (Paolo *et al.*, 2018). The acids in tomato are generally citric and malic acids, organic acids contain about 15% of the dry content of tomato fruit. calcium (Ca), potassium (K), magnesium (Mg), and phosphorus (P) are normally found in tomato fruit and may reach to 8% of the dry matter. These minerals directly effect on pH and titratable acidity which influence the taste of tomatoes. Free amino acids form about 2 - 2.5% of the total dry matter of tomatoes (Ahmet and Seniz 2009).

### 2.4.3 Molecular markers

Plant breeding activities that objective to create varieties with high quality, yield, resistance to biotic and abiotic stresses, among other attributes of economic importance, place the greatest emphasis on understanding the genetic diversity of a germplasm (Herison *et al.*, 2017). DNA markers are a common method for evaluating genetic diversity because they provide a more thorough assessment of a species' genome, are minimally impacted by the environment, and reveal variation at the DNA level (Ansari *et al.*, 2016; Vargas *et al.*, 2020). The genetic base of the cultivated tomato has been reduced outstanding to the continuous selection processes caused by domestication and genetic improvement. It is essential to include wild species in breeding programs because they provide variability and have valuable genes that can be used to improve cultivated species (Bergougnoux, 2014). Molecular characterization can play a role in discovery of the history and estimating the diversity, distinctiveness, and population structure. It can also serve as an aid in the genetic management of small populations, to avoid excessive inbreeding. Numerous investigations have been defined within and between-population diversity (Ramesh *et al.*, 2020).

Two main kinds of DNA-based marker systems have been discovered: hybridization-based (non-PCR) markers and PCR-based markers.

### 2.4.3.1 Non-PCR based techniques

Molecular markers based on restriction-hybridization techniques were employed relatively early in the field of plant studies and combined the use of restriction endonucleases and the hybridization method. Restriction fragment length polymorphism (RFLP) can be defined as a common non-PCR technique (Jonah *et al.*, 2011). RFLP and variable numbers of tandem repeats (VNTRs) markers are examples of molecular markers based on restriction-hybridization techniques. In RFLP, DNA polymorphism is detected by hybridizing a chemically-labelled DNA probe to a southern blot of DNA digested by restriction endonucleases, consequential in differential DNA fragment profile (Naeem, 2014). The RFLP markers are comparatively highly polymorphic, codominantly inherited, highly replicable and consent the simultaneously screening of several samples (Kumar *et al.*, 2009). DNA blots can be evaluated repeatedly by stripping and reprobing (typically eight to ten times) with different RFLP probes. However, this technique is not very commonly used as it is time-consuming, involves expensive and radioactive/toxic reagents and requires large quantities of high-quality genomic DNA (Mondini *et al.*, 2009). These limitations led to the development of a new set of less technically complex methods known as PCR-based techniques.

### 2.4.3.2 PCR-based techniques

The use of this kind of marker has been widely used, since the polymerase chain reaction (PCR) has been invented. This technique comprises in the amplification of some discrete DNA products, deriving from regions of DNA which are flanked by regions of high homology with the primers. These regions must be close enough to one another to permit the elongation phase (Kumar *et al.*, 2009). PCR-based techniques are divided into arbitrarily primed PCR-based techniques or sequence non-specific techniques; and sequence targeted PCR-based techniques. Primers in the first category are designed arbitrarily/or semi-arbitrarily, there is no information about the flanking sequence of the region which is amplified (Agarwal *et al.*, 2008). The examples of molecular markers of this technique are RAPD (random amplified polymorphic DNA), ISSR (inter simple sequence repeats), AFLP (amplified fragment length polymorphism), and DNA amplification fingerprinting (DAF). These molecular markers are also described as dominant molecular markers. Primers at the second category target a single known site, such as a gene. Microsatellites or Simple sequence repeats (SSR), PCR-DNA sequencing, sequence tagged microsatellites (STMs), single nucleotide polymorphisms (SNPs) are the examples for this category (Mondini *et al.*, 2009; Soriano, 2020; Amiteye, 2021).

Different molecular marker systems have been applied to assess the genetic variation in tomato plant such as simple sequence repeats (SSRs), random amplified polymorphic DNA (RAPD), amplified fragment length polymorphism (AFLP), sequence-related amplified polymorphism (SRAP), inter simple sequence repeat (ISSR) and single nucleotide polymorphism (SNP) (García-Martínez *et al.*, 2006; Henareh *et al.*, 2016; Mata-Nicolás *et al.*, 2020; Brake *et al.*, 2021;

Caramante *et al.*, 2021; El-Mansy *et al.*, 2021), gene-targeted markers, a novel marker system, such as conserved DNA-derived polymorphism (CDDP) and start codon targeted polymorphism (SCoT) ( Collard and Mackill 2009b; Abdeldym *et al.*, 2020).

### Inter simple sequence repeat (ISSR)

Inter simple sequence repeat (ISSR) maker was developed such that no sequence information was required. Primers based on a repeat sequence, such as (CA)n, can be made with a degenerate 3′-anchor, such as (CA)$_8$ RG or (AGC)$_6$ TY. The resultant PCR reaction amplifies the sequence between two SSRs, yielding a multilocus marker system useful for fingerprinting, diversity analysis and genome mapping (Godwin *et al.*, 1997).

ISSR marker technique is very simple, fast, cost effective, highly discriminative, reliable, requires a small quantity of DNA sample, does not need any prior primer sequence information and non-radioactive (Bhatia *et al.*, 2009).

Regarding tomato investigations in terms of diversity using ISSR marker, Kiani *et al.* (2018) evaluated the genetic diversity and relationships between 12 tomato genotypes, using 12 ISSR primers, as a result, 69 bands were produced, while 53 bands were polymorphic, with the average 0.29 polymorphism index content (PIC). A higher degree of polymorphism among tomato accessions was revealed, and four main groups were produced. Vargas-Ponce *et al.*, (2011) used six ISSR primers on eight Mexican tomato husk species, to determine their utility for interspecific taxonomic discrimination and to assess their potential for inferring interspecific relationships. A total of 101 bands were amplified, with 100% polymorphism across samples. The number of bands per primer diverse was from 10 to 21. All primers produced different fingerprint profiles for each species, confirming the ISSR value in taxonomic discrimination. In another study, 11 ISSR markers were used for variation analysis among 41 tomato accessions. As a result, 50 bands were generated, whereas 32 of them were polymorphic, representing 63.3% of all the amplified loci. Polymorphism percentage ranged from 50 to 90% and an average number of polymorphic bands of 4.0 was detected. The average genetic diversity index was 0.61, and the genotypes divided into six clusters (Sharifova *et al.*, 2017).

### Start codon targeted (SCoT) marker

Start codon targeted (SCoT) polymorphisms are reproducible markers that are based on the short-conserved region in plant genes surrounding the ATG translation start (or initiation) codon. SCoT markers have been successfully carried out to estimate genetic diversity and structure, identify cultivars, and for quantitative trait loci (QTL) mapping and DNA fingerprinting in diverse species (Etminan *et al.*, 2016). SCoT markers were usually reproducible but exceptions indicated that primer length and annealing temperature are not the sole factors determining reproducibility. SCoT marker PCR amplification profiles indicated dominant markers like RAPD markers (Collard and Mackill 2009a; Thakur *et al.,* 2021).

In the case of using the SCoT marker to diversity in tomato plant, 11 tomato genotypes were carried out using SCoT markers, for selecting genotypes in the breeding programs, a total 94 bands with a mean 11.75 bands per primer were produced while seven SCoT primers were applied. Among these bands, 84 bands were polymorphic, and 14 marker loci was revealed (Habiba *et al.*, 2020). In the study which was conducted by Abdein *et al.* (2018), 63 bands were produced while using seven SCoT primers, among them, 38 (60.3%) bands were polymorphic between eight tomato genotypes. At the same time, they used six ISSR primers, 55 bands were amplified with 26 polymorphic bands. According to the data of PIC in both markers, all traits revealed to be higher in SCoT system. Subsequently, SCoT markers would be a better choice compared to ISSR markers in classification of tomato genotypes. Another study on SCoT marker were evaluated for the diversity in pepper plant which is a member of Solanaceae family, the results revealed five groups in the dendrogram which generated by 10 SCoT markers among 15 pepper accessions and the principal component analysis also identified five genetic clusters. Furthermore, the SCoT markers detected 64 polymorphic loci (NPL), the percentage polymorphic loci (PPL) ranged from 80.00-95.73%, and estimation of gene flow was 3.84. This study showed that SCoT markers may be more useful and informative in measuring genetic diversity and differentiation of the accessions of the genus *Capsicum* (Igwe *et al.*, 2019).

### *Conserved DNA-derived polymorphism (CDDP) marker*

Conserved DNA-derived polymorphism (CDDP) markers are designed specifically to target conserved sequences of plant functional genes. A single primer similar to ISSR and SCoT primers is required for amplification (Collard and Mackill, 2009b). These shorter conserved gene sequences can be found at several regions inside the plant genomes, which giving numerous primer binding sites and is based on sequences encoding short conserved amino acid chains within plant proteins, and this marker can be applied to characterize the genetic relatedness among different genotypes and predicts relations among the conforming phenotypes (El-Mogy *et al.*, 2022). This method has previously demonstrated its efficacy in estimating genetic diversity in a number of plant species (Hajibarat *et al.*, 2015).

The CDDP marker was used for grouping between two tomato cultivars and four wild tomatoes, this tool was helpful to predict the physiological and agronomical behavior of grafting on different tomato rootstocks (El-Mogy *et al.*, 2022).

Abdeldym *et al.* (2020) found that the CDDP marker was more specific to cluster the five accessions of tomato which characterized by better performance under salinity condition. This relatively high degree of accuracy may be because CDDP employs conserved regions of well-known plant gene families mainly involved in response to abiotic and biotic stresses.

El-Mogy *et al.* (2022) for the first-time employed SCoT and CDDP markers to describe the genetic relatedness among some tomato accessions in relation to their drought tolerance. They revealed that SCoT and CDDP systems as effective markers can successfully differentiate between the five tomato genotypes with dominance of SCoT over CDDP. They indicated that SCoT successfully categorized 44% polymorphism, while CDDP determined only 28% polymorphism among the investigated genotypes. The higher polymorphism recorded upon utilizing SCoT may be attributed to the abundance of binding sites targeted by its primers, compared with CDDP which was restricted to conserve regions within certain gene families (Abdeldym *et al.*, 2020). The CDDP revealed the same grouping pattern for genotypes that appeared in the heatmap that was established using phenotypic, physiological, and agronomical criteria under water-deficient conditions. The high precision may be associated with the dependence of CDDP on conserved regions of gene families involved in response to different types of stresses (El-Mogy *et al.*, 2022).

## 2.5 Drought Stress in Tomato Plant

Drought is an important natural phenomenon which affects morphological, physiological, biochemical and yield attributes of tomato plants leading to death. Many abiotic stressors, including as drought, extreme temperature, heavy metals, and high salinity, have significant impacts on tomato production. They result in yield losses of up to 70% (Rajarajan *et al.*, 2023). Among abiotic stresses, drought is considered the most important growth-limiting factor, particularly in arid and semiarid regions (Tahiri *et al.*, 2022). Drought occurs once the plant water necessity cannot be fully supplied. Water is the utmost abundant constituent of the plant body containing 80–95% of fresh biomass and plays a vital role in approximately all physiological features of plant metabolism, growth and development (Kapoor *et al.*, 2020). Generally, drought stress is usually linked to decreased cell growth and proliferation rates, decreased leaf size and shoot height, altered stomatal activity, and restricted nutrient absorption leading to decreased plant production (Kumawat and Sharma, 2018). Drought stress influences physiological activities such as photosynthesis, relative water content, and osmotic adjustment in tomatoes (Rajarajan *et al.*, 2023). Additionally, water deficiency usually reduces the turgor pressure of guard cells, which leads to stomatal closure and subsequent membrane damage. Furthermore, abnormally functioning enzymes, particularly those involved in ATP generation, lower photosynthetic activity (Sharma *et al.*, 2020).

ROS are produced in response to drought stress conditions, which affect cellular redox regulation processes (Ibrahim *et al.*, 2020a). Water deficit conditions trigger several defense responses to influence water use efficiency and to mitigate drought-induced damages (Kapoor *et al.*, 2020). In addition, stressed plants gradually develop advanced drought tolerance strategies including encouragement of biosynthesis of compatible solutes and enhancement of the enzymatic and non-enzymatic components of the antioxidant apparatus. Moreover, stressed plants increasingly adopt sophisticated drought resistance techniques, such as promoting the biosynthesis of compatible solutes and enhancing the enzymatic and non-enzymatic antioxidant apparatus components (Ibrahim *et al.*, 2020b). At the morphological level, the root is the major driver of water in most plant forms; hence, it is an important feature encouraging plant response to drought stress (Salehi-Lisar *et al.*, 2016).

During water stress, many physiological and molecular processes are disturbed such as root-shoot growth, water relation, mineral absorption, leaf expansion and direction, stomatal closure, transpiration rate, photosynthesis and respiration rates, solute translocation, etc (Jangid *et al.*, 2016). Toxic elements such as ROS, produced during stress period create oxidative damage to the cellular organization. Tomato like other plants has its antioxidant system to scavenge such harmful elements and accumulate osmoprotectants such as proline, glycine betaine, etc to maintain osmotic adjustment (Khan *et al.*, 2015).

George *et al.* (2013) examined tomato to drought stress using 4% PEG6000 on 10 different tomato varieties at seedling stage, several parameters were evaluated, for example germination percentage was slightly influenced by stress, while growth was significantly affected. Relative water content significantly reduced from 89.28% to 87.73%, under control and drought conditions, respectively, proline content was increased from 4.4 µmoles g of fresh weight under controlled condition to 5.8 µmoles/g of fresh weight under drought condition.

Cui *et al.* (2020), evaluated the response of tomato plant to drought stress in China. Tomato plants were drip-irrigated to 100% field capacity at all growth stages, with treatment (control) receiving half the amount of irrigation when the soil water content reached 70% field capacity, the vegetative phase (stage I), the flowering and fruit development phase (stage II), and the fruit ripening phase (stage III). Drought stress at stages II and III reduced yield by 13% and 26%, respectively, when compared to the control treatment. Fruit hardness and color index were favorably affected by drought stress, although fruit water content and shape index showed no differences between treatments. In response to limited water supply, taste and nutritional quality measures such as total soluble solids, soluble sugar, organic acids, and ascorbic acid improved. Despite having a negative effect on fruit yield, drought stress applied at stage III tended to improve fruit quality traits. They also discovered that applying drought stress at stage I can be a positive management approach because it saves water and has fewer negative effects than applying drought stress at other critical growth stages, thereby minimizing the adverse effects of drought stress (Cui *et al.*, 2020).

### 2.5.1 Drought stress defense mechanisms

Drought tolerance is the result of a combination of three different defense mechanisms, notably escape, avoidance, and tolerance (Kumar *et al.*, 2012b; Chatterjee and Solankey, 2015). The escapement mechanism is accomplished by plants' ability to complete their life cycle prior to the initiation of drought stress. This response includes shorter time periods for the various phenological stages, which means that plants mature in less time (Kumar *et al.*, 2012b). The avoidance mechanism, on the other hand, is achieved through increased water absorption and reduced water losses from cells during drought periods, resulting in high water potential in plant tissues. This is accomplished through a variety of processes, including the reduction of canopy and leaf area, which results in a reduced perception of solar radiation and, as a result, reduced transpiration. This mechanism also includes stomatal closure, the formation of cuticular wax, and changes in root density and length (Giordano *et al.*, 2021). Finally, plants can withstand drought stress if they maintain cellular turgor and water loss in the face of low water potential and moisture deficiency. This can be accomplished through solute accumulation in the cytoplasm, which increases the elasticity of cell membranes, as well as cell size reduction (Kumar *et al.*, 2012b). Plant defense mechanisms against oxidative stress include bioactive molecules such as tocopherols, ascorbate, glutathione, carotenoids, and flavonoids, as well as enzymes such as superoxide dismutase (SOD), catalase (CAT), and others (Raza *et al.*, 2020), additionally, abscisic acid (ABA), salicylic acid (SA), jasmonic acid (JA), and ethylene are examples of phytohormones. Phytohormones regulate a wide range of physiological and developmental processes via signaling pathways (Raza *et al.*, 2019; Giordano *et al.*, 2021). Abscisic acid, for example, was found in high concentrations in plants exposed to abiotic stressors. Under water-stress conditions, ABA induces stomatal closure, while the same hormone also regulates transpiration and the activity of some genes via a pathway involving *SnRK2/OST1, PP2C* (protein phosphatases), and *PYR/PYL/RCAR* proteins (Raza *et al.*, 2019). Moreover, salicylic acid regulates the activity of other stress hormones involved in stress, whereas ethylene influences seed germination and plant growth under abiotic and biotic stresses (Giordano *et al.*, 2021).

**2.5.2 The effect of biostimulants on the drought stress**

An agronomic tool of increasing interest is the use of different formulations of certain organic materials and microorganisms, defined by the term biostimulants. Biostimulants are plant-promoting substances/microorganisms derived from organic materials that are applied to soil to increase nutrient uptake, stimulate plant growth, increase tolerance to abiotic and biotic stresses, and improve product quality (Bradáčová *et al.*, 2016). Biostimulants are usually grouped into different families based on the raw materials used for their production: humic substances, complex organic materials, beneficial chemical elements (e.g., silicon), inorganic salts, algae and plant extracts, protein hydrolysates, chitin and chitosan derivatives, antiperspirants (e.g., kaolin), amino acids and other compounds (Rouphael and Colla 2020). Seaweed extracts, humic acid (HA), fulvic acid (FA), phosphite, arbuscular mycorrhizal fungi, and/or plant growth-promoting rhizobacteria (PGPR) are used as biostimulants to increase plant production and mitigate the effects of abiotic stresses (Colla and Rouphael, 2015; Colla *et al.,* 2015). These substances can improve plant stress tolerance, crop nutrient use efficiency, the bioavailability of nutrients in the soil or rhizosphere and quality traits. For the above-mentioned reasons, biostimulants can benefit crops when applied under optimal environmental conditions and in states of abiotic and biotic stress (Del Buono *et al.,* 2023). The PGPR family such as *Entrobacte, Bacillus, Azospirillum, Serratia, Burkholderia, Arthrobacter, Azotobacter, Klebsiella, Pseudomonas* and *Alcaligenes*, effectively promote plant growth and development (Turan *et al.,* 2021)

New management strategies are needed to resolve and mitigate abiotic stresses, particularly, drought stress. Biofertilizers are very important to improve soil quality, plant growth, and rationalizing water use. The role of beneficial rhizobacteria, and mycorrhizal fungi, in improving tomato growth and yield, and drought tolerance are crucial (Anli *et al.*, 2020). Tahiri *et al.* (2022) indicated that, in tomato plants, water stress negatively affected on plant growth and yield traits, and unbalanced the antioxidant enzymes. Application of biofertilizers decreased the negative effects of drought stress. For instance, significant increases in shoot biomass and fruit number per plant were resulted compared to the control (Tahiri *et al.*, 2022).

Regarding the tomato fruit quality, biofertilizers have positive effects on sugar and protein contents. According to the antioxidant enzymes, significant reductions in polyphenol oxidase, peroxidase, catalase, and superoxide dismutase activities in roots were recorded. Tahiri *et al.* (2022) stated that beneficial microorganisms enhanced the water stress tolerance of tomato plants by improving plant growth, osmolyte accumulation, and mineral accumulation (Tahiri *et al.*, 2022).

Turan *et al.* (2021) indicated that the different commercial plant biostimulants (Powhumus® (PH), Huminbio Microsense Seed® (SC), Huminbio Microsense Bio® (RE), and Fulvagra® (FU)) were used as seed coatings and/or drench solutions in tomato plant, all biostimulants improved the plant growth and yield compared with the control and had positive effects on the growth of cherry tomato in fertile soil and under stressed conditions (Turan *et al.*, 2021). Another studies revealed that seed germination, seedling growth, plant height, shoot weight, nutrient content, bloom period, and chlorophyll content were increased while tomato plants were inoculated with PGPR (Yildirim, 2007; Yildirim *et al.*, 2015).

### 2.5.3 The role of biotechnology in the resistance tomato plants to drought stress

Drought tolerance enhancement through traditional genetic improvement techniques has less effective, it needs a long time, and requires huge resources (Rajarajan *et al.*, 2023). Recently, the omics approaches were discovered to be the most effective tool for more precisely investigating the mechanism of drought tolerance (Chaudhary *et al.*, 2019). Modern sequencing technologies have substantially accelerated tomato genomes and transcriptomics research. Transcriptomic analysis, in particular, can be more effective in understanding the genes and pathways involved in stress tolerance (Dai *et al.*, 2017). It also enables for the identification of more microsatellites, which are characteristics that can assist in large-scale genotyping. New genomics-based breeding technologies even as genotyping by sequencing, genome-wide association studies, genomic selection, and SNPs are effective tools for genotyping the genetic resources for diverse trait improvements, including drought stress tolerance  (Sim *et al.*, 2012). These approaches can also identify drought-tolerant QTLs for efficient QTL introgression to elite lines. Furthermore, genome-editing techniques such as Crisper/Cas9 and RNAi have emerged as important tools for enhancing drought stress tolerance in tomato at multiple levels (Liu *et al.*, 2020). Consequently, overexpression of drought-responsive genes and transcription factors (TFs) had a significant impact in resistance development in tomato varieties (Rajarajan *et al.*, 2023).

In various species, genes responsible for changes in physiological and morphological traits during drought stress have been identified. For example, the activity of many genes and the expression of dominant alleles of those genes determine root length and number, whereas recessive alleles determine root thickness (Kumar *et al.*, 2012b).

Giordano *et al.* (2021) indicated that in tomato plants, *DREBs/CBFs* and *ABF3* genes, encode transcription factors that confer tolerance to drought, cold, and salt stress. The *SNAC1* gene encodes transcription factors involved in the increased sensitivity of stomata to less water loss. Under drought conditions, the *ERA1* gene reduces stomatal conductance. The *Mn-SOD* gene is involved in the synthesis of Mn-superoxide dismutase, which confers stress tolerance. The *AVP1* gene is involved in root development. *P5CS* and *mtlD* genes contribute to osmotolerance by accumulating proline and mannitol. The *GF14l* gene contributes to an increase in photosynthetic rate and tolerance to water deficit during drought stress. The *NADP-Me* gene is involved in reducing stomatal conductance and increasing water use efficiency (WUE). The *wilty* gene is involved in the wilting of tomato leaves under drought stress (Giordano *et al.*, 2021).

In the case of the genes related to drought stress in tomato, Liu *et al.* (2021) indicated that repressing *SlGRAS4* (*SlGRAS4*-RNAi) increased sensitivity to drought stress, whereas overexpressing *SlGRAS4* (*SlGRAS4*-OE) in tomato enhanced tolerance to this stress. Under stress conditions, *SlGRAS4*-OE plants accumulated much less ROS than wild-type and *SlGRAS4*-RNAi plants. Zhang *et al.* (2011) found that drought stress induced the accumulation of *Sly-miR169* in tomato plants, and that over-expression of a *miR169* family member, *Sly-miR169c*, in tomato plants can efficiently down-regulate the transcripts of the target genes. Transgenic plants over-expressing *Sly-miR169c* demonstrated reduced stomatal opening, decreased transpiration rate, decreased leaf water loss, and improved drought tolerance when compared to non-transgenic plants. Another study found that the *sly-miR159* regulatory function in tomato plants' responses to different stresses may be mediated by stress-specific *MYB* transcription factor targeting. The accumulation of the osmoprotective compounds, proline and putrescine, which promote drought tolerance, was associated with *sly-miR159* targeting of the *SlMYB33* transcription factor transcript. This highlights the potential role of *sly-miR159* in tomato plant adaptation to water deficit conditions (López-Galiano *et al.*, 2019). When plants are subjected to drought stress, the metabolism of tomato fruits changes. *sly-miR10532* and *sly-miR7981e* inhibit the expression of mRNAs encoding galacturonosyltransferase-10, the main enzyme in pectin biosynthesis, whereas *sly-miR171b-5p* targets -1,3-glucosidase mRNAs involved in glucan degradation. These results allow for the systematic characterization of miRNA and their target genes in tomato fruit under drought stress conditions (Asakura *et al.*, 2022).

Zhu *et al.* (2018) stated that the basic leucine zipper transcription factor *SlbZIP1* plays an important role in salt and drought stress tolerance by modulating an ABA-mediated pathway, and that *SlbZIP1* may have applications in the engineering of salt- and drought-tolerant tomato cultivars. Abdellatif *et al.* (2023) indicated that phytochromes (*PHYS*) are essential photoreceptors in plants that regulate plant growth and development and are implicated in plant stress response. Drought tolerance was demonstrated by *phyA* and *phyB (B1 and B2)* mutants, as inhibition of electrolyte leakage and malondialdehyde accumulation, indicating decreased membrane damage in the leaves. Both *phy* mutants also reduced oxidative damage by increasing the expression of ROS scavenger genes, inhibiting hydrogen peroxide ($H_2O_2$) accumulation, and increasing the percentage of antioxidant activities as measured by the 2,2-diphenyl-1-picryl-hydrazyl-hydrate (DPPH) test.

## 2.6 Heavy Metal Stress in Plants

There has been a worldwide problem with increased heavy metal pollution in farmland over the last half-century. This is largely due to anthropogenic practices such as specific industrial enterprises, mining, and the inappropriate use of agricultural products and practices that release large amounts of heavy metals, such as wastewater irrigation, the use of factory liquid wastes, and the addition of chemical fertilizers, pesticides, and sewage sludge (Eid *et al.*, 2021).

Heavy metals such as copper (Cu), manganese (Mn), lead (Pb), cadmium (Cd), nickel (Ni), cobalt (Co), iron (Fe), zinc (Zn), chromium (Cr), iron (Fe), arsenic (As), silver (Ag) and the platinum (Pt) accumulated in soils through industrial wastes and sewage disposals. Even though some of these metals are essential micronutrients responsible for many regular processes in plants, influence the plant growth, and plant metabolism. Plants have different mechanisms to fight stress, and they are responsible to maintain homeostasis of essential metals required by plants (Ghori *et al.*, 2019).

### 2.6.1 Cadmium stress in plants

Cadmium (Cd) is one of the heavy metals that disrupt plant biophysiological functions and the most stressful heavy metals to plants. Large amounts of Cd affect the physiological functions and morphological characteristics of most crop species. Many studies have found that Cd contamination reduces photosynthetic efficiency, nutrient uptake, biomass production, and crop yield. Furthermore, high Cd levels have a negative impact on root growth and architecture, which affects nutrient uptake and mobility in plant tissues (Rizwan *et al.*, 2017; Chtouki *et al.*, 2021).

Cadmium occurs in a variety of forms in soil; furthermore, many of them are not available for plant uptake. Cd must be available for uptake in order to be absorbed, which is determined by the metal's speciation, plant species, and soil physicochemical conditions. This metal is easily absorbed and transported to the aerial parts of the plant. Because of morphological variation, physiological characteristics of the plant, and plant growth stages and age, the ability of Cd to be absorbed varies among plant species and genotypes (Asati *et al.*, 2016; El Rasafi *et al.*, 2022). There have been numerous studies on the role of Cd on fresh and dry mass accumulation, height, root length, leaf area, and other plant characteristics. In majority of plants, Cd toxicity has a negative impact on plant growth (height) and chlorophyll content (SPAD values). Many studies have revealed that Cd is highly phytotoxic, limiting plant growth and sometimes leading to plant death. Cd inhibits plant growth and development by raising the dry to fresh mass (DM/FM) ratio in all organs. In addition, Cd toxicity causes a loss of yield and a reduction in plant productivity (Farid *et al.*, 2013). Another study has stated that this metal may prevent seed germination, reduce total plant length, suppress root elongation, and decrease the number of leaves per plant, ultimately leading to plant death (El Rasafi *et al.*, 2022).

Cd is phytotoxic, even at low concentrations that can be easily transferred from contaminated soil to various plant organs, and its accumulation causes toxic effects on plants as well as animals and humans via the food chain. Cd accumulation in plants may reduce chlorophyll content, thereby inhibiting photosynthesis and restricting plant growth. Cd changes the redox potential of the cell by increasing the formation of ROS, which causes oxidative damage to cell membranes and other biomolecules. Many studies have been conducted to investigate the effects of Cd on morphological, physiological, and biochemical changes in various plant species (Shanmugaraj *et al.*, 2019).

Under Cd stress, physiological and biochemical changes occur frequently. The most obvious changes are variations in the components of gas exchange characteristics, proline, malondialdehyde (MDA), sugars, proteins, and enzyme activity. The change is caused by an excess of free radicals, enzyme inhibition, and/or nutrient deficiency (Rizwan *et al.*, 2017; Hasanuzzaman *et al.*, 2018).

Cadmium stress in plants also causes significant changes in enzyme activity as a result of oxidative stress in plant cells. Excess Cd causes cells to overproduce ROS such as $H_2O_2$, $O_2$, and OH. Free radical formation increases the activity of several enzymes, including peroxidase (POD), superoxide dismutase (SOD), ascorbate peroxidase (APX), and catalase (CAT) (Gupta *et al.*, 2019; Pandey and Dubey, 2019).

Plants' antioxidant system is an important defense strategy for dealing with increased toxic metal levels, including Cd. Under Cd exposure, this system undergoes a number of changes that confer Cd tolerance and the ability to detoxify Cd. Catalase, ascorbate peroxidase, monodehydroascorbate reductase, superoxide dismutase, dehydroascorbate reductase, glutathione reductase, glutathione s-transferase, and glutathione peroxidase are examples of enzymatic antioxidants, while glutathione, α-tocopherols, phenolic compounds, ascorbate, non-protein amino acids, and alkaloids are non-enzymatic antioxidants. This complex system is activated to protect the plant from ROS by converting them to fewer toxic products, thereby assisting in the maintenance of plant cell redox equilibrium. The antioxidant defense system also removes free radicals, protecting the structure and functions of the plant cell membranes (Song *et al.*, 2017).

Plant hormones influence plant growth and development and also serve as a warning sign of metal toxicity. Auxin (IAA), gibberellins (GA$_3$), cytokinin (CK), salicylic acid (SA), ethylene (ET), abscisic acid (ABA), jasmonic acid (JA), brassinosteroids (BRs), and strigolactones (SLs) are the major groups of phytohormones that respond to Cd toxicity. Plant hormones are active at low concentrations and play an important role in plant color, taste, and smell development as well as cellular process regulation. Plant hormones are also required for stress adaptation and defense against abiotic stressors such as metal toxicity (El Rasafi *et al.*, 2022).

**2.6.2 Cd stress in tomato**

Cadmium (Cd) contamination endangers human health. To limit human Cd intake, screening and breeding low-Cd absorption tomato cultivars is critical. The effect of cadmium (Cd) on the vegetative and reproductive growth of tomato was shown by Rehman *et al.* (2011), they used 10, 20, 30, and 40 µg cadmium concentrations twice during the pre-flowering and post-flowering stages, and the results showed that higher doses of Cd increased total chlorophyll content while decreasing plant biomass. The concentration of cadmium had a negative correlation with the number and area of leaves. Shekar *et al.* (2011) indicated that Cd has an effect on the tomato at different stages of growth and development. A lower cadmium concentration increased the percentage of germination, survival percentage, plant height, root length, early flowering, more pollen viability, and total chlorophyll content, while a higher concentration inhibited all biomass characteristics. When ten tomato cultivars (K-25, K-21, NTS-9, Kaveri, NBR-Uday, Swarnodya, Sarvodya, NBR-Uttam, Malti, and S-22) were exposed to different concentrations of $CdCl_2$ (0.0, 50, 100, or 150 µM), all growth and photosynthetic characteristics were reduced. After 30 days, results showed that the tolerance of all cultivars to Cd stress varied. The genotypes of K-25, K-21 and NTS-9 displayed the maximum resistance to cadmium stress, while the genotypes of Sarvodya, NBR-Uttam, and Malti experienced severe damages (Hasan *et al.,* 2009). In the case of Cd accumulation in tomato plant parts, Gratão *et al.* (2015) stated that the majority of Cd accumulated in the root when compared to the leaves. Nogueirol *et al.* (2016) evaluated the response of the production, nutritional, and enzymatic antioxidant systems of two tomato genotypes (Calabash Rouge and CNPH 0082) by using different Cd levels (0, 3, 6, and 12 mg/kg of soil). Cadmium treatment resulted in decreased biomass of shoots and roots in both genotypes and nutritional imbalances, mainly in terms of nitrogen (N), phosphorus (P), and manganese (Mn) metabolism. Cd exposure increased the content of malondialdehyde (MDA) and hydrogen peroxide ($H_2O_2$) in tomato plant tissues, as well as the activity of catalase, ascorbate peroxidase, and guaiacol peroxidase (Nogueirol *et al.*, 2016).

Hana *et al.* (2008) showed the effect of Cd on anti-oxidative enzymes in tomato by using different $CdCl_2$ concentrations (0, 20, 40, 80, 100, and 200 µM). Ascorbate peroxidase (APX) and guaiacol peroxidase (GPX) activities showed an increase below 100 µM concentration after treatment. However, at concentrations greater than the determined level, a significant decrease in enzyme activity was observed. The increase in enzymatic activity can be associated with the induction of oxidative stress by cadmium treatment.

In the case of genes related to Cd stress tolerance in tomato, in response to cadmium (Cd) stress, *SlRING1* expression was highest. Under Cd stress, silencing *SlRING1* significantly reduced chlorophyll content and biomass accumulation. $H_2O_2$ and malondialdehyde levels were significantly higher in *SlRING1*-silenced plants under Cd stress compared to non-silenced tomato plants. Furthermore, Cd accumulation in shoots and roots was significantly higher in *SlRING1*-silenced tomato plants than in non-silenced tomato plants. Overall, *SlRING1* plays an important role in tomato plant tolerance to Cd stress (Qi *et al.*, 2020; Ahammed *et al.*, 2021). *HsfA1a,* a transcription factor, attained Cd tolerance to tomato plants, in part by provoking melatonin biosynthesis in response to Cd stress. The analysis of leaf phenotype, chlorophyll content, and photosynthetic efficiency revealed that silencing the *HsfA1a* gene decreased Cd tolerance while overexpression increased Cd tolerance (Cai *et al.,* 2017).

## 2.7 Plant Parasitic Nematodes (PPNs)

Nematode is one of the largest and widely distributed groups of animals in marine, freshwater, and terrestrial environments. Their numerical dominance, often exceeding 1 million individuals per square meter and accounting for roughly 80% of all individual animals on the planet (Bird and Bird, 2012). Their diverse lifestyles and presence at various trophic levels indicate that they play an important role in many ecosystems. *Caenorhabditis elegans* is its most well-known representative: the first animal whose genome was completely sequenced. Aside from bacterivorous nematodes like *C. elegans*, there is a diverse range of trophic ecologies present, including fungal feeding, predation, and parasitism of plants, invertebrates, higher animals, and humans (Félix and Braendle, 2010).

Plant parasitic nematodes (PPNs) attack the majority of economically important crops, causing global yield losses of up to 12.3% on average. Certain crops may experience losses of up to 30%. PPNs are obligate biotrophs that feed on nearly all plant tissues, including flowers, roots, stems, and leaves, but the majority of PPN species feed on roots. Regardless of their feeding habits, all PPNs have a specialized mouth spear called a stylet that allows them to penetrate cell walls and feed on plant cells (Holbein *et al.*, 2016).

The most well-known plant parasitic nematodes are cyst (*Globodera and Heterodera* spp.) and RKN (*Meloidogyne* spp.), which cause significant damage to crops such as soybean, potato, tomato, and sugar beet (Holterman *et al.*, 2006). The two most important groups are root-knot nematodes (*Meloidogyne* spp.) and lesion nematodes (*Pratylenchus* spp.), which can infect, feed on, and reproduce on a variety of crops and plant species. *Meloidogyne incognita*, a tropical RKN, is a polyphagous species that has been dubbed the world's most damaging crop pathogen. Crop losses caused by nematodes are difficult to quantify, with global estimates ranging from $US80 billion to $US157 billion per year. Nematodes were estimated to cause annual crop losses of $US10 billion

using reliable data from the United States, compared to $US6.6 billion for insect pest losses (Coyne *et al.*, 2018).

Nonparasitic nematodes are useful indicators of soil biological condition because this ecologically diverse group exhibits a wide range of sensitivity to environmental stresses and plays important roles in the soil food web (Holterman *et al.*, 2006).

### 2.7.1 Tomato root-knot nematodes (RKN)

Many pests, including plant pathogenic nematodes, attack tomatoes and cause severe growth retardation. Root-knot nematode (*Meloidogyne* spp.) is ranked first among major plant pathogens and first among the world's ten most important genera of plant parasitic nematodes. The RKN has a wide geographic distribution, a diverse host range, and a high destructive potential. In spite of being one of the most important plant parasitic nematodes that is associated with low tomato production (Mukhtar, 2018). *Meloidogyne* is a genus of over 80 species that includes the plant parasitic nematodes that are the most economically damaging to crop production on a global scale. The most common species of this genus are *Meloidogyne incognita*, *Meloidogyne javanica*, *Meloidogyne arenaria, Meloidogyne chitwoodi, Meloidogyne fallax,* and *Meloidogyne hapla* which account for more than 95% and these are the most widely distributed species (Adam *et al.*, 2007). *Meloidogyne* species have been associated with tomatoes for centuries. Several studies have been conducted to evaluate the root-knot nematode's potential damage on various tomato cultivars; its yield loss potential ranges from 25% to 100%. Commercial cultivars and rootstocks containing the *Mi* gene have been successfully used to manage *Meloidogyne incognita, M. javanica,* and *M. arenaria* (Seid *et al.*, 2015; Regmi *et al.*, 2020). Bozbuga *et al.* (2020) found that the *Mi* gene is very important to control root-knot nematodes, for this purpose, 99 tomato genotypes were screened for *Mi* gene resistance against *Meloidogyne incognita.* They indicated that only one genotype, among 99 tomato genotypes, was determined this gene and this genotype showed an immune reaction against nematode.

Peng and Kaloshian (2014) stated that both genes of *SlSERK3A* and *SlSERK3B* in tomato plants have an important role in defense system to invade the RKN and silencing either *SlSERK3A* or *SlSERK3B* resulted in increased susceptibility to this parasite. Vos *et al.* (2013) findings were indicated that arbuscular mycorrhizal fungi (AMF) have excessive potential as biocontrol organisms against the RKN, *Meloidogyne incognita* which causes severe root gall formation in plants, however, knowledge of the underlying molecular mechanisms involved in nematode biocontrol is limited.

According to the use of plant extracts against nematode infection on tomato, fresh leaf extracts of *Azadirachta indica* (Neem), *Allium sativum* (Garlic) and *Tagetes erecta* (African marigold) were used against *Meloidogyne incognita* on tomato under *in vitro*, pots, and field conditions. The results showed that neem leaf extract had the most effect on immobilized juveniles (J2), and garlic leaf extract proved to be the best control to reduce root galls by 57% in pots under greenhouse conditions and 33% in field conditions, while increasing fruit yield by 47%. When compared to these plant extracts, using the nematicide resulted in the greatest reduction of nematode populations (Abo-Elyousr *et al.*, 2010).

# CHAPTER THREE
# MATERIALS AND METHODS

## 3.1 Plant Materials

The mature fruits of 64 tomato accessions were collected during summer 2019, from six provinces of Kurdistan region: Sulaymaniyah, Erbil, Duhok, Garmian, Raparin, and Halabja. Morphological markers such as size, form, color of fruit and plant size were used to identify and collect different accessions (Mazzucato *et al.*, 2008). At the time of collection, mature fruits were harvested from each accession, and then seeds were gathered and dried (Dias *et al.*, 2006). Each accession was coded based on the name of the collection site (Table 3.1).

**Table 3.1 Codes, names, and locations of collection of 64 tomato accessions.**

| Accession code | Local name | Province | Accession code | Local name | Province |
|---|---|---|---|---|---|
| AC1 | Barsim | Garmian | AC33 | Sunweak | Halabja |
| AC2 | Kurdi Gawray Swr | Garmian | AC34 | Pamayi Wrd | Halabja |
| AC3 | Dhahabi | Garmian | AC35 | Mulayin | Halabja |
| AC4 | Pamayi Wrd | Garmian | AC36 | Rozh | Halabja |
| AC5 | Karazi | Garmian | AC37 | Sirin | Halabja |
| AC6 | Swri Wrd | Sulaymaniyah | AC38 | Israili Sharazur | Halabja |
| AC7 | Sirin | Sulaymaniyah | AC39 | Pamayi Wasat | Halabja |
| AC8 | Israili | Sulaymaniyah | AC40 | Roma | Halabja |
| AC9 | Wrdi Be Tow | Sulaymaniyah | AC41 | Wrdi Gallapan | Halabja |
| AC10 | Dhahabi | Sulaymaniyah | AC42 | Swri Wrd | Halabja |
| AC11 | Bakrajo | Garmian | AC43 | Pamayi Kurdi | Garmian |
| AC12 | Slemani | Garmian | AC44 | Sewi Hawler | Erbil |
| AC13 | Braw | Garmian | AC45 | Dhahabi Wasat | Sulaymaniyah |
| AC14 | Amad | Garmian | AC46 | Kurdi Pamayi Gawra | Sulaymaniyah |
| AC15 | Pamayi Wasat | Garmian | AC47 | Pamayi Wrd | Sulaymaniyah |
| AC16 | Charmo | Garmian | AC48 | Gallapan | Sulaymaniyah |
| AC17 | Sangaw | Garmian | AC49 | Pamayi Wasat | Sulaymaniyah |
| AC18 | Kurdi Gawray Swr | Garmian | AC50 | Balami Raniya | Raparin |
| AC19 | Swri Hanjiri | Duhok | AC51 | Pamayi Sarsawz | Raparin |
| AC20 | Hanin | Duhok | AC52 | Pamayi Gawra | Raparin |
| AC21 | Jersey | Duhok | AC53 | Kurdi Gawray Swr | Raparin |
| AC22 | Barsim Towdar | Duhok | AC54 | Balami Qaladze | Raparin |
| AC23 | Wrdi Heshui | Duhok | AC55 | Kurdi Pamayi | Raparin |
| AC24 | Sewi | Duhok | AC56 | Sewi Qaladze | Raparin |
| AC25 | Kurdi Gawray Swr | Erbil | AC57 | Pamayi Wasat | Raparin |
| AC26 | Dhahabi | Erbil | AC58 | Kurdi Balakayati | Raparin |
| AC27 | Pamayi Wrd | Erbil | AC59 | Barsim | Raparin |
| AC28 | Kurdi Gawray Swr | Halabja | AC60 | Kurdi Pshdar | Raparin |
| AC29 | Ibrahim | Halabja | AC61 | Raza Pashayi | Garmian |
| AC30 | Yadgar | Halabja | AC62 | Helakyi Raq | Halabja |
| AC31 | Balami | Halabja | AC63 | Sandra | Halabja |
| AC32 | Super | Halabja | AC64 | Balami Sharazur | Sulaymaniyah |

### 3.1.1 Seedling preparation and plant growth under greenhouse conditions

Tomato seedlings of 64 accessions were prepared by sowing the seeds in plastic trays filled by peatmoss and growing the seedlings in a greenhouse. After one month, the seedlings were transferred into the greenhouse, and the plants grew during the spring and summer seasons of 2020 at the university of Sulaimani, college of Agricultural Engineering Sciences, in Bakrajo. The plants of all accessions were irrigated and pruned as necessary (Appendix 1). Table 3.2 depicts the greenhouse soil properties.

**Table 3.2 Greenhouse soil properties.**

| Soil properties | Unites | Value |
|-----------------|--------|-------|
| Sand | | 142.4 |
| Silt | $g\ kg^{-1}$ | 430.6 |
| Clay | | 427.0 |
| Texture | ----- | Silty Clay |
| pH | ----- | 7.95 |
| EC | $dS\ m^{-1}$ | 1.1 |
| Total N | $g\ kg^{-1}$ | 15.6 |
| Available P | $mg\ kg^{-1}$ | 4.43 |
| Soluble K | $g\ kg^{-1}$ | 65.6 |
| Available Fe | $mg\ kg^{-1}$ | 3.08 |
| Available Zn | | 1.74 |
| Organic Matter | $g\ kg^{-1}$ | 12.3 |
| CaCO₃ | | 256 |

### 3.1.2 Morphological characteristics and tomato fruit traits

Morphological and fruit characteristics were recorded on eight random tomato plants, such as plant height (PH-cm), root length (RL-cm), plant dry weight (DW-g) leaf area (LA-cm²) (using Digimizer software) , total chlorophyll content (TCC-SPAD) (using CCM-200 Plus, OPTI-Sciences) (Jiang *et al.*, 2017), fruit weight (FW-g), fruit size (FS-cm³), fruit thickness (FT-mm), fruit diameter (FD-mm), total fruit weight per plant (TFW-g), fruit moisture content (MC-%), total solids (TS-%), total soluble solids (TSS-%), non-soluble solids (NSS-%) and fruit firmness (FF-g/cm²). The firmness was determined using Brookfield equipment (Brookfield CT3 Texture Analyzer, USA), The moisture contents of the samples were determined using the oven drying method. Before drying, the weight of the fruit samples and empty glass petri-dishes were recorded, as was the weight of the glass petri-dishes and samples after 72 hours of oven drying at 70 °C. The following formula was used to calculate the moisture contents of the fruits:

$$MC\% = \frac{FW - DW}{FW} X\ 100$$

Where FW and DW are the weights of fresh and dried fruits, respectively. TSS was determined using a standard procedure. Fruit juice was recovered by pulping and crushing the fruits. A handheld refractometer (ATAGO Pocket PAL-2, Japan) was used to measure the TSS in the juice. Following the cleaning and calibration of the refractometer, a known volume of juice (drop) was put on top of the refractometer at the designated spot. The Brix unit was used to express TSS results (Xu *et al.*, 2013; Tatelbaum, 2014; Eltom *et al.*, 2017; Dono *et al.*, 2020). The TS and NSS were determined by the following equations:

**TS (%) = 100 – MC%**

**NSS (%) = TS% – TSS%**

### 3.1.3 Plant material preparation for DNA extraction

Seeds of all tomato accessions were sown in small pots at the College of Agricultural Engineering Sciences, University of Sulaimani in March of 2019, under glasshouse conditions. Fresh leaves were harvested after four weeks and, then ground with liquid nitrogen. DNA from all accessions was extracted according to the cetyl trimethyl ammonium bromide (CTAB) protocol (Tahir, 2015). The quality and quantity of extracted DNA were checked on a 1% agarose gel and by a nanodrop spectrophotometer (NanoPLUS-MAANLAB AB, SWEDEN).

### 3.1.4 ISSR, SCoT, and CDDP assays

Thirteen ISSR primers (Isshiki *et al.*, 2008; Lata *et al.*, 2010; Sharifova *et al.*, 2017), 26 SCoT primers (Feng *et al.*, 2018; Ahmed *et al.*, 2020), and 15 CDDP primers (Collard and Mackill 2009b; Ahmed *et al.*, 2020; Rasul *et al.,* 2022; Tahir *et al.,* 2023) were used for genetic diversity analysis in tomato accessions. These primers were chosen based on their polymorphism rates in previous studies in tomato and other crops (Table 3.3). PCR amplification was performed in a reaction containing 4 µL of template DNA (100 ng), 10 µL PCR master mix (AddStart Taq Master, Addbio, Korea), 4 µL primer (Addbio, Korea), and 7 µL of deionized water. The PCR reaction was performed in an Applied Biosystems thermocycle machine as follows: initial denaturation at 94 °C for 9 minutes, followed by 36 cycles of denaturation at 94 °C for 1 minute, annealing at a specific temperature (depending on primer) for 1 min, and extension at 72 °C for 2 minutes. A final extension cycle at 72 °C for 8 minutes was followed. PCR products were separated on 1.6% agarose gels and stained with ethidium bromide.

### 3.1.5 Statistical data analysis for morphological and fruit traits, and molecular data

For morphology and fruit characteristics data of tomato, XLSTAT version 2019 was used for one-way analysis of variance, and Duncan's multiple range test was used to analyze differences between means (P ≤ 0.01). Principal component analysis (PCA) and hierarchical cluster plots (using the Ward method) were created from fruit traits data by XLSTAT version 2019 and JMP Pro 16 software, respectively. Regarding molecular analysis, the amplification bands of each primer were scored and coded manually by recording 0 and 1 for the absence and presence of bands, respectively. PowerMarker version 3.25 software was used for calculating polymorphism information content (PIC) and gene diversity (GD). Using XLSTAT version 2019 and CLC sequence viewer 8, an unweighted pair-group technique with arithmetic averages (UPGMA) was used to define accessions groups and calculating the genetic distance. For the analysis of population structure, a Bayesian model-based analysis was performed using STRUCTURE 2.1 software (Pritchard *et al.*, 2000). A Mantel statistic test was used to compare the similarity matrices using XLSTAT version 2019, and finally, GenAlEx 6.5 software was applied for genetic variation within and among accessions. Gene flow was calculated via PhiPT value by using the following formula: [(1/PhiPT)-1]/4, where PhiPT is the population variance divided by the total genetic variations (Mekonnen *et al.*, 2020).

**Table 3.3 ISSR, SCoT and CDDP primers used in this study, and their sequences and annealing temperatures.**

| ISSR Primers | Sequence of primer (5′ – 3′) | Annealing temperature (°C) | SCoT Primers | Sequence of primer (5′ – 3′) | Annealing temperature (°C) |
|---|---|---|---|---|---|
| UBC-808 | AGAGAG AGAGAGAGAGC | 50.00 | SCoT1 | CAACAATGGCTACCACCA | 49.86 |
| UBC-810 | GAGAGAGAGAGAGAGAT | 50.00 | SCoT2 | CAACAATGGCTACCACCC | 50.70 |
| UBC-812 | GAGAGAGAGAGAGAGAA | 50.40 | SCoT3 | CAACAATGGCTACCACCG | 51.27 |
| UBC-814 | CTCTCTCTCTCTCTCTA | 50.00 | SCoT4 | CAACAATGGCTACCACCT | 49.50 |
| UBC-815 | CTCTCTCTCTCTCTCTG | 50.00 | SCoT5 | CAACAATGGCTACCACGA | 50.10 |
| UBC-818 | CACACACACACACACAG | 52.80 | SCoT6 | CAACAATGGCTACCACGC | 52.05 |
| UBC-822 | TCTCTCTCTCTCTCTCTCA | 50.00 | SCoT7 | CAACAATGGCTACCACGG | 51.27 |
| UBC-823 | TCTCTCTCTCTCTCTCC | 50.00 | SCoT10 | CAACAATGGCTACCAGCC | 51.20 |
| UBC-825 | ACACAC ACACACACACT | 50.00 | SCoT11 | AAGCAATGGCTACCACCA | 51.40 |
| UBC-826 | ACACACACACACACACC | 50.00 | SCoT12 | ACGACATGGCGACCAACG | 55.93 |
| UBC-834 | AGAGAGAGAGAGAGAGGT | 50.00 | SCoT13 | ACGACATGGCGACCATCG | 55.39 |
| UBC-888 | CGTCGTCGTCACACACACACACA | 52.00 | SCoT14 | ACGACATGGCGACCACGC | 58.60 |
| UBC-891 | ACTACTACTTGTGTGTGTGTGTG | 52.00 | SCoT15 | ACGACATGGCGACCGCGA | 59.90 |
| CDDP Primers | Sequence of primer (5′ – 3′) | | SCoT16 | ACCATGGCTACCACCGAC | 54.05 |
| ABP1-1 | ACSCCSATCCACCGC | 50.00 | SCoT19 | ACCATGGCTACCACCGGC | 57.10 |
| ERF1 | CACTACCCCGGSCTSCG | 50.00 | SCoT20 | ACCATGGCTACCACCGCG | 57.50 |
| ERF2 | GCSGAGATCCGSGACCC | 50.00 | SCoT21 | ACGACATGGCGACCCACA | 56.70 |
| Knox1 | AAGGGSAAGCTSCCSAAG | 50.00 | SCoT22 | AACCATGGCTACCACCAC | 51.85 |
| Knox2 | CACTGGTGGGAGCTSCAC | 50.00 | SCoT23 | CACCATGGCTACCACCAG | 52.43 |
| Knox3 | AAGCGSCACTGGAAGCC | 50.00 | SCoT24 | CACCATGGCTACCACCAT | 51.60 |
| MADS-1 | ATGGGCCGSGGCAAGGTGC | 50.00 | SCoT29 | CCATGGCTACCACCGGCC | 57.90 |
| Myb1 | GGCAAGGGCTGCCGC | 50.00 | SCoT32 | CCATGGCTACCACCGCAC | 55.90 |
| Myb2 | GGCAAGGGCTGCCGG | 50.00 | SCoT33 | CCATGGCTACCACCGCAG | 55.60 |
| WRKYF1 | TGGCGSAAGTACGGCCAG | 50.00 | SCoT34 | ACCATGGCTACCACCGCA | 56.30 |
| WRKYR1 | GTGGTTGTGCTTTGCC | 50.00 | SCoT35 | CATGGCTACCACCGGCCC | 57.90 |
| WRKY-R2 | GCCCTCGTASGTSGT | 50.00 | SCoT36 | GCAACAATGGCTACCACC | 51.50 |
| WRKY-R3 | GCASGTGTGCTCGCC | 50.00 | | | |
| WRKY-R2B | TGSTGSATGCTCCCG | 50.00 | | | |
| WRKY-R3B | CCGCTCGTGTGSACG | 50.00 | | | |

## 3.2 Drought Experiments

### 3.2.1 *In vitro* tests of all tomato accessions to drought stress by PEG-6000

### 3.2.1.1 Evaluation of morphological parameters of tomato seedlings

Polyethylene glycol-MW 6000 (PEG-6000) was used to determine the tolerance to drought stress during germination and seedling growth. The tomato seeds were sterilized by soaking in a 4% sodium hypochlorite (bleach) solution for 6 minutes and then washed seven times with distilled water. Two filter papers and a disposable plastic petri dish (9 cm in diameter) were used. Twenty-five seeds of each accession with five replications were transferred to each petri dish. 10 mL of distilled water, 7.5% PEG, and 15% PEG were applied to each petri dish as the control (T0), treatment 1 (T1), and treatment 2 (T2), respectively. All samples were placed in an incubator (Daihan LabTech Co., Ltd., Korea) and kept at a temperature at 23±2 °C. The seeds with radical lengths of 2 mm or more were measured as germinated seeds. After 14 days of growing the seedlings inside the petri dishes, the seedlings were taken out to evaluate the morphological parameters, such as germination percentage (GP), root length (RL), shoot length (SL), fresh weight (FW), and dry weight (DW). The GP was calculated using the following equation:

$$\text{Germination (\%)} = \frac{\textbf{Number of germinated seeds}}{\textbf{Total number of used seeds}} \text{ x 100}$$

After that, all samples were collected and powdered with liquid nitrogen, stored at -20 °C, and used for the phytochemical tests.

### 3.2.1.2 Phytochemical tests

*Proline content (PC) determination*

The proline content (PC) of the tomato seedling was estimated following the method of (Lateef *et al.*, 2021). 0.1 g of ground seedling tissue was homogenized in 1.4 mL of 3% (w/v) sulfosalicylic acid and centrifuged at 10000 rpm for 15 min. All the supernatant was taken and mixed with 2 mL of acid ninhydrin reagent (2.63 g of ninhydrin dissolved in 63 mL of glacial acetic acid and 42 mL of 6 M phosphoric acid) and 2 mL of glacial acetic acid in glass tubes. The samples were incubated in a water bath at 92 °C for 60 minutes. To each sample, 4 mL of toluene was added and thoroughly mixed. A UV-visible spectrophotometer was used to read the supernatant (toluene layer) at 520 nm against a blank that only contained toluene (UVM6100, MAANLAB AB, Sweden). The standard curve of proline (1 mg/mL) was prepared by taking different concentrations of L-proline. L-proline standard solutions (0.0, 50, 100, 150, 200, 250, 300, 350, 500, and 700 µg) were applied to the stopper tubes and all were diluted with distilled water up to 1 mL. A linear regression between the absorbance values at 520 nm and the L-proline content was detected. The proline content of the fresh leaf sample was estimated from this typical curve. Values were the results of three replicates and are represented as µg/g of fresh leaves. Proline content was calculated by the following formula:

$$PC\ (\mu g/gFW) = \frac{V}{W} \times C$$

Where V is the volume of extract (mL), W is the fresh weight of the leaves (g), and C is the concentration of proline obtained from the standard curve (Figure 3.1).

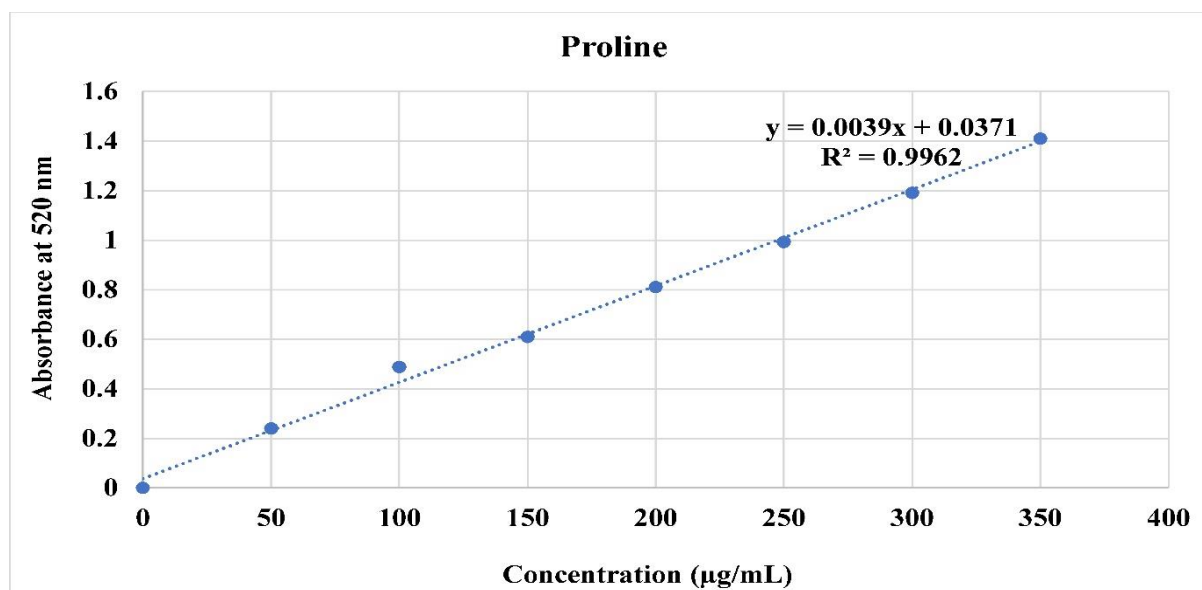

**Figure 3.1 Standard curve of proline.**

### *Estimation of soluble sugar content (SSC)*

Soluble sugar content was estimated following the method defined by (Lateef *et al.*, 2021). 0.1 g of ground seedling tissue put in an Eppendorf tube, then added 700 µL distilled water, and shaking for 20 min. The samples were boiled at 92 °C for 30 min, and cooled by cold water and centrifuged for 12 min at 8000 rpm, the supernatant was collected. Anthrone reagent was prepared by dissolving 0.41 g of anthrone with 44 mL distilled water and then added 231 mL $H_2SO_4$. 25 µL of the supernatant was mixed with 2000 µL of anthrone reagent. The solution mixture was incubated at 95 °C for 7 min, the colour of the solution was changed to dark green. The solution of samples was cooled, and read against the blank (anthrone reagent solution) at 620 nm and a UV-visible spectrophotometer (UVM6100, MAANLAB AB, Sweden) was used. Soluble sugar content was calculated by the following formula:

$$SSC(\mu g/gFW) = \frac{V}{W} \times C$$

Where V is the volume of extract (mL), W is the fresh weight of the seedling sample (g), and C is the concentration of glucose obtained from the standard curve (Figure 3.2). A stock solution of standard compound (glucose) was prepared by adding 10 mL of deionized water to 10 mg of glucose to get a final concentration of 1 mg/mL. A series of dilutions of glucose (0, 4, 10, 20, 30, 50, 80, 160, 320, 640 µg/mL) was prepared. Linear regression was observed between the absorbance values at 620 nm and the glucose concentrations.

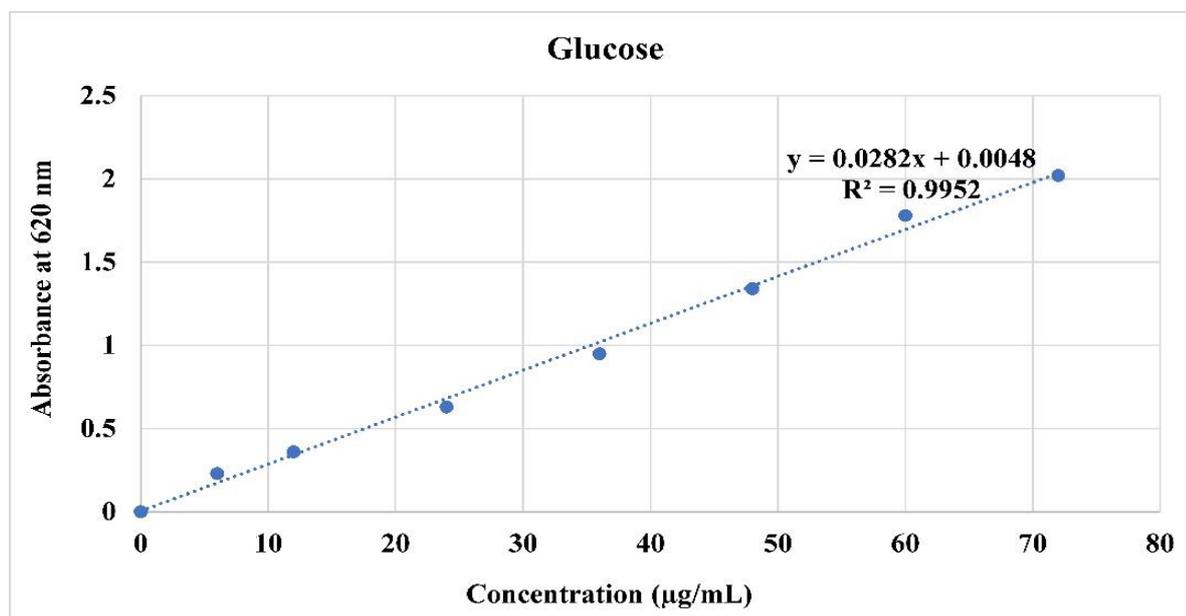

**Figure 3.2 Standard curve of glucose.**

### *Total phenolic content (TPC)*

According to (Lateef *et al.*, 2021), total phenolic content (TPC) was evaluated in tomato seedling tissue. 0.1 g of ground tissue was mixed with 1000 µL of 60% (v/v) of acidic methanol (99% Methanol+1% HCl), and shaking for 40 minutes, then all samples were incubated overnight at 5 °C. the sample mixture was centrifuged at 12000 rpm for 15 minutes and the supernatant was collected for TPC analysis. 150 µL of the supernatant mixed with 1050 µL of 1: 9 Folin–Ciocalteu reagent: water (v/v) after 7 min added 850 µL 10% $Na_2CO_3$ and incubated in dark for 30 minutes. After reaction started, the colour of mixture solution was changed to light blue and read at 750 nm against the blank (150 µL $dH_2O$ mixed with 1050 µL 1: 9 Folin–Ciocalteu reagent: water (v/v) and 850 µL 10% $Na_2CO_3$), a UV-visible spectrophotometer (UVM6100, MAANLAB AB, Sweden) was used. Gallic acid (GAE) was employed as a standard, the standard solution was prepared by dissolving 9 mg of gallic acid in 9 mL of methanol to attain a final concentration of 1 mg/mL. A sequence of dilutions of gallic acid (0, 50, 100, 150, 200, 250, 300 µg/mL) had been used to produce a standard curve and linear association between the absorbance values at 750 nm and the gallic acid content was observed. The total phenolic content in each sample was determined using the standard curve (Figure 3.3). The following equation was used to calculate the TPC:

**TPC (µg GAE/gm FW)** $= \dfrac{v}{w} \times C$

Where V is the volume of extract (mL), W is the fresh weight of the sample (g), and C is the concentration of gallic acid collected from the standard curve.

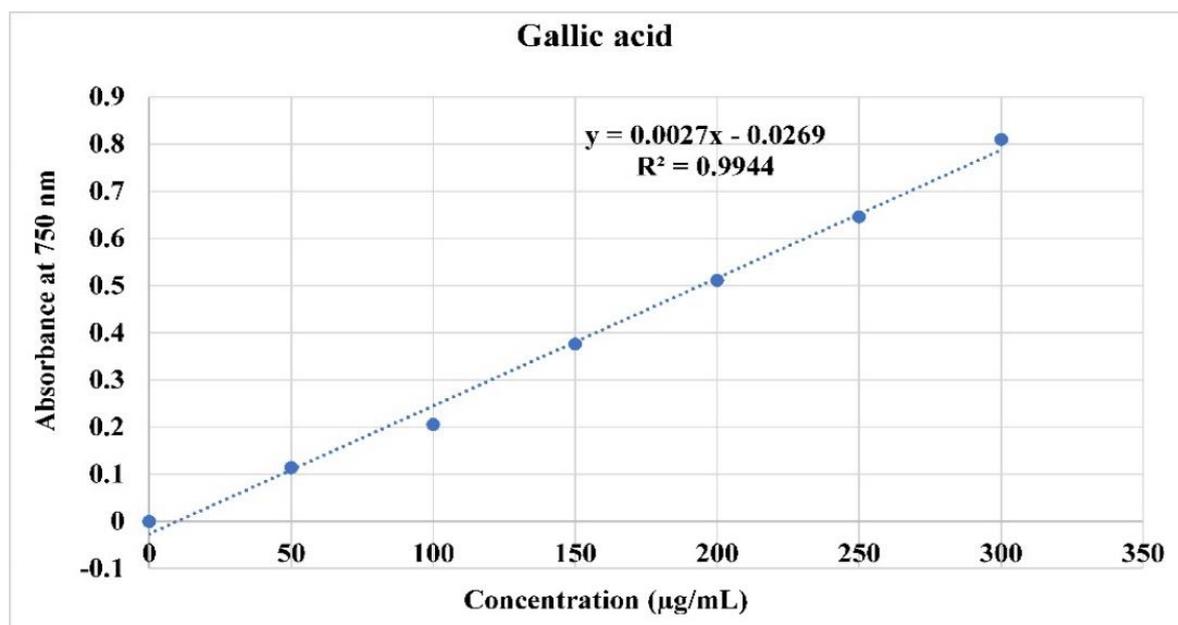

**Figure 3.3 Standard curve of gallic acid.**

### Antioxidant capacity (AC)

Antioxidant capacity was determined by using 0.1 g of ground seedling tissue, added into 1 mL of 60% (v/v) of acidic methanol (99% Methanol+1% HCl), and shaken for 40 minutes, then all samples were incubated overnight at 5 °C, the mixture was centrifuged at 12000 rpm for 15 min and the supernatant was taken. 100 μL of extract was mixed with 1.9 mL of 1-diphenyl-2-picrylhydrazyl (DPPH) solution (0.01g DPPH dissolved in 260 mL of %95 methanol). The sample mixtures were incubated in dark for 30 minutes at room temperature, absorbed the samples at 517 nm against the blank (95% methanol) using a UV-visible spectrophotometer (UVM6100, MAANLAB AB, Sweden) was used (Lateef *et al.*, 2021).

The standard compound, 6-hydroxy-2,5,7,8-tetramethylchroman-2-carboxylic acid (Trolox), was used to build the calibration curve. Trolox (12 mg) was combined with 12 mL of 75% ethanol (v/v) solvent and diluted to achieve concentrations of (0.00, 0.33, 0.66, 1.320, 2.00, 2.7, and 3.4 μg/mL) (Figure 3.4). Linear regression was found between the absorbance values at 517 nm and the varied Trolox concentrations. The following equation was used to estimate the antioxidant capacity:

**Antioxidant capacity by DPPH (μg Trolox/g FW)** $= \frac{V}{W} \times C$

Where V is the volume of extract (mL), W is the fresh weight of the sample (g), and C is the concentration of Trolox determined from the standard curve.

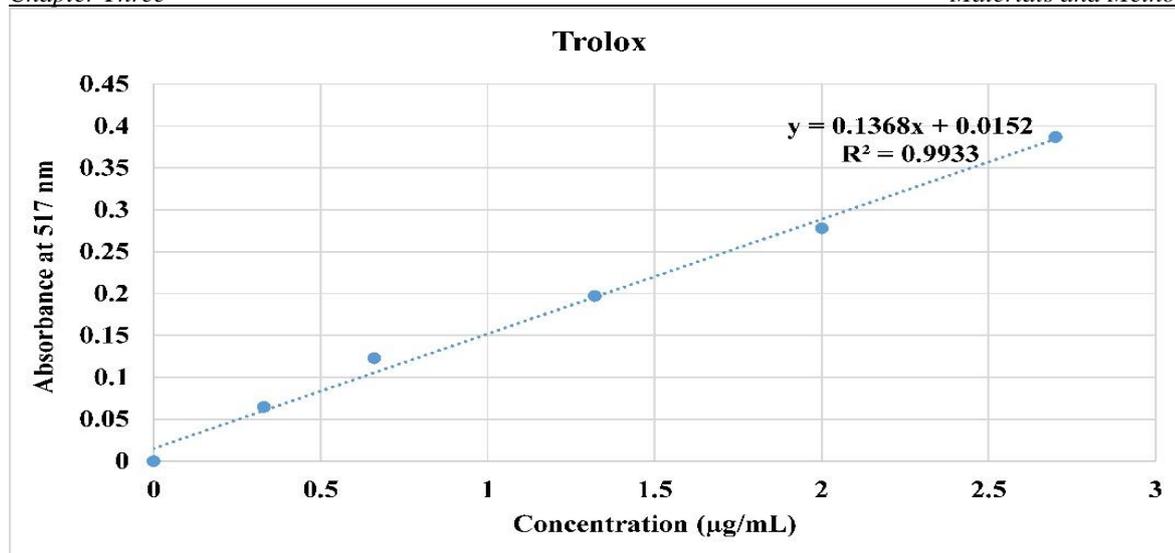

**Figure 3.4 Trolox standard calibration curve.**

### *Guaiacol peroxidase activity (GPA)*

Guaiacol peroxidase activity (GPA) was measured using 0.1 g of ground fresh tissue mixed with 900 µL of phosphate buffer (38.5 mL $KH_2PO_4$ (1M) + 61.5 $K_2HPO_4$ (1M) mixed with 800 µL $dH_2O$, pH was adjusted to 7, then the solution buffer volume completed to 1000 µL by $dH_2O$. All samples were shaken for 20 minutes and centrifuged at 1000 rpm for 20 minutes, the supernatant was taken for analysis. 200 µL of supernatant mixed with 1800 µL of phosphate buffer (pH 7), then 100 µL of (20 mM) guaiacol was added and shaken. After that 200 µL of 40 mM $H_2O_2$ (Hydrogen peroxide) was added, the reaction started and the absorbance was recorded, and allowed the sample to complete reaction after 1 minute, the reaction was stopped and the absorbance recorded again at 470 nm. Guaiacol peroxidase activity was calculated by the following formula (Lateef *et al.*, 2021).

$$\textbf{GPA(units/min/gFW)} = \left(\frac{\textbf{35.86}}{\Delta\textbf{t}}\right)\textbf{x}\left(\frac{\textbf{1}}{\textbf{1000}}\right)\textbf{x}\left(\frac{\textbf{TV}}{\textbf{VU}}\right)\textbf{x}\left(\frac{\textbf{1}}{\textbf{FWT}}\right)$$

Where extinction coefficient = 35.86 mM$^{-1}$cm$^{-1}$; $\Delta$t = time change in minute; TV = total volume of the extract (mL); VU = volume used (mL); FWT = weight of the fresh tissue (g).

### *Catalase activity (CAT)*

Catalase activity (CAT) was estimated using 0.1 g of ground fresh tissue mixed with 900 µL of CAT buffer (250 mL CAT buffer prepared by dissolving 1.51 g of Trish-HCl, 1250 µL Triton X-100, 250 µL of EDTA (0.5M), and 5 g PVP (Polyvinylpyrrolidone) in 250 mL $dH_2O$. The samples were shaken for 20 minutes and centrifuged at 1000 rpm for 15 minutes, supernatant was collected. 100 µL of supernatant mixed with 1 mL of phosphate buffer (38.5 mL $KH_2PO_4$ (0.5M) + 61.5 $K_2HPO_4$ (0.5M) mixed with 800 mL $dH_2O$, pH was adjusted to 7, then the solution buffer volume completed to 1000 mL by $dH_2O$. The reaction was started by adding 1 mL of 40 mM of hydrogen peroxide ($H_2O_2$) and the absorbance was recorded at 240 nm, then the reaction was completed after 1 minute, the absorbance recorded again (Lateef *et al.*, 2021). The catalase activity was measured by the following formula:

$$CAT((units/min)/gFW) = \left(\frac{\text{Change in absorbance (min)} \times \text{Total volume of extraction (mL)}}{\text{Extinction coefficient} \times \text{volume of sample taken (mL)}}\right) \times \left(\frac{1}{\text{FWT}}\right)$$

Where extinction coefficient = $6.93 \times 10^{-3}$ mM$^{-1}$ cm$^{-1}$ and FWT = weight of the fresh tissue.

### *Lipid peroxidation assays (LP)*

This experiment was initiated by mixing an amount of ground powder tissue (0.4 g) with 2 mL of Tris-HCl buffer solution (pH 7.4) comprising 1.5% (w/v) of polyvinylpyrrolidone (PVP). Then the mixture was shaken well for the duration of 10 minutes. Afterward, the solution mixture was centrifuged at 10000 rpm for half an hour. All the upper layers were then taken and transferred to glass tube. Following that, 2 mL of 0.5% (w/v) thiobarbituric acid (TBA) and 20% (w/v) trichloroacetic acid (TCA) was mixed with the supernatant and boiled for 35 minutes at 95 °C in a water bath. After the heating, the samples were immediately placed in a cold-water to stop the reactions, and the pinkish color appeared among the samples. The reaction mixture, after centrifugation at 4000 rpm for 12 minutes, was measured at two different wavelengths, 532 and 600 nm. The first measurement is a true measurement of the sample, while the second is for correcting unclear turbidity by subtracting the value of absorbance at 600 nm. The concentration of lipid peroxidation (LP) was stated in nmol g$^{-1}$ seedling fresh weight:

$$LP = \frac{AB532 - AB600 \times 1000 \times VL}{EC \times WE}$$

Where AB532 is the absorbance at 532 nm, AB600 is the absorbance at 600 nm, VL is the volume of extract (mL), WE is the fresh weight of the sample (g), and EC is the extinction coefficient of 155 mM$^{-1}$cm$^{-1}$ (Buege and Aust, 1978; Tahir *et al.,* 2022).

### 3.2.1.3 Statistical data analysis

Using XLSTAT software version 2020, one-way ANOVA-CRD and Duncan's new multiple range tests were utilized to evaluate significant differences (P ≤ 0.01 and P ≤ 0.001) among tomato accessions (Addinsoft, New York, USA). Utilizing XLSTAT software, the box chart and principal component analysis plot were produced. Additionally, the ranking approach, utilizing several calculated characters, was employed to identify the best accessions in accordance with the indicated strategy. The average number of ranks (ASRs) and the stress tolerance index (STI) were developed as selected criteria for the best accessions across all traits.

**3.2.2 Drought experiments under greenhouse condition**

Based on the results of *in vitro* tests of 64 tomato accessions to drought stress by polyethylene glycol (MW 6000) (PEG-600) (3.2.1), this study used two sensitive tomato accessions, AC13 (Braw) and AC30 (Yadgar), and two tolerant tomato accessions, AC61 (Raza Pashayi) and AC63 (Sandra).

**3.2.2.1 Experimental design components, plant treatments, and growth conditions**

To conduct this investigation, a factorial completely randomized design (CRD) with two factors was applied. The first factor represented tomato accessions (two sensitive and two tolerant), and the second factor represented the treatment group, which consisted of irrigated plants (SW), stressed plants that were treated (SS), stressed plants that were treated with oak leaf powder (SOS), stressed plants that were treated with oak leaf powder and oak leaf extract (SOES), and stressed plants that were treated with oak leaf powder and biofertilizers (SOBS). Seeds of four accessions were planted in plastic trays in a greenhouse. Oak leaves (*Quercus aegilops* Oliv.) were gathered in (May 2021), dried, and ground into powder for the SOS, SOES, and SOBS treatments. The seedlings (after one month) were transplanted into the plastic pots (40 cm height and 18 cm diameter). The pots for SW and SS treatments contained only 10 kg of soil, whereas the pots for SOS, SOES, and SOBS contained 10 kg of soil and 80 g of oak leaf powder. Each treatment composed of 8 replications (8 plants).

To make the extract of oak leaf, 60 g of powdered oak leaves were dissolved in 1 L of distilled water, shaken for 3 hours, and then incubated overnight at 5 °C. After centrifuging for 30 minutes at 4000 rpm, the supernatant was collected and diluted (1: 29 v/v) with distilled water. This extract was applied four times by foliar spray before flowering (first stress stage), and fruiting (second stress stage) with three-days intervals. For biofertilizer treatment, 40 mg per plant of Fulzyme Plus (JH Biotech., Inc., USA) was applied as fertigation at three times in 15 days. This biofertilizer consisted of beneficial bacteria like *Bacillus subtilis* and *Pesudomonas putida* ($2 \times 10^{10}$ g), enzymes like protease, amylase, lipase, and chitinase, and hormones like gibberellin (0.3%) and cytokinin (0.3%). Water stress at 40% of field capacity was applied before flowering (first stress stage) and fruiting (second stress stage) and the combination of first and second stress (Jangid *et al.*, 2016). The plants grew over the spring and summer season (April to September) of 2021. The average daytime and nighttime relative humidity in the greenhouse during the experiment was 42.84,17.17%, and the average temperature was 39.55/23.59 °C. Plants were kept in a regular photoperiod with 14 hours of natural light per day. Weeds were physically eliminated in the pots during the growing season, and unhealthy or dried leaves were taken out.

## 3.2.2.2 Evaluation of morphological and physiological parameters

Plant morphological data such as shoot length (SL-cm), shoot fresh weight (SFW-g), shoot dry weight (SDW-g), root length (RL-cm), root fresh weight (RFW-g), root dry weight (RDW-g), and fruit weight per plant (FWT-g) were measured at the end of the stress period. The total chlorophyll content (TCC-SPAD) was determined using a SPAD-meter at the end of the stress period. Using the method outlined by (Lateef *et al.*, 2021), the relative water content (RWC-%) of the leaves was estimated using six leaves from eight tomato plants harvested at the end of the stress period.

### 3.2.2.3 Tomato leaves and fruits collection

At the end of the stress period, fresh tomato leaves were collected, ground using liquid nitrogen, and frozen at -20 °C for use in biochemical investigations. Tomato fruits were hand-harvested at full maturity and stored at -20 °C for use in tomato fruit quality tests.

### 3.2.2.4 Moisture content, titratable acidity, and total soluble solid measurement

The moisture content (MC) was estimated by weighing 10 g of fresh tomato fruit and then drying the samples at 70 °C for 72 hours until a constant weight was achieved. The weight of the dry samples was determined, and the MC percentage was calculated using the following equation (Rahman *et al.*, 2017).

$$\text{MC (\%)} = \frac{\text{FW} - \text{DW}}{\text{FW}} \times 100$$

Where MC is the moisture content of tomato fruit, FW is the fresh weight of tomato fruit, DW is the dry weight of tomato fruit.

Titratable acidity (TA) was determined by combining 3 mL of tomato juice with 2 to 3 drops of phenolphthalein, and the mixture with 0.1N NaOH was titrated. TA was computed using the following formula:

$$\text{TA (\%)} = \frac{\text{Volume of titrant x N (NaOH)x Acid equivilent}}{\text{Volume of used juice x 1000}} \times 100$$

Total soluble solids (TSS, Brix) was determined by using digital refractometer (Rahman *et al.*, 2017).

**3.2.2.5 Measurement of biochemical traits**

*Ascorbic acid (ASC content)*

Ascorbic acid (ASC) content was determined by combining 0.4 g of powdered tomato fruit tissue with 1300 µL of 1% (w/v) HCl and vigorously shaking the mixture for 30 minutes. The mixture was centrifuged for 10 minutes at 13000 rpm, and the supernatant was collected. The supernatant was mixed with 1900 µL of 1% (v/v) HCl and measured at 243 nm against a blank containing 1% (v/v) of HCl (Abbasi *et al.*, 2019). The ASC was defined as µg/g of fresh flesh weight using the following formula:

$$\text{ASC (µg/g FW)} = \frac{\text{Volume of juice (mL)}}{\text{Fresh weight of flesh (g)}} \text{ x Concentration from standard curve of ascorbic acid (µg/mL)}$$

The standard curve of ascorbic acid was prepared by using 0, 5, 10, 15, 20, 25, 30 mg/mL of ascorbic acid (Figure 3.5).

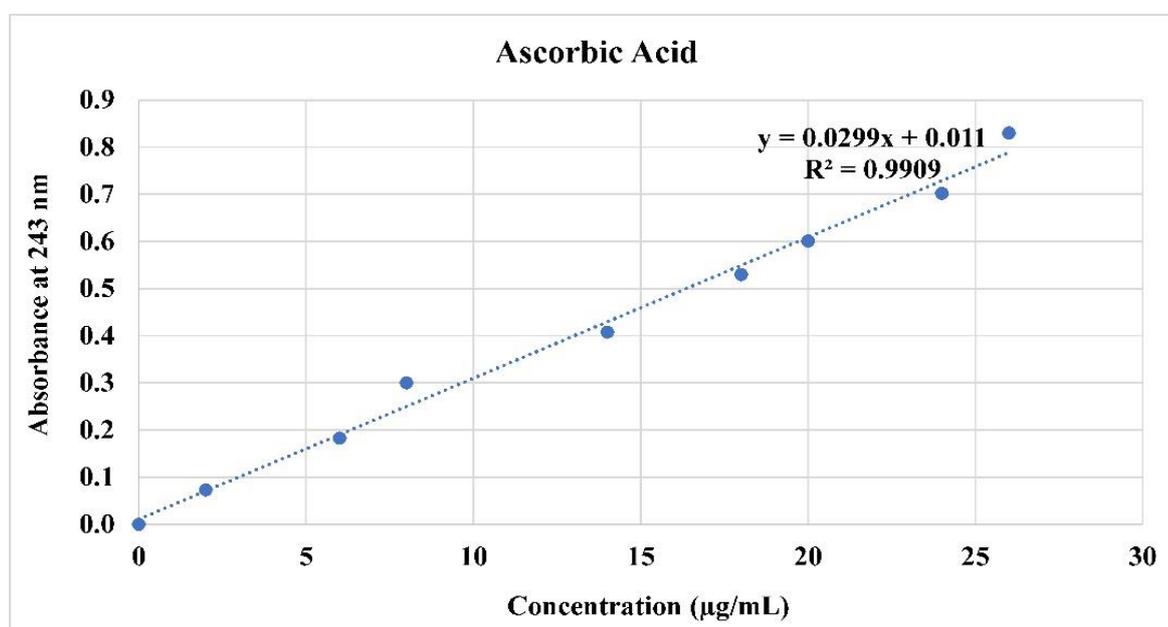

**Figure 3.5 Standard curve of ascorbic acid.**

*Carotenoid content (CAC)*

One gram of powdered tomato fruit tissue was mixed with 1000 µL of 100% methanol, and the mixture was incubated overnight at 5 °C. After centrifuging the samples for 8 minutes at 13000 rpm, 500 µL of the supernatant was collected and mixed with 1500 µL of 100% methanol. At 470 nm, the sample was read against a blank of 100% methanol (Ferrante *et al.*, 2008). and the carotenoid concentrations were expressed as µg per gram of fresh flesh weight and estimated by this formula:

$$\text{CAC (µg/g FW)} = \frac{\text{Absorbance reading x Total volume of juice (mL) x 10000}}{\text{Carotene extinction coefficient in methanol (2210) x Fresh weight of flesh (g)}}$$

To estimate the soluble sugar content (SSC) and total phenolic content (TPC) in fresh leaves and tomato fruit using the same procedure which described in (**3.2.1.2**) and also the proline content (PC), antioxidant compound capacity (AC), the activities of guaiacol peroxidase (GPA), catalase (CAT) and lipid peroxidation assays (LP) were determined in fresh leaves using the procedures reported (**3.2.1.2**).

### 3.2.2.6 GC-MS analysis of oak leaf extract

The chemical components of oak leaf extract were identified using an Agilent 7890 B gas chromatograph and an Agilent 5977 mass spectrometer, both manufactured by MSD, USA. HP-5MS UI capillary column (30 m × 0.25 × 0.25 mm) fused with 5% phenyl methyl siloxane and a splitless injector were used in a gas chromatograph. The initial temperature in the column oven was 40 °C, held steady for 60 second, and then increased to 300 °C at a rate of 10 °C per minute. To do this, we used a constant flow rate of 1 mL/minute of helium as the carrier gas and heated the injector to 290 °C. In the splitless model, the injection volume was 1 mL, the purge flow was 3 mL/minute, the total flow was 19 mL/minute, and the pressure was 7.0699 psi. The mass spectrometer was run with the help of the Mass Hunter GC/MS Acquisition software and the Mass Hunter qualitative program, which scanned fragments in the range of 35 m/z to 650 m/z. The interface temperature (MSD transfer line) was set at 290 °C, the ionization source temperature was set at 230 °C, and the quad temperature was set at 150 °C. The solvent cut time began at 4 minutes and ended between 35 and 40 minutes.

### 3.2.2.7 Statistical data analysis for field drought experiments

XLSTAT version 2019.2.2 (Boston, USA) was used to run statistical analyses (two-way analysis of variance, Duncan's multiple range tests, and principal component analysis (PCA)) for assessing the data obtained in this study at $P \leq 0.05$. The trait index was calculated by the following formula (Tahir *et al.*, 2022):

$$\textbf{Trait index } (\%) = \frac{\textbf{(Mean of treated and stressed plants} - \textbf{Mean of irrigated plants)}}{\textbf{Mean of irrigated plants}} \; x \; \textbf{100}$$

The values of all studied traits were represented by the mean ± standard deviation (SD). Each value is the average of three replications for physicochemical parameters and eight replications for morpho-physiological traits.

## 3.3 Heavy Metal Experiments

### 3.3.1 *In vitro* tests of all tomato accessions to heavy metal using cadmium (Cd)

#### 3.3.1.1 Evaluation of morphological parameters of tomato seedlings

Cadmium chloride hemi-pentahydrate ($CdCl_2$) was used to determine the tolerance to heavy metal during germination and seedling growth. The tomato seeds and samples were prepared as described in the drought section (**3.2.1**). Twenty-five sterilized (4% sodium hypochlorite) tomato seeds put in each petri dish with two filter papers. Distilled water was used as the control (T0) and three levels of cadmium 150 μM (T1), 300 μM (T2) and 450 μM (T3) were used (Al Khateeb *et al.*, 2014). Five replications were used in each treatment and 10 mL of the solution were applied for each petri dish. All samples were placed in an incubator (Daihan LabTech Co., Ltd., Korea) and kept at a temperature of about 23±2 °C. After 14 days, the germination percentage (GP-%), root length (RL-cm), shoot length (SL-cm), seedling fresh weight (FW-g) and dry weight (DW-g) were evaluated, then the samples were ground with liquid nitrogen and stored at -20 °C, which prepared for phytochemical tests.

#### 3.3.1.2 Seedling biochemical tests

The stored ground fresh seedling was used to determine the biochemical contents such as proline content (PC), soluble sugar content (SSC), total phenolic content (TPC), antioxidant capacity (AC), guaiacol peroxidase (GPA), catalase (CAT), and lipid peroxidation assays (LP). The methods of estimation of these biochemicals were described previously in the drought section (3.2.1.2).

### 3.3.2 Evaluation of heavy metal stress under greenhouse condition

In this experiment, two susceptible tomato accessions, (Super) AC32 and (Sewi Qaladze) AC56, and two tolerant tomato accessions, (Karazi) AC05 and (Sirin) AC07, based on the results of *in vitro* tests of 64 tomato accessions to heavy metal stress by cadmium (Cd) were subjected to heavy metal stress by cadmium (Cd) under greenhouse conditions.

#### 3.3.2.1 Experimental design components, plant treatments, and growth conditions

To conduct this investigation, a factorial completely randomized design (CRD) with two factors was applied. The first factor represented tomato accessions (two sensitive and two tolerant), and the second factor represented the treatment group, which consisted of control plants, soil treated with Cd (Cd+Soil) and plants that had been treated with Cd and oak leaf residue (Cd+Soil+Oak). Seeds of four accessions were planted in plastic trays in a greenhouse. Oak leaves (*Quercus aegilops* Oliv.) were gathered in (May 2021), dried, and ground for the treatments of (Cd+Soil+Oak). The oak leaf residue was prepared by grinding the oak leaf, and 200 g of ground oak leaf was dissolved in 1 liter of distilled water, to which 20 g of NaOH was added. The mixture was then shaken well and incubated for 24 hours at room temperature. The suspensions were filtered through fine mesh to remove the water and obtain the plant residues. After this process with distilled water, the plant residues were repeatedly washed until the pH decreased to near neutral (7.0). Then, the residues dried at room temperature. The seedlings were transplanted in plastic pots (40 cm height and 18 cm diameter). The pots for the control contained only 10 kg of soil, whereas the pots for Cd+Soil and Cd+Soil+Oak contained 10 kg of soil, and 350 mg of Cd was added to each pot for both treatments, and for the (Cd+Soil+Oak) treatment, 100 g of oak leaf residue was also added. Each treatment consisted of 10 replications (10 plants).

The plants grew during the spring and summer seasons of 2021. The greenhouse condition was the same as in the drought section.

### 3.3.2.2 Evaluation of morphological and physiological parameters

Plant morphological data such as shoot length (SL-cm), shoot fresh weight (SFW-g), shoot dry weight (SDW-g), root length (RL-cm), root fresh weight (RFW-g), root dry weight (RDW-g), and fruit weight per plant (FWT-g) were measured at the end of the growing season.

### 3.3.2.3 Tomato leaf biochemical tests

At the end of the growing season, fresh tomato leaves were collected, ground using liquid nitrogen, and frozen at -20 °C for use in biochemical tests. The biochemical parameters such as proline content (PC), soluble sugar content (SSC), total phenolic content (TPC), antioxidant capacity (AC), guaiacol peroxidase (GPA), and catalase (CAT) were tested according to the methods described previously in the drought section (3.2.1.2).

### 3.3.2.4 Cadmium determination in root, stem, leaf, and fresh fruit

To determine the cadmium concentration in root, stem, leaf and fruit, atomic absorption spectrometry (AAS) was applied. The root, stem, and leaf sample of all treatments and accessions were dried and ground, and tomato fruits were used freshly. The samples were digested with a 5: 1 concentrated $HNO_3$: $HClO_4$ solution. To find out if the digested samples were contaminated with cadmium or no, the concentrations of Cd were measured four times with flame atomic absorption spectrometry (AAS) (Tahir et al., 2023b).

**3.3.2.5 Statistical data analysis**

XLSTAT version 2019.2.2 (Boston, USA) was used to run statistical analyses (two-way analysis of variance, Duncan's multiple range tests, and principal component analysis (PCA) for assessing the data obtained in this study at P ≤ 0.05.

**3.4 Nematode Resistance Experiment**

Four tomato accessions were chosen based on morphological characteristics and molecular markers as described in (3.1.2 and 3.1.4) to evaluate the resistance to nematode infection. The seedlings of these accessions were prepared in plastic trays, and after one month, the seedlings were transferred to a greenhouse and planted in plastic pots with 10 kg of soil. Three treatments of control, nematode+oak and nematodes with 10 replications were applied. The pots for the control and nematode treatments contained only 10 kg of soil, whereas the pots for the nematode+oak treatment contained 10 kg of soil and 80 g of oak leaf powder. After four weeks of seedling planting, approximately 15000 eggs (Ehwaeti *et al.*, 1998) of root-knot nematodes (*Meloidogyne* spp.*)* (Mahmood, 2017) were added to each pot of both nematode+oak and nematode treatments. Eggs were collected from the galls of plant roots which infected by nematodes from the field in Sharazur. The infected roots were cleaned and washed then chopped by electronic mixer and a solution were made which included nematode eggs. The density of eggs was measured under microscope (Figure 3.6).

**3.4.1 morphological parameters and disease severity**

At the end of the growing season, shoot length (SL), shoot dry weight (SDW), root length (RL), root fresh weight (RFW), total fruit weight per plant (TFW), and disease severity (DS%) were measured (Jaiteh *et al.*, 2012)..

**3.4.2 Assessment of the tomato plants for root-knot nematode infection.**

The roots of the harvested tomato plants at the end of growing season, were each washed separately and dabbed dry with tissue paper. Galling was scored on scale of 0-10 rating (Bridge and Page, 1980; Jaiteh *et al.*, 2012). Disease severity was measured by the following equations:

**Disease severity** % = $\frac{\text{Sum of the individual disease assessments}}{\text{Number of plants observed} \times \text{Maximum disease grade}} \times \textbf{100}$

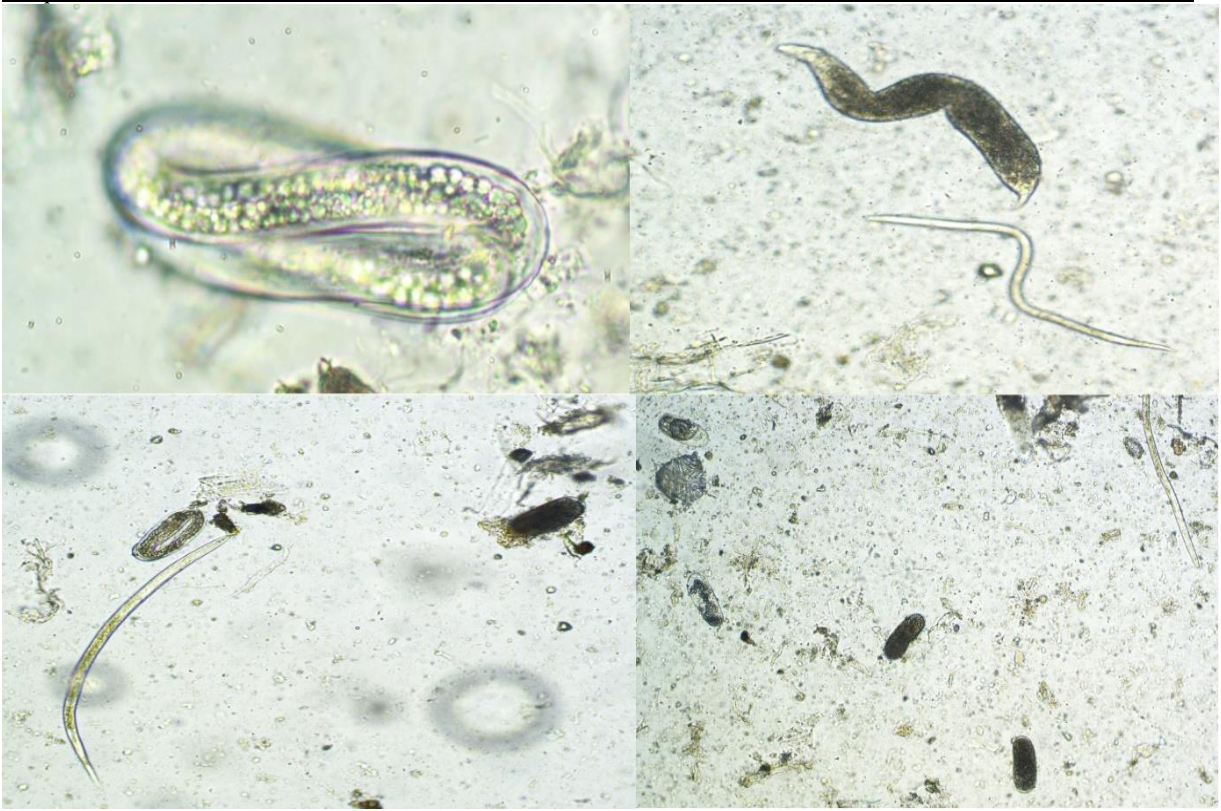

**Figure 3.6 Root-knot nematode eggs (*Meloidogyne* spp.) which used in this study.**

# CHAPTER FOUR

# RESULTS AND DISCUSSION

## 4.1 Results

### 4.1.1 Assessment of morphological and molecular markers

#### 4.1.1.1 Morphology of tomato plant.

Data in (Appendix 2) confirmed the presence of highly significant differences among all accessions for all morphological characteristics. According to the results of morphology of tomato plant, it can be seen the maximum and the minimum plant height was revealed in AC39 (334.33 cm) and AC8 (84.33 cm), respectively. The longest root recorded by AC60 which was 51 cm, and the shortest root with 23.33 cm was recorded by AC21. The maximum and minimum plant dry weights were recorded by AC51 (454.80 g) and AC8 (44.29 g) respectively. The largest leaf area was recorded by AC52 which was 384.88 cm$^2$, and the smallest leaf area showed in AC32 by 71.80 cm$^2$. The highest and the lowest total chlorophyll contents were recorded by AC13 (93.4 SPAD) and AC6 (11.03 SPAD), respectively.

#### 4.1.1.2 Assessment of fruit characteristics in tomato accessions

Data in Table 4.1 and Appendix 3 demonstrate significant differences among 64 accessions for 10 investigated variables. AC43 had the highest value of fruit weight and fruit size, which were 285.20 g and 295.10 cm$^3$, respectively, while AC11 had the lowest values, which were 6.39 g and 6.90 cm$^3$, respectively. AC21 had the thickest fruit, measuring 76.92 mm, while AC14 had the thinnest fruit, which was measured 26.27 mm. AC18 and AC11 had the biggest and smallest fruit diameter measurements, with 86.99 mm and 20.92 mm, respectively. AC63 had the largest fruit output per plant of 2935.65 g, while AC13 had the lowest fruit yield of 694.93 g. AC39 (95.74%) and AC11 (89.99%) had the highest and lowest percentages of moisture contents, respectively, while AC11 and AC39 had the maximum and minimum percentages of total solids, 4.25 and 10.01%, respectively. AC5 had the greatest total soluble solids (Brix) value of 7.77, while AC40 had the lowest value of 2.93, while AC11 had the largest percentage of non-soluble solid (3.48%) and AC38 seemed to have the lowest value of 0.52%. AC22 had the most fruit firmness (3193.00 g/cm$^2$), whereas AC16 had the least value (907.67 g/cm$^2$).

**Table 4.1 Descriptive statistics of fruit characters in tomato accessions.**

| Traits | Min (Accession) | Max (Accession) | Mean | SD | VC (%) | F | P-value |
|--------|-----------------|-----------------|------|-----|--------|---|---------|
| FW | 6.39 (AC11) | 285.20 (AC43) | 106.05 | 68.39 | 64.48 | 84.91** | < 0.0001 |
| FS | 6.90 (AC11) | 295.10 (AC43) | 110.79 | 70.95 | 64.04 | 114.04** | < 0.0001 |
| FT | 26.27 (AC14) | 76.92 (AC21) | 44.78 | 11.95 | 26.68 | 54.13** | < 0.0001 |
| FD | 20.92 (AC11) | 86.99 (AC18) | 57.38 | 17.05 | 29.71 | 58.53** | < 0.0001 |
| FWP | 694.93 (AC13) | 2935.63 (AC63) | 1718.63 | 613.69 | 35.71 | 2310.50** | < 0.0001 |
| MC | 89.99 (AC11) | 95.74 (AC39) | 93.70 | 1.00 | 1.07 | 59.60** | < 0.0001 |
| TS | 4.25 (AC39) | 10.01 (AC11) | 6.30 | 1.00 | 15.92 | 59.60** | < 0.0001 |
| TSS | 2.93 (AC40) | 7.77 (AC5) | 4.66 | 0.97 | 20.87 | 250.37** | < 0.0001 |
| NSS | 0.52 (AC38) | 3.48 (AC11) | 1.64 | 0.77 | 47.24 | 32.79** | < 0.0001 |
| FF | 907.67 (AC16) | 3193 (AC22) | 2209.40 | 579.90 | 26.25 | 67.73** | < 0.0001 |

**FW: fruit weight (g), FS: fruit size (cm³), FT: fruit thickness (mm), FD: fruit diameter (mm), FWP: total fruit weight per plant (g), MC: fruit moisture content (%), TS: total solids (%), TSS: total soluble solids (brix), NS: non-soluble solids (%), FF: fruit firmness (g/cm²), Min: minimum, Max: maximum, SD: standard deviation, VC: variation coefficient, \*\*: highly significant.**

### 4.1.1.3 Multivariate analysis of fruit traits in tomato accessions

Principal component analysis (PCA) is a multivariate statistical methodology used for evaluating and understanding the complex and huge datasets. The pattern of variability in tomato accessions was analyzed using PCA based on the correlation between the traits and extracted clusters to assess the variety of the accessions and their relationship with the observed traits. Following the PCA result, it was determined that the two principal components (F1 and F2) described 67.25% of the total quality variance (Figure 4.1). Furthermore, the first principal component (F1) explained 48.49% of the overall variation; it was positively linked with FS, FF, FT, FD, and MC, but negatively with NSS and TS; the second principal component (F2) clarified 18.76% of the total variability, and was positively associated with FWP and TSS. As a result, the plot formed by the first two components could distinguish the tomato accessions based on their major determining features. Based on 10 fruit trait datasets, the PCA plot divided 64 accessions into 5 clusters (CL1-CL5). In our data set, five groups of attributes (GrI-GrV) were defined using the PCA biplot while simultaneously considering F1 and F2. The FS, FF, FT, and FD traits were assigned to group I (GrI), while FWP formed the second group (GrII). TSS and TS were placed in group III (GrIII), whereas NSS and MC were allocated to groups IV (GrIV) and V (GrV), respectively. The PCA biplot revealed that GrI and GrIV traits, which were the major contributors in the first component, were highly related to cluster 4 (CL4) accessions, whereas group II traits, which were also contributors in the second component, were related to rowing cluster 2 (CL2) accessions. The characteristics of GrIII that contributed to PC2 were the most strongly related to the accessions of cluster 3 (CL3), while another variable (NSS) of GrIV was closely associated with the accessions of cluster 4 (CL4). The PCA-biplot also displayed the cluster centroids and estimated distances between them.

Seven clusters were formed in the case of hierarchical clustering of fruit characteristics data from all tomato accessions (Figure 4.2). The first cluster (in red) had eight accessions: AC1, AC21, AC22, AC24, AC40, AC44, AC53, and AC62, while the second cluster (in green) had twelve: AC12, AC60, AC25, AC39, AC45, AC8, AC64, AC35, AC36, AC13, AC26, and AC31. However, AC3, AC19, AC20, AC54, and AC9 were the five accessions in Cluster 3 (in blue). AC10, AC52, AC32, AC33, AC46, AC17, AC50, AC56, AC55, AC57, AC58, AC29, AC30, AC59, AC18, AC28, AC37, and AC43 comprised the fourth cluster (in brown). AC4, AC34, AC12, AC27, AC38, AC7, AC47, AC61, AC15, AC63, and AC41 were parts of cluster five (in teal), while AC6, AC42, AC49, AC51, AC16, AC23, and AC48 were parts of cluster six (in purple). AC5, AC 11, and AC14 comprised the final cluster (in olive).

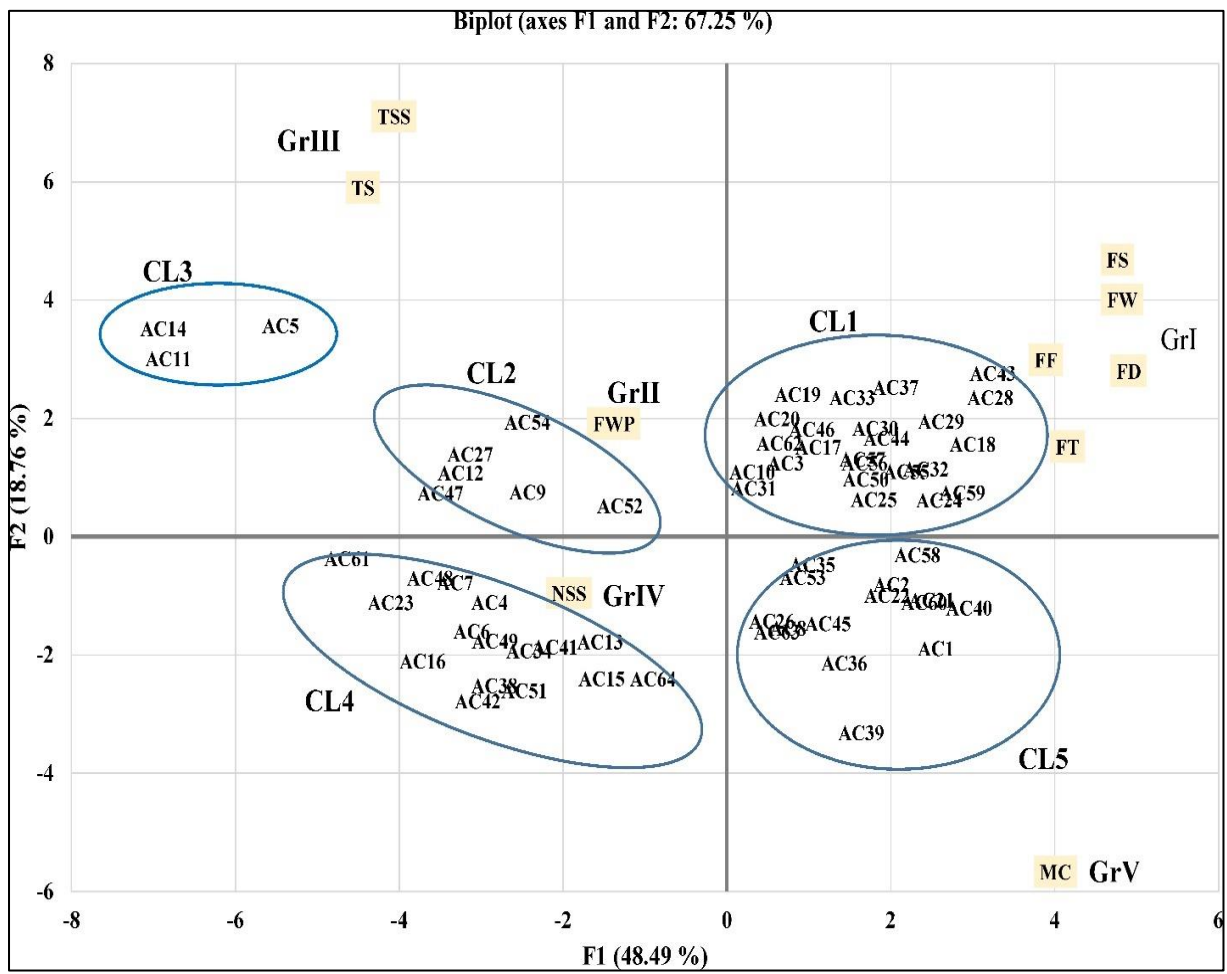

**Figure 4.1 A PCA-biplot of tomato fruit attributes and accessions. Accessions and traits were distributed in distinct ordinates based on their dissimilarity. The angles between the vectors produced from the middle point of biplots show whether the investigated features interact positively or negatively. FW: fruit weight, FS: fruit size, FT: fruit thickness, FD: fruit diameter, FWP: total fruit weight per plant, MC: fruit moisture content, TS: total solids, TSS: total soluble solids, NS: non-soluble solid, and FF: fruit firmness. Table 3.1 represents informational details about accessions. The number of clades of accessions was denoted by CL1-CL5. GrI-GrV symbolized the number of trait groups**

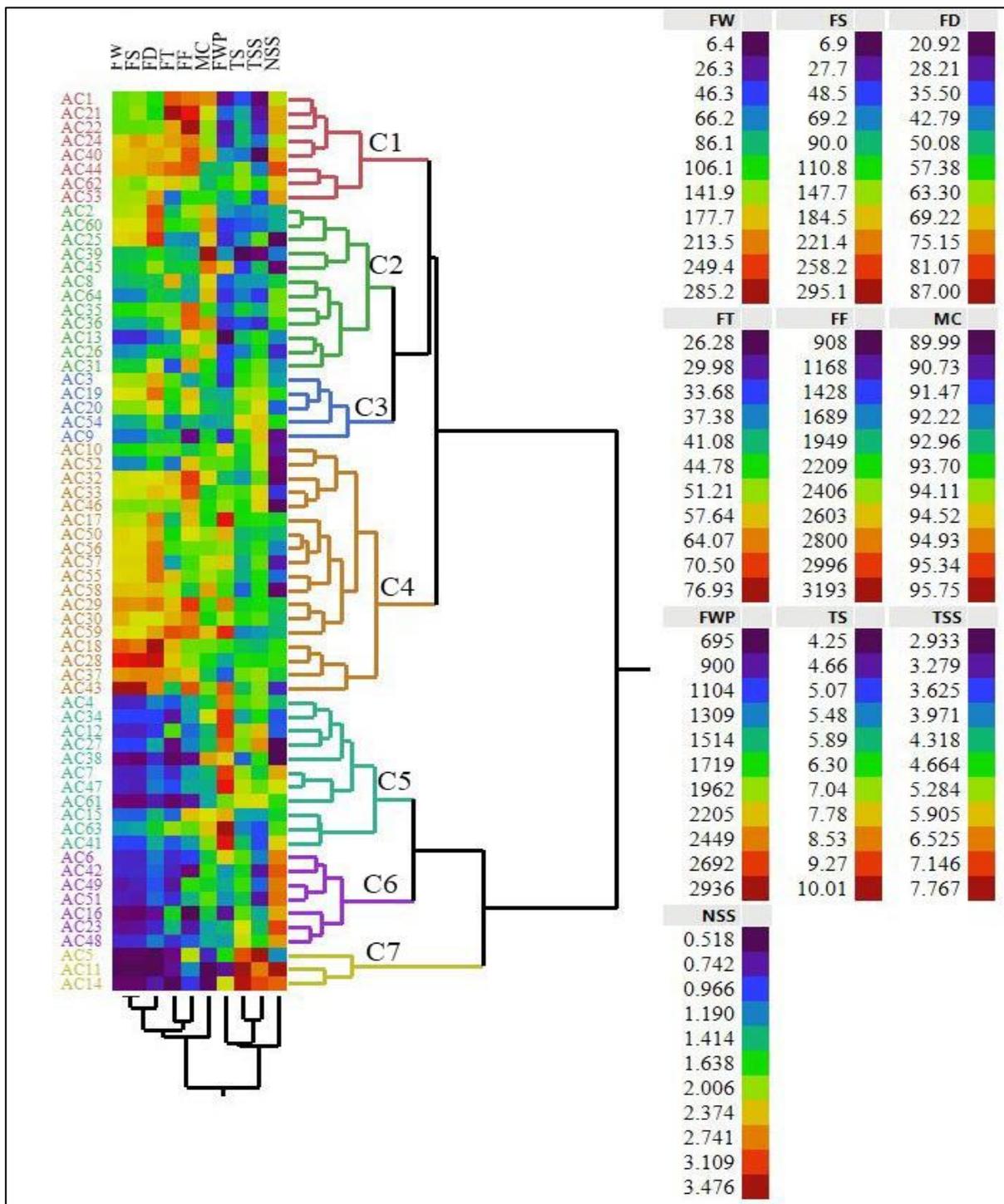

**4.2 A hierarchical cluster created by JMP Pro 16 software based on fruit attributes in 64 tomato accessions. FW: fruit weight, FS: fruit size, FT: fruit thickness, FD: fruit diameter, FWP: total fruit weight per plant, MC: fruit moisture content, TS: total solids, TSS: total soluble solids (Brix), NS: non-soluble solids, and FF: fruit firmness. The number of accession clusters was expressed by C1-C7. For accessions (AC) details, see Table 3.1.**

#### 4.1.1.4 Polymorphism and discriminatory characteristics in ISSR, SCoT, and CDDP markers

Thirteen ISSR primers were used to study genetic diversity among 64 tomato accessions. A total of 121 polymorphic bands, with a mean of 9.31 bands per primer, were generated. Gene diversity (GD) ranged from 0.42 (UBC-810) to 0.98 (UBC-815), with an average of 0.82. Polymorphism information content (PIC) values for 13 ISSR primers ranged from 0.39 (UBC-810) to 0.98 (UBC-815) with an average of 0.80 per primer (Table 4.2 and Appendix 4).

A set of 26 SCoT primers were investigated for PCR optimization, description, and amplification in 64 diverse tomato accessions. All primers produced polymorphic and reliable amplification profiles, resulting in 294 unblurred and brilliant bands (Table 4.3 and Appendix 4). Although the number of bands per primer spanned from 4.0 to 19.0, with an average of 11.31 bands per primer, SCoT13 had the most (19 bands) banding patterns, while SCoT7 had the minimum (4 bands). For the twenty-six SCoT primers, the averages for GD and PIC were 0.85 and 0.84, respectively, with SCoT21 having the least values of 0.52 and 0.50 for GD and PIC, respectively.

All fifteen CDDP primers yielded reproducible polymorphic bands, resulting in 183 amplified polymorphic bands among the accessions (Table 4.4 and Appendix 4). The number of polymorphic bands ranged from 5 (WRKY-R2B) to 17 (Myb2), with a mean of 11.33 bands per primer. Furthermore, the GD differed from 0.39 (Knox2) to 0.98 (Myb2), with a mean of 0.85 per primer. Knox2 and Myb2 primers had minimum (0.36) and maximum (0.98) PIC values, respectively.

**Table 4.2 Polymorphism characteristics of 13 ISSR primers used in this study.**

| Primers | NPB | GD | PIC |
|---------|-----|-----|-----|
| UBC-808 | 7.00 | 0.69 | 0.64 |
| UBC-810 | 3.00 | 0.42 | 0.39 |
| UBC-812 | 9.00 | 0.88 | 0.87 |
| UBC-814 | 8.00 | 0.85 | 0.84 |
| UBC-815 | 18.00 | 0.98 | 0.98 |
| UBC-818 | 10.00 | 0.95 | 0.95 |
| UBC-822 | 8.00 | 0.90 | 0.89 |
| UBC-823 | 10.00 | 0.92 | 0.91 |
| UBC-825 | 11.00 | 0.95 | 0.94 |
| UBC-826 | 6.00 | 0.46 | 0.44 |
| UBC-834 | 9.00 | 0.94 | 0.93 |
| UBC-888 | 12.00 | 0.95 | 0.95 |
| UBC-891 | 10.00 | 0.78 | 0.76 |
| Total | 121.00 | 10.67 | 10.49 |
| Mean | 9.31 | 0.82 | 0.81 |

NPB: number of polymorphism bands, GD: gene variability, PIC: polymorphism information content.

In terms of the combination of ISSR, SCoT, and CDDP markers, 585 polymorphic bands were produced (Tables 4.2, 4.3, and 4.4). SCoT13 (19 bands) and UBC-810 (3 bands) revealed the greatest and smallest number of polymorphic bands, respectively. The PIC data revealed that seven primers (UBC-815, SCoT13, SCoT14, SCoT15, SCoT34, SCoT35, and Myb2) had the greatest PIC value (0.98), while primer Knox2 had the lowest (0.36).

**Table 4.3 Number of polymorphism bands (NPB), gene variability (GD), and the polymorphism information content (PIC) in tomato accessions acquired using 26 SCoT markers.**

| Primer name | NPB | GD | PIC |
|-------------|-----|-----|-----|
| SCoT1 | 8.00 | 0.65 | 0.64 |
| SCoT2 | 7.00 | 0.75 | 0.72 |
| SCoT3 | 10.00 | 0.94 | 0.93 |
| SCoT4 | 9.00 | 0.90 | 0.89 |
| SCoT5 | 10.00 | 0.89 | 0.87 |
| SCoT6 | 5.00 | 0.72 | 0.70 |
| SCoT7 | 4.00 | 0.75 | 0.72 |
| SCoT10 | 5.00 | 0.76 | 0.73 |
| SCoT11 | 12.00 | 0.88 | 0.87 |
| SCoT12 | 18.00 | 0.97 | 0.97 |
| SCoT13 | 19.00 | 0.98 | 0.98 |
| SCoT14 | 16.00 | 0.98 | 0.98 |
| SCoT15 | 16.00 | 0.98 | 0.98 |
| SCoT16 | 8.00 | 0.91 | 0.91 |
| SCoT19 | 11.00 | 0.97 | 0.97 |
| SCoT20 | 16.00 | 0.98 | 0.98 |
| SCoT21 | 13.00 | 0.52 | 0.50 |
| SCoT22 | 8.00 | 0.64 | 0.63 |
| SCoT23 | 11.00 | 0.92 | 0.91 |
| SCoT24 | 13.00 | 0.72 | 0.69 |
| SCoT29 | 13.00 | 0.97 | 0.96 |
| SCoT32 | 9.00 | 0.65 | 0.64 |
| SCoT33 | 11.00 | 0.92 | 0.91 |
| SCoT34 | 17.00 | 0.98 | 0.98 |
| SCoT35 | 17.00 | 0.98 | 0.98 |
| SCoT36 | 8.00 | 0.84 | 0.82 |
| Total | 294.00 | 22.15 | 21.86 |
| Average | 11.31 | 0.85 | 0.84 |

**Table 4.4 Polymorphism parameters of different CDDP primers collected in tomato accessions.**

| Primer name | NPB | GD | PIC |
|---|---|---|---|
| ABP1-1 | 16.00 | 0.92 | 0.92 |
| ERF1 | 13.00 | 0.97 | 0.97 |
| ERF2 | 11.00 | 0.81 | 0.79 |
| Knox1 | 11.00 | 0.90 | 0.90 |
| Knox2 | 6.00 | 0.39 | 0.36 |
| Knox3 | 12.00 | 0.93 | 0.92 |
| MADS-1 | 17.00 | 0.90 | 0.89 |
| Myb1 | 12.00 | 0.96 | 0.96 |
| Myb2 | 17.00 | 0.98 | 0.98 |
| WRKYF1 | 9.00 | 0.83 | 0.82 |
| WRKYR1 | 7.00 | 0.89 | 0.89 |
| WRKY-R2 | 9.00 | 0.78 | 0.76 |
| WRKY-R3 | 11.00 | 0.87 | 0.86 |
| WRKY-R2B | 5.00 | 0.61 | 0.58 |
| WRKY-R3B | 14.00 | 0.96 | 0.96 |
| Total | 170.00 | 12.70 | 12.56 |
| Average | 11.33 | 0.85 | 0.84 |

NPB: number of polymorphism bands, GD: gene variability, PIC: polymorphism information content.

### 4.1.1.5 Cluster analysis of tomato accessions using ISSR, SCoT, and CDDP information

Using 13 ISSR primers, UPGMA clustering produced two primary groupings of tomato accessions (Figure 4.3A). Cluster 1 (in green) contained 55 accessions; this cluster was separated into two subclusters, with 54 tomato accessions forming the first subcluster and only the accession AC11 forming the second subcluster. Cluster 2 (in yellow) had 9 tomato accessions, which were separated into two subclusters: the first subcluster included 7 accessions from AC2, AC3, AC37, AC38, AC40, AC57, and AC63, and the second subcluster included accessions from AC8 and AC21. The maximum distance coefficient (0.91) was found in the current study between AC38 (Israili) and AC2 (Kurdi Gawray Swr), while AC60 (Kurdi Pshdar) and AC61 (Raza Pashayi) had the smallest distance with a 0.16 coefficient.

UPGMA clustering of tomato accessions using SCoT markers resulted in two major groupings (Figure 4.3B). The first group (green color) had 51 accessions, while the second group contained 13 accessions (yellow color). The first group was divided into three subclusters, as follows: three accessions produced the first subcluster: AC1, AC23, and AC30. AC26 created the second subcluster, and the third subcluster was formed by 47 accessions. The second group was further subdivided into two subclusters. The first subcluster, on the other hand, was made up of four accessions: AC4, AC8, AC63, and AC64, whilst the second was mainly composed of nine accessions: AC47, AC57, AC2, AC3, AC43, AC18, AC37, AC38, and AC40. By merging the amplified bands obtained from the SCoT primer, the binary matrix was generated. This purpose-designed matrix was used to explore genetic differences between 64 tomato accessions. The maximum distance coefficient (0.78) was obtained in the current study between AC36 (Rozh) and AC40 (Roma). AC53 (Kurdi Gawray Swr) and AC54 (Balami Qaladze) had the least distance with a 0.21 coefficient.

The UPGMA approach was employed for cluster analysis of tomato accessions based on CDDP data, and these markers produced two primary groups (Figure 4.3C). The first group (green color) had 58 accessions, while the second group (yellow color) had 6 accessions. In the first group, two sub-clusters were generated as follows: AC13, AC28, AC36, and AC38 created the first subcluster, and the second subcluster contained 54 accessions. The second group was separated into two subclusters, with AC59. Both AC61 and AC62 made the first subcluster, and AC60, AC61, and AC63 formed the second. In the current research, the greatest distance coefficient (0.85) was reported between AC5 (Karazi) and AC38 (Israili), with a 0.12 coefficient, AC53 (Kurdi Gawray Swr) and AC54 (Balami Qaladze) had the smallest distance.

The cluster analysis, which was based on the combination of three types of marker data, classified the 64 accessions into two major groupings (Figure 4.3D). The first group (in green) consisted of 51 accessions, while the second group (yellow color) consisted of 13 accessions. Each cluster was further divided into two subclusters.

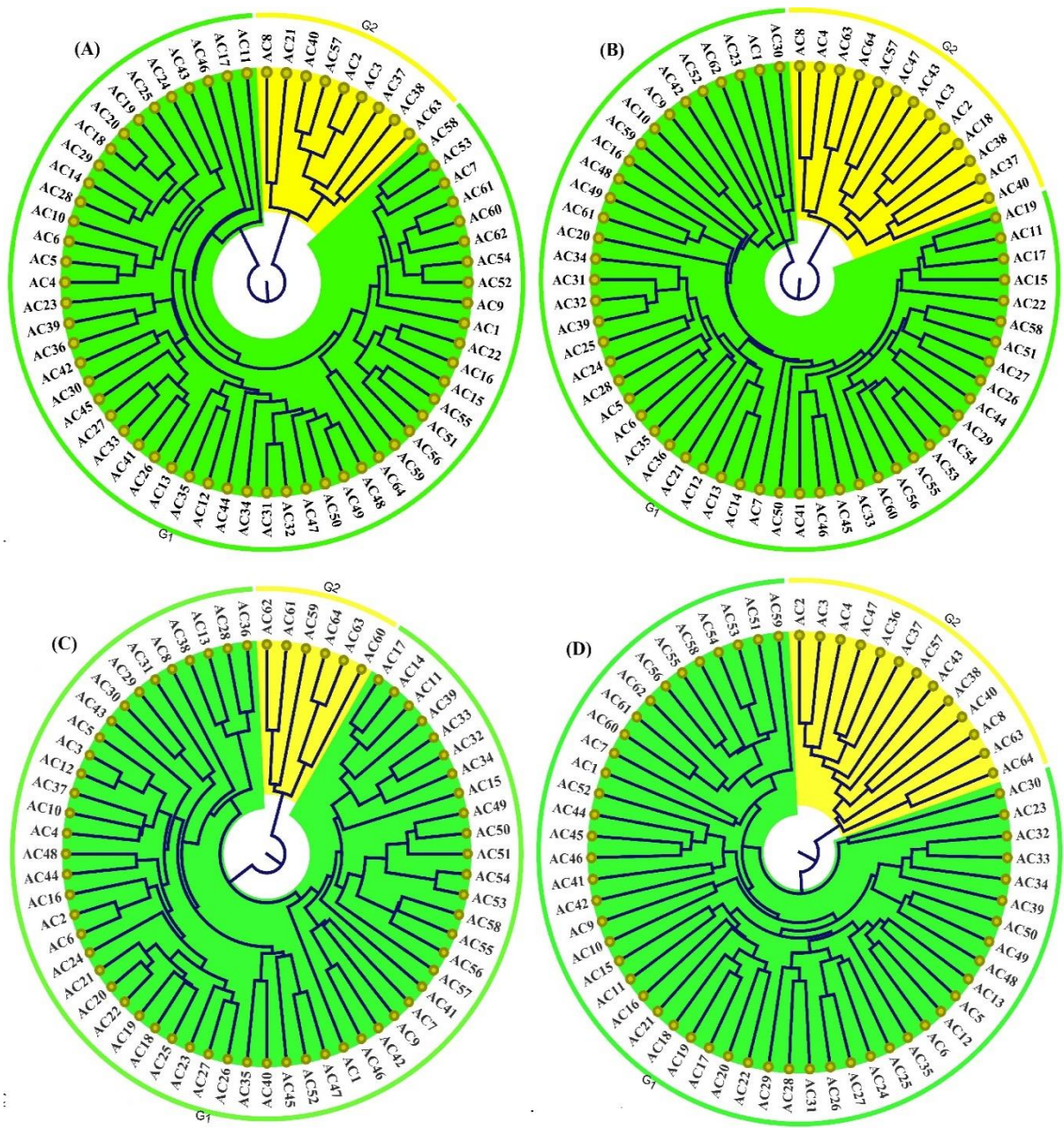

**Figure 4.3 UPGMA dendrogram illustrating the clustering of 64 tomato accessions based on genetic dissimilarities of three markers. (A) ISSR dataset; (B) SCoT dataset; (C) CDDP dataset; (D) combined ISSR+SCoT+CDDP dataset.**

### 4.1.1.6 STRUCTURE analysis of tomato accessions based on ISSR, SCoT, and CDDP datasets

The population stratification of 64 tomato accessions based on ISSR markers was studied using STRUCTURE analysis. The K-value was used to determine the number of clusters of accessions based on genotypic data throughout the entire genome. To determine the ideal K-value, the number of clusters (K) was plotted versus K, which revealed a sharp peak at K = 2 (Figure 4.4A). The ideal K-value suggested that two populations (population 1 in red and population 2 in green) had the highest chance of population clustering, with 46 and 6 accessions, respectively (Figure 4.4B). The threshold of membership probability was 0.80. Each subgroups' accessions were assigned individually, and fractions less than 0.20 were deemed admixed. Twelve accessions, including AC4, AC5, AC8, AC11, AC12, AC21, AC28, AC35, AC36, AC39, AC47, and AC57, were admixed between the two populations, indicating that these accessions are not pure. The STRUCTURE results calculated the fixation index (Fst) for each population and indicated significant divergence within both populations. For population 1 and population 2, Fst values of 0.39 and 0.55 were obtained, respectively. The values of expected heterozygosity were between 0.20 and 0.23 (Table 4.5).

The tomato accessions in the SCoT dataset were assigned using Bayesian clustering based on their population structure. The estimated membership fraction ranged from K 1 to K 9, and the maximum ad hoc delta K value was recorded at K = 2, indicating that the K2 provided the best credible probabilities, and 64 accessions were categorized into two groups (Figure 4.4C). The membership probability threshold was found to be 0.80. The accessions were assigned individually to each group, and fractions less than 0.20 were considered admixed. The first population had 11 tomato accessions, including AC2, AC3, AC4, AC8, AC36, AC37, AC38, AC40, AC47, AC57, and AC63, whereas the second population contained 33 tomato accessions. There were twenty admixed accessions detected between the two populations (Figure 4.4D). The Fst calculated for two populations demonstrated a significant deviation from zero, suggesting a strong level of differentiation within each populations' individuals. The expected heterozygosity was 0.22 for both populations (Table 4.5).

Using the STRUCTURE HARVESTER analysis, the ideal K-value was established based on CDDP data, and the number of clusters (K) was plotted vs K, exhibiting a sharp peak at K = 2 (Figure 4.5A). The 64 tomato accessions were only separated into two genetic groups based on CDDP data. The first major group (red color) had 29 accessions from diverse sources. The second group (green color) included 21 accessions of varied origins and morphological characteristics (Figure 4.5B). Fourteen accessions from the two groups were considered to have admixed genetic makeup. Based on CDDP data, the within-population Fst varied from 0.30 to 0.44, indicating significant gene flow and negligible population difference. The expected heterozygosity ranged between 0.19 and 0.23 (Table 4.5).

For a better understanding of tomato accessions clustering, STRUCTURE analysis was done across all tomato accessions using pooled data from three markers. The number of genetic groupings (K) reached a high point at K2 (Figure 4.5C), indicating the presence of two populations. This detection indicated that in the first and second populations, respectively, 10 and 34 pure accessions were distributed (Figure 4.5D). Between the two groups, a total of 20 accessions were admixed. Fst values of 0.39 and 0.33 were obtained from the combined data for population 1 and population 2, respectively (Table 4.5).

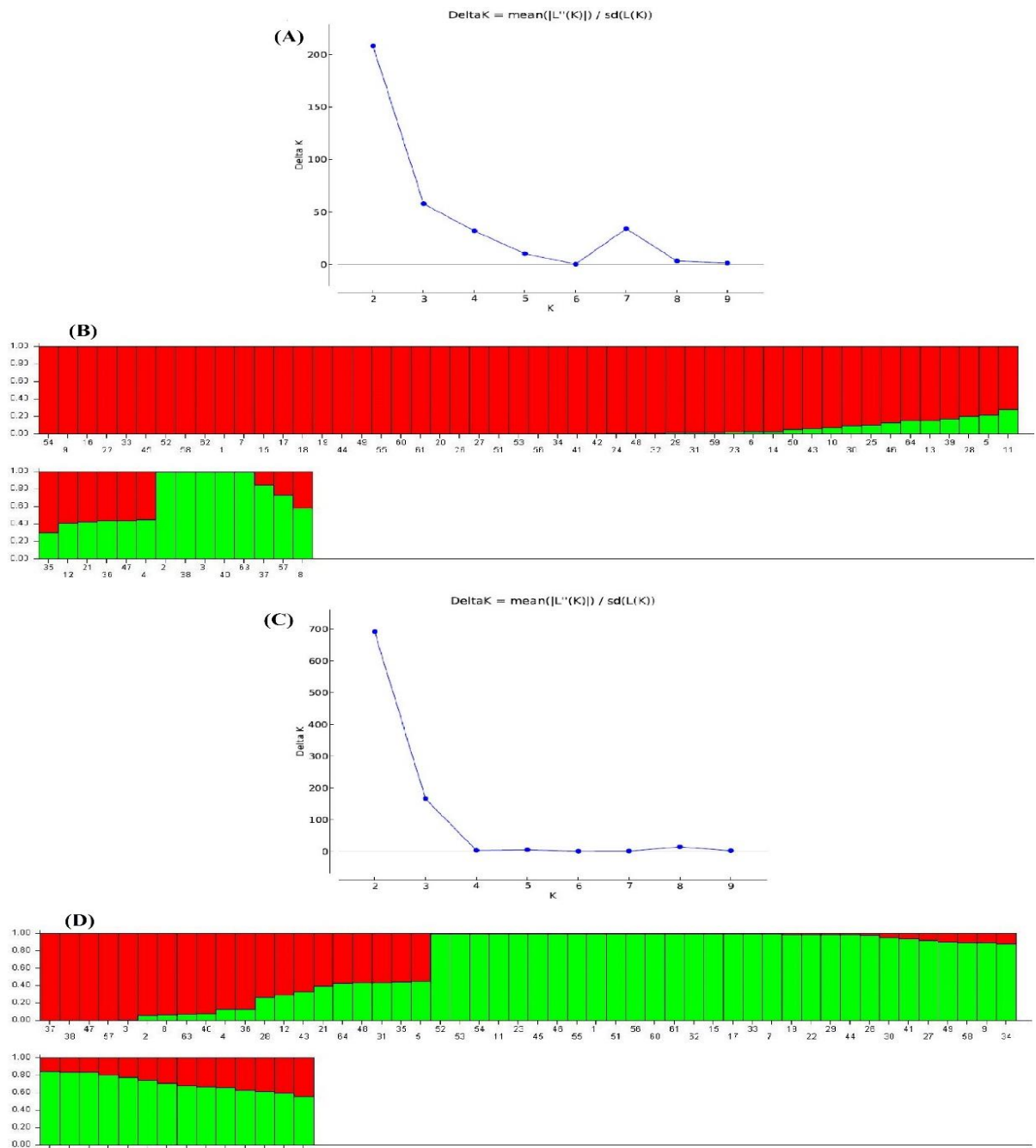

**Figure 4.4 (A) Delta K for different population numbers (K) derived from ISSR data; (B) Estimated population structure of 64 tomato accessions on K = 2 derived from the ISSR dataset; (C) Delta K for different population numbers (K) derived from SCoT data; (D) Estimated population structure of 64 tomato accessions on K = 2 derived from the SCoT dataset. Accessions in red were categorized as population 1, whereas those in green were classified as population 2.**

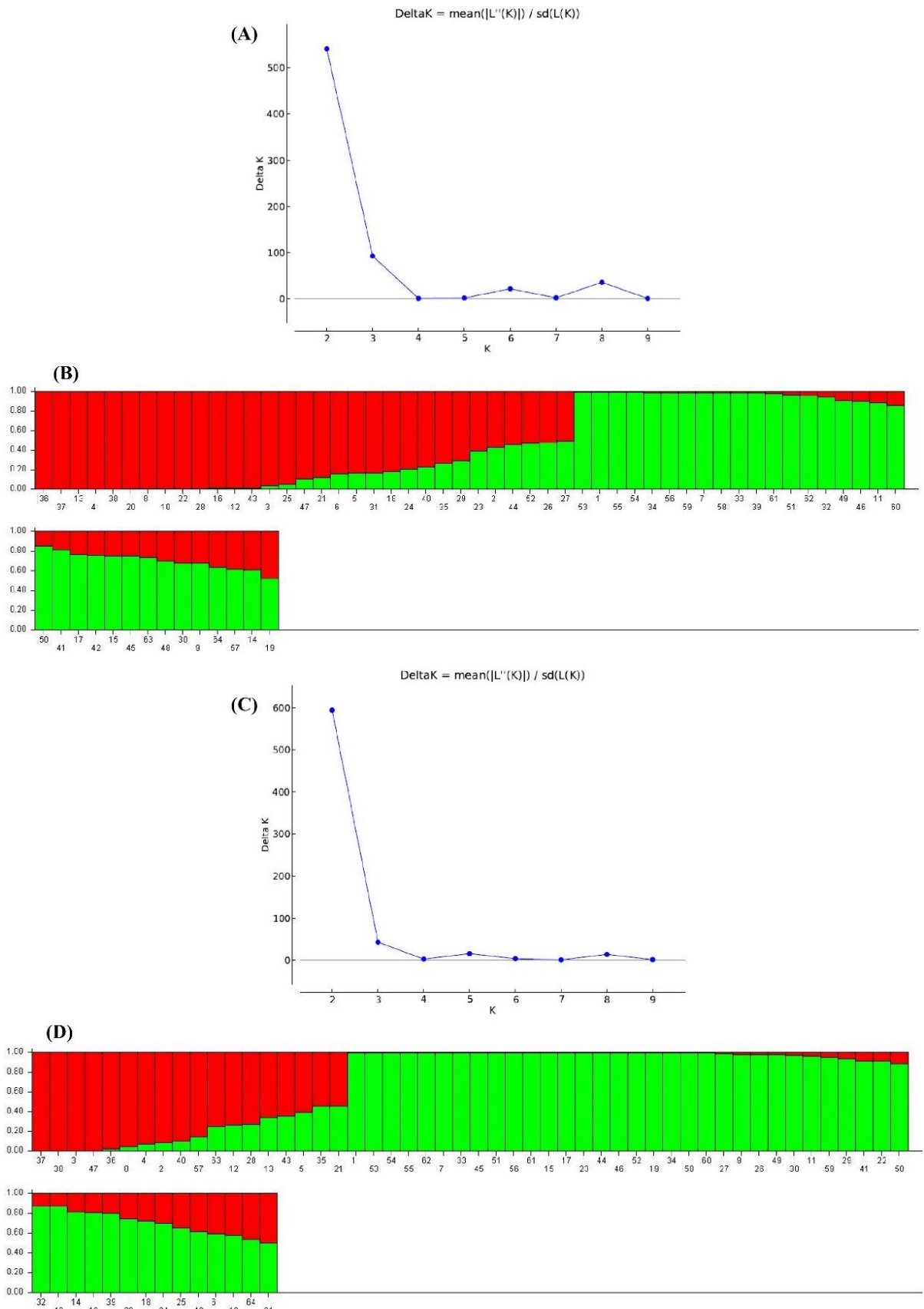

**Figure 4.5 (A) Delta K for various population numbers (K) derived from CDDP data; (B) Estimated population structure of 64 tomato accessions on K = 2 derived from the CDDP dataset; (C) Delta K for various population numbers (K) derived from ISSR+SCoT+CDDP data; (D) Estimated population structure of 64 tomato accessions on K = 2 derived from the ISSR+SCoT+CDDP data. Accessions marked in red were classified as population 1, whereas those labeled in green were rated as population 2.**

**Table 4.5 Structure parameters of tomato accessions based on ISSR, SCoT, CDDP, and combined (ISSR+SCoT+CDDP) data.**

| Marker type | Population | Inferred cluster | Expected heterozygosity | Fixed index |
|---|---|---|---|---|
| ISSR | Population 1 | 0.82 | 0.23 | 0.39 |
| | Population 2 | 0.18 | 0.20 | 0.55 |
| SCoT | Population 1 | 0.68 | 0.22 | 0.39 |
| | Population 2 | 0.32 | 0.22 | 0.40 |
| CDDP | Population 1 | 0.48 | 0.19 | 0.44 |
| | Population 2 | 0.52 | 0.23 | 0.30 |
| ISSR+SCoT+CDDP | Population 1 | 0.30 | 0.22 | 0.39 |
| | Population 2 | 0.70 | 0.23 | 0.33 |

## 4.1.1.7 Genetic variation of populations

The analysis of molecular variance (AMOVA) was used to estimate genetic variation among and within populations based on molecular data obtained by three markers (ISSR, SCoT, and CDDP) (Table 4.6). The significant PhiPT values were 0.08 with a p-value of 0.001 for ISSR, 0.05 with a p-value of 0.001 for SCoT, 0.12 with a p-value of 0.001 for CDDP, and 0.08 with a p-value of 0.001. The results for ISSR, SCoT, and CDDP data, as well as ISSR+SCoT+CDDP data, revealed that variation between populations accounted for 8.24, 5.23, 12.41, and 7.88% of overall variance, respectively. However, the highest variation within the population was created, with 91.76, 94.77, 87.59, and 92.12% for ISSR, CDDP, SCoT, and ISSR+SCoT+CDDP, respectively (Table 4.6). The gene flow values calculated from three types of markers were 2.78, 4.53, 1.76, and 2.92 for ISSR, SCoT, CDDP, and combined data, respectively (Figure 4.6).

**Table 4.6 Analysis of molecular variance for 64 tomato accessions by three different markers.**

| Method | Source | df | SS | MS | Est. Var. | % |
|---|---|---|---|---|---|---|
| ISSR | Among populations | 5.00 | 180.09 | 36.02 | 1.68 | 8.24** |
| | Within populations | 58.00 | 1086.26 | 18.73 | 18.73 | 91.76** |
| | Total | 63.00 | 1266.34 | | 20.41 | 100.00 |
| SCoT | Among populations | 5.00 | 357.28 | 71.46 | 2.52 | 5.23** |
| | Within populations | 58.00 | 2645.19 | 45.61 | 45.61 | 94.77** |
| | Total | 63.00 | 3002.47 | | 48.12 | 100.00 |
| CDDP | Among populations | 5.00 | 280.65 | 56.13 | 3.24 | 12.41** |
| | Within populations | 58.00 | 1325.48 | 22.85 | 22.85 | 87.59** |
| | Total | 63.00 | 1606.13 | | 26.09 | 100.00 |
| ISSR+SCoT+CDDP | Among populations | 5.00 | 818.95 | 163.79 | 7.46 | 7.88** |
| | Within populations | 58.00 | 5056.92 | 87.19 | 87.19 | 92.12** |
| | Total | 63.00 | 5875.87 | | 94.65 | 100.00 |

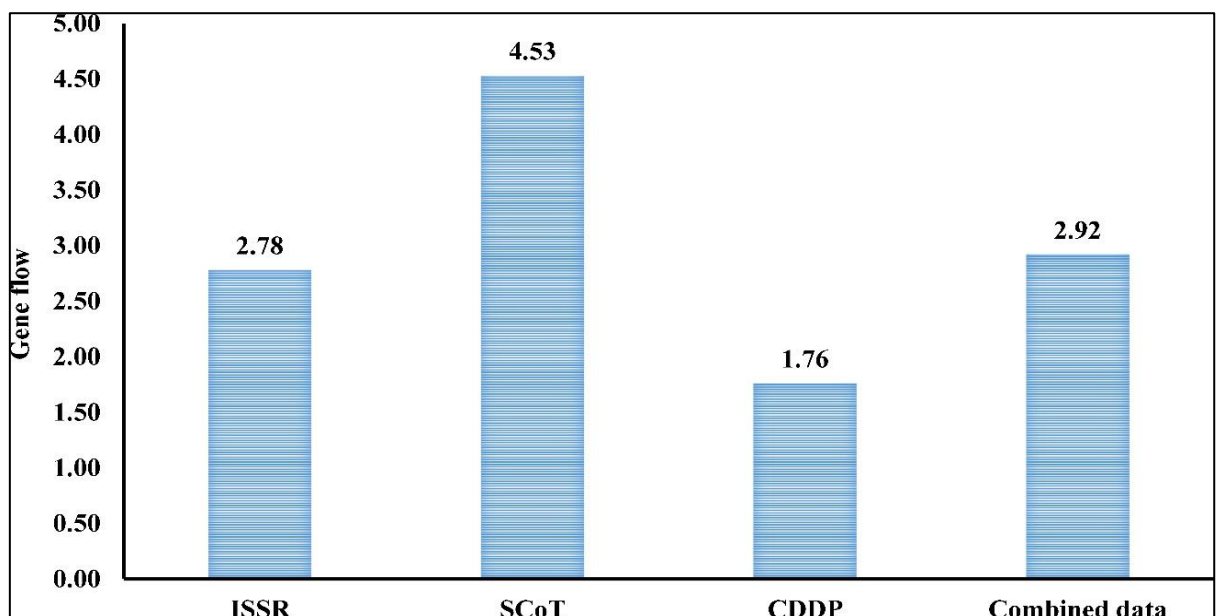

**Figure 4.6 Gene flow determined via PhiPT based on ISSR, SCoT, CDDP, and combined data.**

### 4.1.1.8 Correlation between dissimilarity matrices of three markers

The Mantel test was used to evaluate the association between three different genetic dissimilarity matrices (Table 4.7). A positive and significant association was found between ISSR and SCoT (0.76, p-value = 0.001), ISSR and CDDP (0.46, p-value = 0.001), SCoT and CDDP (0.61, p-value = 0.001), and among three distinct markers (0.68, p-value = 0.001).

**Table 4.7 Mantel coefficient values for distance matrices generated from different molecular markers.**

|  | ISSR | SCoT | CDDP | ISSR+SCoT+CDDP |
|---|---|---|---|---|
| ISSR | - | 0.76** | 0.46** | |
| SCoT | | - | 0.61** | 0.68** |
| CDDP | | | - | |

### 4.1.2 Assessment of tomato plants to drought stress

### 4.1.2.1 *In vitro* responses of tomato accessions to drought stress using PEG

***The effect of drought stress on seedling morphological traits***

According to the ANOVA analysis and box charts, there was a significant difference between the accessions under both control and drought-induced (PEG treatment) conditions (Table 4.8 and Figure 4.7). Table 4.9 shows that there were highly significant differences between accessions, PEG concentrations, and interactions between them for all morphological traits (GP, RL, SL, FW, and DW) (P ≤ 0.001). As PEG concentration increased, these traits significantly decreased for all accessions in all morphological characteristics except DW, which was increased (Table 4.8). Under the control condition, the scored data of GP for the accessions ranged from 55% (AC13) to 100% (AC4, AC6), with a mean of 90.10%; while RL from 3.12 (AC13) to 10.93 (AC29) cm, with an average of 8.21 cm; and SL from 4.50 (AC1) to 8.41 (AC63) cm, with an average of 7.08 cm; also, FW from 29.55 (AC11) to 68.97 (AC20) mg, with an average of 47.35 mg, and DW from 0.92 (AC11) to 2.83 (AC20) mg, with an average of 1.95 mg (Table 4.8 and Appendix 5).

At T1 (7.5% PEG) stress, the mean values of GP ranged from 55% (AC13) to 100% (AC4, AC6), with an average of 86.34%; and RL from 2.2 (AC13) to 7.98 (AC54) cm with an average of 6.14 cm, SL from 3.98 (AC18) to 7.20 (AC43) cm with an average of 5.52 cm, FW from 25.31 (AC18) to 56.78 (AC10) mg with a mean of 39.24 mg, and DW from 1.35 (AC42) to 3.92 (AC10) mg with an average of 2.61 mg (Table 4.7 and Appendix 6). The mean values of the accessions varied between 20.00% (AC13) to 94.67% (AC61) with an average of 80.04%, 0.83 (AC13) to 8.02 (AC56) cm with a mean of 4.65 cm, 1.00 (AC13) to 6.07 (AC63) cm with a mean of 3.63 cm, 16.72 (AC5) to 38.39 (AC60) mg with an average of 30.25 mg and 1.69 (AC42) to 5.00 (AC42) mg with a mean of 3.08 mg for GP, RL, SL, FW and DW, respectively, under induced drought stress T2 (15% PEG) (Table 4.7 and Appendix 7). As shown for each trait by the lower and upper box plot limits, the box charts (Figure 4.7) of all traits showed significant variations between T0 (Control), T1 (7.5% PEG), and T2 (15% PEG), all tomato accessions under control condition had significantly higher trait values when compared to stressed plants, except DW trait value which was lower compared to T1 (7.5% PEG), and T2 (15% PEG). These results indicate that T2 (15% PEG) had more effectiveness for decreasing the seedling growth.

The mean value of all accessions' morphological traits under all drought conditions, was indicated in Appendix 8 and the results showed high significant variation among accessions. The highest and lowest values of GP revealed for AC6 and AC13 which were 97.22% and 43.33% respectively, while AC50 and AC8 showed the longest and shortest root length (RL) values by 8.09 cm and 4.05 cm, respectively. The longest and shortest shoot lengths (SL) were recorded by AC63 (7.10 cm) and AC37 (4.33 cm), respectively. The highest value of FW and DW were indicted by AC30 and AC8 with 53.16 mg and 3.71 mg, respectively, while the lowest value of these traits recorded in AC6 with 27.82 mg and AC11 by 1.38 mg.

The interaction values between the tomato accessions and PEG induced was shown in Appendix 9, and there was a highly significant difference among them. The combination of AC4 + Control and AC6 + Control recorded the highest values of germination percentages (GP) by 100%, while the lowest value was revealed under AC13 + PEG-15, with 20%. The longest root length (RL) and shoot length (SL) recorded by the interaction of AC29 + Control and AC63 + Control, which were 10.93 cm 8.41cm, respectively, while the shortest values of these traits were indicated under the combination of AC13 + PEG-15 with 0.83 cm and 1.00 cm, respectively. The highest and the lowest values of FW revealed under the interaction of AC20 + Control and AC5 + PEG-15 which were 68.97 mg and 16.72 mg, respectively. The interaction of AC11 + Control indicated the lowest value of DW by 0.92 mg, whereas the combination of AC9 + PEG-15 recorded the highest value of DW which was 5.00 mg.

**Table 4.8 Descriptive statistics of morpho-chemical traits under different PEG concentrations.**

| Traits | T0 (Control) | | | | |
|--------|-----|-----|------|----|-----|
| | Min | Max | Mean | F | P>F |
| GP | 55 | 100 | 90.1 | 12.20*** | < 0.0001 |
| RL | 3.12 | 10.93 | 8.21 | 15.10*** | < 0.0001 |
| SL | 4.5 | 8.41 | 7.08 | 5.62*** | < 0.0001 |
| FW | 29.55 | 68.97 | 47.35 | 7.14*** | < 0.0001 |
| DW | 0.92 | 2.83 | 1.95 | 6.68*** | < 0.0001 |
| PC | 268.87 | 1697.59 | 854.45 | 104.83*** | < 0.0001 |
| SSC | 66.42 | 214.57 | 129.97 | 47.89*** | < 0.0001 |
| TPC | 49.96 | 131.8 | 78.33 | 40.42*** | < 0.0001 |
| AC | 461.89 | 653.78 | 557.82 | 89.32*** | < 0.0001 |
| GPA | 0.09 | 0.33 | 0.2 | 6.64*** | < 0.0001 |
| CAT | 25.97 | 149.35 | 88.88 | 8.06*** | < 0.0001 |
| LP | 2.98 | 6.82 | 4.55 | 110.74*** | < 0.0001 |
| | T1 (7.5% PEG) | | | | |
| GP | 55 | 100 | 86.34 | 8.24*** | < 0.0001 |
| RL | 2.21 | 7.98 | 6.14 | 11.00*** | < 0.0001 |
| SL | 3.98 | 7.2 | 5.52 | 7.02*** | < 0.0001 |
| FW | 25.31 | 56.78 | 39.24 | 6.64*** | < 0.0001 |
| DW | 1.35 | 3.92 | 2.61 | 6.58*** | < 0.0001 |
| PC | 617.08 | 2961.18 | 1447.35 | 677.37*** | < 0.0001 |
| SSC | 124.75 | 339.88 | 189.86 | 515.62*** | < 0.0001 |
| TPC | 92.85 | 505.58 | 269.15 | 2522.69*** | < 0.0001 |
| AC | 534.19 | 738.92 | 663.47 | 474.26*** | < 0.0001 |
| GPA | 0.19 | 0.51 | 0.35 | 41.63*** | < 0.0001 |
| CAT | 84.42 | 305.19 | 174.21 | 52.79*** | < 0.0001 |
| LP | 3.77 | 7.97 | 5.78 | 693.51*** | < 0.0001 |
| | T2 (15% PEG) | | | | |
| GP | 20 | 94.67 | 80.04 | 11.84*** | < 0.0001 |
| RL | 0.83 | 8.02 | 4.65 | 17.23*** | < 0.0001 |
| SL | 1 | 6.07 | 3.63 | 9.79*** | < 0.0001 |
| FW | 16.72 | 38.39 | 30.25 | 7.63*** | < 0.0001 |
| DW | 1.69 | 5 | 3.08 | 8.09*** | < 0.0001 |
| PC | 520.15 | 4714.77 | 2295.87 | 4439.84*** | < 0.0001 |
| SSC | 84.32 | 396.36 | 220.88 | 1334.72*** | < 0.0001 |
| TPC | 189.1 | 617.19 | 389.86 | 6249.08*** | < 0.0001 |
| AC | 570.68 | 839.59 | 732.29 | 105.95*** | < 0.0001 |
| GPA | 0.22 | 0.65 | 0.4 | 163.68*** | < 0.0001 |
| CAT | 71.43 | 350.65 | 178.77 | 115.61*** | < 0.0001 |
| LP | 4.18 | 9.19 | 6.39 | 524.02*** | < 0.0001 |

GP: germination percentage (%), RL: root length (cm), SL: shoot lenght (cm), FW: fresh weight (g), DW: dry weight (g), PC: proline content (µg/g FW), SSC: soluble sugar content (µg/g FW), TPC: total phenolic content (µg/g FW), AC: antioxidant capacity (µg/g FW), GPA: guaiacol peroxidase activity (units/min/g FW), CAT: catalase (units/min/g FW), and LP: lipid peroxidation (nmol/g FW), ***: highly highly significant.

**Table 4.9 Statistics that describe morphological traits and phytochemical parameters of accessions, concentration and thier combination.**

| Traits | Accessions | Pr > F | Concentration | Pr > F | Accessions* Concentration | Pr > F |
|--------|-----------|--------|---------------|--------|---------------------------|--------|
| GP | 26.51*** | <0.0001 | 184.31*** | <0.0001 | 2.87*** | <0.0001 |
| RL | 29.84*** | <0.0001 | 1717.21*** | <0.0001 | 6.57*** | <0.0001 |
| SL | 12.96*** | <0.0001 | 2319.67*** | <0.0001 | 4.64*** | <0.0001 |
| FW | 14.82*** | <0.0001 | 715.98*** | <0.0001 | 3.21*** | <0.0001 |
| DW | 16.30*** | <0.0001 | 493.62*** | <0.0001 | 2.75*** | <0.0001 |
| PC | 3091.76*** | <0.0001 | 111025.50*** | <0.0001 | 1050.77*** | <0.0001 |
| SSC | 1121.39*** | <0.0001 | 25652.06*** | <0.0001 | 311.35*** | <0.0001 |
| TPC | 5368.06*** | <0.0001 | 631700.60*** | <0.0001 | 1434.61*** | <0.0001 |
| AC | 239.4224*** | <0.0001 | 36960.56*** | <0.0001 | 135.4114*** | <0.0001 |
| GPA | 133.43*** | <0.0001 | 6638.74*** | <0.0001 | 33.83*** | <0.0001 |
| CAT | 112.90*** | <0.0001 | 4799.69*** | <0.0001 | 49.45*** | <0.0001 |
| LP | 1392.67*** | <0.0001 | 28019.48*** | <0.0001 | 114.31*** | <0.0001 |

**GP: germination percentage (%), RL: root length (cm), SL: shoot lenght (cm), FW: fresh weight (g), DW: dry weight (g), PC: proline content (µg/g FW), SSC: soluble sugar content (µg/g FW), TPC: total phenolic content (µg/g FW), AC: antioxidant capacity (µg/g FW), GPA: guaiacol peroxidase activity (units/min/g FW), CAT: catalase (units/min/g FW), and LP: lipid peroxidation (nmol/g FW).**

### *Drought stress tolerance through biochemical characteristics*

The ANOVA analysis (Table 4.8) and box charts (Figure 4.8) showed that there were significant differences among the accessions in the phytochemical results of (PC, SSC, TPC, AC, GPA, CAT and LP) from the control (T0) and both PEG treatments (T1: 7.5% PEG and T2: 15% PEG) that were carried out in drought conditions. Furthermore, all biochemical traits were significantly different among accessions, PEG concentrations and their interactions (P ≤ 0.001) (Table 4.9).

According to (Table 4.8), all biochemical traits were increased with an increase in the PEG concatenations that were induced, stressed tomato seedlings produced a high level of these chemical compounds as compared to control tomato seedlings. In response to drought stress, plants produce and store the necessary solutes, such as amino acids, polyols, and carbohydrates, to support osmotic balance and the absorption and retention of water. To assess the tomato's response to the low water potential produced by PEG, proline content (PC) and soluble sugar content (SSC) were evaluated. Proline content (PC) was increased gradually as PEG concentration increased. The mean of all accessions under control condition (0.0% PEG) was 854.45 µg/g FW and PC was increased significantly under T1 (7.5% PEG) and T2 (15% PEG), which were 1447.34 µg/g FW and 2295.87 µg/g FW (Table 4.8). The minimum and maximum PC under control conditions were recorded by AC1 and AC42 (Appendix 10), which were 268.87 µg/g FW and 1697.59 µg/g FW, respectively, while the highest PC values under T1 (7.5% PEG) and T2 (15% PEG) were indicated by AC61 (2961.18 µg/g FW) (Appendix 11) and AC27 (4714.77 µg/g FW) (Appendix12), respectively.

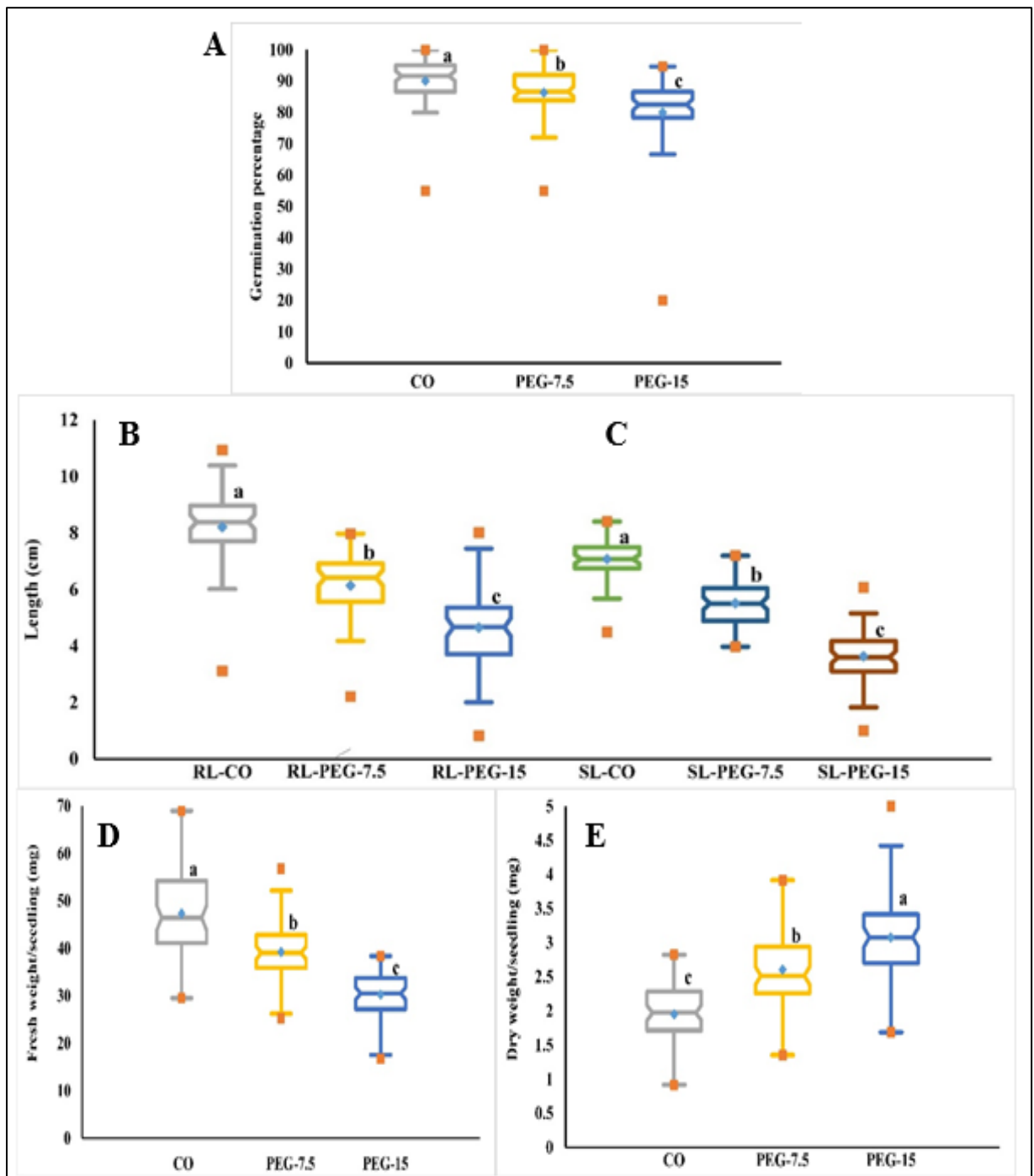

**Figure 4.7 Graph box illustrating the variation in phenotypic traits under control and stress conditions. (A) Germination percentage (GP), (B) Root length (RL), (C) Shoot length (SL), (D) Seedling fresh weight (FW), and (E) Seedling dry weight (DW). The values specified are the mean values determined for the three measurements collected for (T0= 0.00%PEG) control (CO) and PEG concentrations (T1= 7.5% PEG and T2= 15%PEG). Different letters represent a significant difference between the mean values according to Duncan's Multiple-Range Test (P ≤ 0.01). A blue dot in the box indicates the mean, while orang dots represent the minimum and maximum values.**

The lowest values of this chemical trait under both induced PEG treatments were shown by AC11 (Appendix 11) and AC30 (Appendix12) by 617.08 µg/g FW and 520.15 µg/g FW, respectively.

These results demonstrate that the T2 (15% PEG) was more effective on sensitive accessions compared to tolerant accessions. The average values of soluble sugar contents (SSC) were 129.97 µg/g FW, 189.86 µg/g FW, and 220.88 µg/g FW under T0 (Control), T1 (7.5% PEG), and T2 (15% PEG) conditions, respectively. The highest and the lowest values of SSC under control conditions were 66.42 µg/g FW and 214.57 µg/g FW; at T1 (7.5% PEG) were 124.75 µg/g FW and 339.88 µg/g FW; and at T2 (15% PEG) were 84.32 µg/g FW and 396.36 µg/g FW, respectively (Table 4.8, Appendices, 10, 11 and 12).

Total phenolic contents (TPC) and antioxidant capacity (AC) were affected by drought stress using PEG concentrations compared to control conditions in seedlings of all tomato accessions. The average values of these chemical parameters were increased by increasing induced PEG concentrations (Table 4.8). Under control condition, the average of TPC and AC were 78.33 µg/g FW and 557.82 µg/g FW, respectively, while the minimum values of these traits were recorded by AC28 and AC60 with 49.96 µg/g FW and 461.89 µg/g FW, respectively, and the maximum values were indicated by AC6 and AC11 with 131.80 µg/g FW and 653.78 µg/g FW, respectively (Table 4.8 and Appendix 10). The mean values of TPC and AC were 269.15 µg/g FSW and 663.47 µg/g FW, respectively, while 7.5% PEG was induced. The lowest and the highest values of TPC under T1 (7.5% PEG) were recorded by AC11 with 92.85 µg/g FW and AC61 with 505.58 µg/g FW, respectively. The minimum and the maximum values of AC at T1 (7.5% PEG) were indicated by AC14 with 534.19 µg/g FW and AC24 with 738.92 µg/g FW, respectively (Table4.8 and Appendix 11). T2 (15% PEG) treatment revealed the highest TPC and AC values when compared to the control condition, with TPC averaging 389.86 g/g FSW and AC averaging 732.29 g/g FW were resulted. The minimum values of TPC and AC for this treatment were recorded at AC11 with 189.10 µg/g FW and AC14 with 570.68 µg/g FW, respectively, while the maximum values of TPC and AC were revealed in AC61 with 617.19 µg/g FW and AC6 with 839.59 µg/g FW, respectively (Table 4.8 and Appendix 12). The enzymatic activities of guaiacol peroxidase activity (GPA), catalase activity (CAT), and lipid peroxidase (LP) of tomato seedlings were gradually impacted by enhanced drought stress. The average GPA values under T0 (Control), T1 (7.5% PEG), and T2 (15% PEG) conditions were 0.20, 0.35, and 0.40 units/min/g FW, respectively, whereas the CAT means were 88.88, 174.21, and 178.77 units/min/g FW, respectively, and the average values of LP were 4.55, 5.78, and 6.39 nmol/g FW, respectively (Table 4.8). The minimum values of GPA at T0 (AC9 and AC19), T1 (AC8), and T2 (AC19) were 0.089, 0.189, and 0.218 units/min/g FW, respectively, and the lowest values of CAT were 25.97, 84.42, and 71.43 units/min/g FW in AC2, AC2, and AC13, respectively, and for LP 2.98, 3.77, and 4.18 nmol/g FW in AC18, AC27, and AC27, respectively were resulted (Table 4.8, Appendices 10, 11 and 12). The maximum value of GPA under control conditions were recorded in AC50 and AC54, which was 0.331 units/min/g FW, and CAT in AC6, AC14, AC20, AC29, AC31, and AC59 with 149.35 units/min/g FW, while the value of LP was 6.82 nmol/g FW at AC5. The highest values of GPA, CAT, and LP were recorded in AC50, AC61, and AC45 with 0.508 units/min/g FW, 305.19 units/min/g FW, and 7.97 nmol/g FW, respectively in T1 (7.5% PEG). Using T2 (15% PEG), the highest values of GPA, CAT, and LP were 0.654 units/min/g FW, 350.65 units/min/g FW, and 9.19 nmol/g FW in AC54, AC61, and AC5 respectively.

The average values of all biochemical traits under drought conditions (T0, T1, and T2) were shown in Appendix 13. The data obtained revealed highest significant differences among all accessions. AC39 and AC11 showed the maximum and the minimum PC values with 2786.48 µg/g FW and 641.61 µg/g FW, respectively. The highest and lowest SSC values were displayed by AC58 and AC57, with (297.18 µg/g FW and 95.53 µg/g FW) and (403.83 µg/g FW and 114.07 µg/g FW), respectively, AC61 and AC11 had the highest and lowest TPC values. The two accessions with the highest and lowest AC values were AC611 and AC58, with 708.51 µg/g FW and 553.78 µg/g FW, respectively. The maximum values of GPA, CAT, and LP were recorded by AC54, AC61, and AC5 with 0.495 units/min/g FW, 242.42 units/min/g FW, and 7.94 nmol/g FW, respectively, while the minimum values of these traits revealed in AC19, AC2, and AC27 with 0.175 units/min/g FW, 69.26 units/min/g FW, and 3.78 nmol/g FW, respectively.

The interactions of accessions with drought conditions for all phytochemicals are indicated in Appendix 14. The highest values of PC and SSC were revealed under the combination of AC27 + PEG-15, which were 4714.77 µg/g FSW and 396.36 µg/g FW, respectively, while the lowest PC was indicated at the interaction of AC1 + Control, with 268.87 µg/g FW and the minimum SSC was revealed at AC8 + Control by 66.42 µg/g FW. In the case of TPC and AC, the maximum value was recorded by the combination of AC61 + PEG-15 (617.19 µg/g FW) and AC6 + PEG-15 (839.59 µg/g FW) respectively, while the minimum values presented by the interaction of AC28 + Control and AC60 + Control with 49.96 µg/g FW and 461.89 µg/g FW, respectively.

The highest values of GPA, CAT and LP were 0.654 units/min/g FW, 350.65 units/min/g FW, and 9.19 nmol/g FW at the interaction of AC54 + PEG-15, AC61 + PEG-15, and AC5 + PEG-15, respectively, meanwhile, the lowest values were recorded by the combination of AC9 + Control (0.089 units/min/g FW), AC2 + Control (25.97 units/min/g FW), and AC18 + Control (2.98 nmol/g FW), respectively.

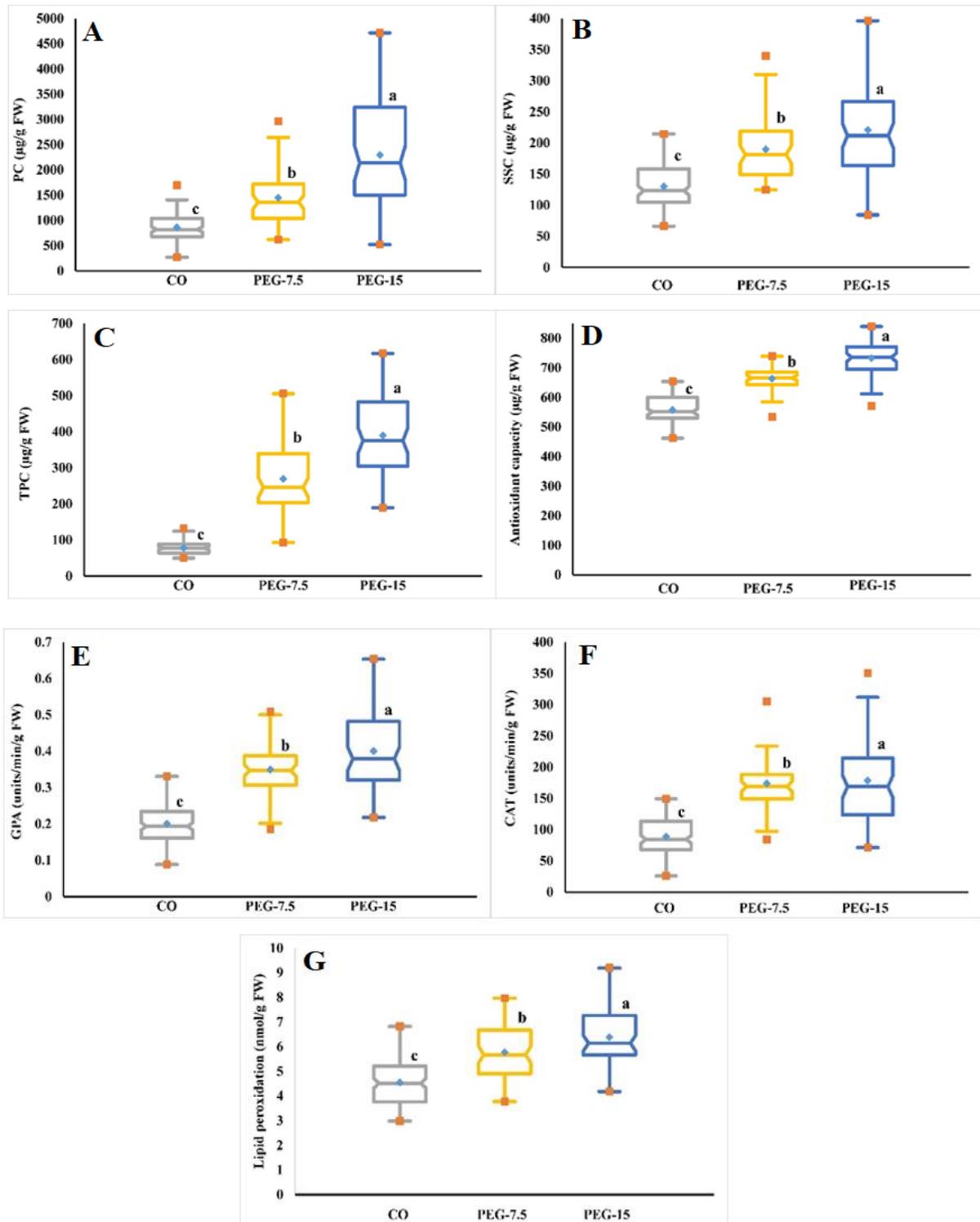

**Figure 4.8 Box chart indicates of distinction of the phytochemical stress markers in tomato seedlings. (A) Proline content (PC), (B) soluble sugar content (SSC), (C) total phenolic content (TPC), antioxidant capacity (AC), (E) ) guaiacol peroxidase activity (GPA), (F) Catalase activity (CAT) and (G) lipid peroxidation (LP) assay in tomato accessions under control and drought stress. The values given are the mean values obtained for the three measurements for (T0= 0.00%PEG) control (CO) and PEG concentrations (T1= 7.5% PEG and T2= 15%PEG). Different letters represent a significant difference between the mean values according to Duncan's Multiple-Range Test (P ≤ 0.01). A blue dot in the box indicates the mean, while red dots represent the minimum and maximum values.**

*Ranking of accessions for germination percentage and seedling elongation*

The best accessions with the tested features were identified using a ranking approach based on germination percentage and seedling growth (root length + shoot length).

The best accessions for germination and seedling growth have been determined using the average sum of rankings (AR), which has been applied as a predictor. The best accession to drought tolerance has been selected based on the highest stress tolerance index (STI) and the lowest average number of ranks (AR). AC4 and AC63 had the highest performances under drought stress treatments T1 (7.5% PEG) and T2 (15% PEG) conditions, respectively, while AC13 had the lowest response to drought under both conditions (Table 4.10). The important ranking for selecting the best accession to drought stress was under the combination of T1 and T2. According to the results (Table 4.10), AC61, AC9, and AC63 had the highest ranking for response to PEG, and it can be assumed that these accessions are the most resistant to drought stress. On the other hand, AC13, AC30, and AC8 were the most susceptible accessions to PEG.

*Omics analysis*

In order to find features that are affected by descriptive variables, statistical differential expression (Omics method) has been applied in the disciplines of genomics and biochemistry. Utilizing the mean values of chemical features obtained in response to various induced PEG (Table 4.11), we were able to determine the degree of tolerance and susceptibility in our case study.

Under the T1 (7.5% PEG) and T2 (15% PEG) applications, three biochemical traits (PC, SSC, and TPC) indicated highly significant responses of tomato accessions. The mean value of PC under T1 ranged between (1642.00 and 821.24) for the high tolerant and low tolerant responses, respectively, while under T2, it was varied between 1880.00 and 881.18, for the high tolerant and low tolerant responses, respectively. In the case of SSC, high tolerate response, recorded the mean value of 204.13 and 196.86 under T1 and T2, respectively, whereas in low tolerate response, the mean values were 142.42 and 143.45 under T1 and T2, respectively. TPC trait, under T1 recorded the mean values for high and low tolerate response which were 302.84 and 151.41, respectively, and indicated the mean values of 353.89 and 241.08 for high and low tolerate response under T2, respectively.

**Table 4.10 Ranking of sixty-four tomato accessions, using the stress tolerance index (STI) and average number of ranks (AR). According to the germination percentage and growth traits of seedlings, grown under the drought conditions of T1 (7.5% PEG), T2 (15% PEG) and both treatments (T1 and T2). The best accessions were those**

with the highest STI and lowest AR, and the lowest rank was assigned to the most consistent performance of each accession.

| Accession Code | T1 | | | T2 | | | T1 and T2 | | |
|---|---|---|---|---|---|---|---|---|---|
| | STI | AR | Rank | STI | AR | Rank | STI | AR | Rank |
| AC1 | 0.82 | 56.45 | 56 | 0.94 | 14.27 | 19 | 0.88 | 44.82 | 51 |
| AC2 | 0.85 | 54.09 | 53 | 0.90 | 32.82 | 27 | 0.87 | 48.64 | 53 |
| AC3 | 0.79 | 57.45 | 57 | 0.72 | 54.82 | 57 | 0.76 | 56.55 | 58 |
| AC4 | 1.15 | 4.64 | 1 | 0.98 | 25.00 | 11 | 1.07 | 14.45 | 8 |
| AC5 | 0.85 | 39.45 | 52 | 0.78 | 40.00 | 54 | 0.82 | 39.27 | 42 |
| AC6 | 1.14 | 5.91 | 2 | 1.01 | 14.64 | 3 | 1.08 | 10.00 | 4 |
| AC7 | 1.01 | 19.73 | 21 | 0.86 | 41.91 | 37 | 0.94 | 31.27 | 32 |
| AC8 | 0.64 | 47.45 | 61 | 0.39 | 62.73 | 63 | 0.52 | 61.09 | 62 |
| AC9 | 1.11 | 5.55 | 6 | 1.00 | 9.73 | 4 | 1.06 | 7.09 | 2 |
| AC10 | 0.90 | 41.45 | 42 | 0.84 | 31.64 | 40 | 0.87 | 35.64 | 38 |
| AC11 | 0.75 | 45.27 | 60 | 0.46 | 61.91 | 62 | 0.60 | 59.82 | 61 |
| AC12 | 0.94 | 35.36 | 36 | 0.90 | 20.18 | 26 | 0.92 | 25.36 | 22 |
| AC13 | 0.37 | 45.64 | 64 | 0.13 | 64.00 | 64 | 0.25 | 63.73 | 64 |
| AC14 | 0.77 | 55.09 | 59 | 0.73 | 45.27 | 56 | 0.75 | 49.91 | 55 |
| AC15 | 0.98 | 39.82 | 28 | 0.98 | 15.82 | 8 | 0.98 | 28.27 | 29 |
| AC16 | 0.94 | 33.55 | 35 | 0.81 | 46.82 | 46 | 0.88 | 41.55 | 47 |
| AC17 | 1.03 | 31.45 | 16 | 0.92 | 34.45 | 23 | 0.98 | 32.09 | 34 |
| AC18 | 0.92 | 47.36 | 40 | 0.87 | 35.73 | 36 | 0.89 | 42.09 | 48 |
| AC19 | 0.87 | 51.18 | 47 | 0.77 | 54.09 | 55 | 0.82 | 53.27 | 57 |
| AC20 | 0.86 | 49.91 | 51 | 0.78 | 48.09 | 53 | 0.82 | 48.55 | 52 |
| AC21 | 0.91 | 32.00 | 41 | 0.79 | 45.36 | 50 | 0.85 | 40.27 | 45 |
| AC22 | 0.95 | 32.27 | 34 | 0.82 | 45.09 | 44 | 0.89 | 38.45 | 41 |
| AC23 | 0.93 | 40.36 | 37 | 0.89 | 27.82 | 30 | 0.91 | 33.27 | 36 |
| AC24 | 0.92 | 29.45 | 39 | 0.85 | 29.18 | 38 | 0.89 | 28.82 | 31 |
| AC25 | 0.95 | 41.82 | 33 | 0.89 | 34.18 | 31 | 0.92 | 40.00 | 44 |
| AC26 | 1.07 | 18.18 | 12 | 0.98 | 16.64 | 10 | 1.03 | 14.82 | 9 |
| AC27 | 1.00 | 16.55 | 22 | 0.95 | 13.00 | 16 | 0.97 | 13.91 | 7 |
| AC28 | 1.00 | 29.18 | 23 | 0.93 | 21.27 | 21 | 0.97 | 22.00 | 17 |
| AC29 | 1.02 | 29.09 | 19 | 0.94 | 24.82 | 18 | 0.98 | 25.45 | 23 |
| AC30 | 0.53 | 62.73 | 63 | 0.48 | 60.45 | 61 | 0.51 | 62.73 | 63 |
| AC31 | 1.00 | 23.27 | 25 | 0.93 | 18.00 | 20 | 0.97 | 17.55 | 13 |
| AC32 | 1.02 | 24.91 | 17 | 0.94 | 21.55 | 17 | 0.98 | 19.45 | 15 |
| AC33 | 0.62 | 46.36 | 62 | 0.51 | 56.64 | 60 | 0.56 | 52.82 | 56 |
| AC34 | 1.01 | 29.09 | 20 | 0.93 | 25.55 | 22 | 0.97 | 25.45 | 24 |
| AC35 | 0.84 | 44.55 | 54 | 0.65 | 57.82 | 59 | 0.75 | 56.73 | 59 |
| AC36 | 0.97 | 38.55 | 31 | 0.89 | 36.73 | 32 | 0.93 | 37.91 | 39 |
| AC37 | 0.86 | 53.18 | 50 | 0.66 | 57.55 | 58 | 0.76 | 56.82 | 60 |
| AC38 | 0.97 | 20.36 | 29 | 0.92 | 16.27 | 24 | 0.95 | 16.91 | 11 |
| AC39 | 1.08 | 18.55 | 9 | 1.00 | 12.00 | 5 | 1.04 | 12.09 | 6 |
| AC40 | 0.95 | 36.55 | 32 | 0.84 | 44.55 | 41 | 0.90 | 39.45 | 43 |
| AC41 | 1.07 | 25.55 | 11 | 0.97 | 26.91 | 12 | 1.02 | 25.27 | 21 |
| AC42 | 1.11 | 9.36 | 7 | 0.95 | 26.73 | 15 | 1.03 | 17.36 | 12 |
| AC43 | 1.12 | 13.55 | 4 | 0.99 | 19.18 | 6 | 1.05 | 14.91 | 10 |
| AC44 | 1.05 | 20.45 | 13 | 0.79 | 51.36 | 52 | 0.92 | 42.27 | 49 |
| AC45 | 0.89 | 28.00 | 43 | 0.82 | 32.73 | 45 | 0.85 | 31.36 | 33 |
| AC46 | 1.02 | 15.00 | 18 | 0.90 | 29.55 | 29 | 0.96 | 22.36 | 18 |
| AC47 | 1.04 | 9.91 | 15 | 0.85 | 39.27 | 39 | 0.95 | 20.00 | 16 |
| AC48 | 1.05 | 11.00 | 14 | 0.83 | 46.91 | 42 | 0.94 | 33.00 | 35 |
| AC49 | 0.79 | 54.55 | 58 | 0.79 | 35.18 | 51 | 0.79 | 43.45 | 50 |
| AC50 | 0.98 | 32.18 | 27 | 0.95 | 12.64 | 14 | 0.97 | 18.09 | 14 |
| AC51 | 0.87 | 39.45 | 48 | 0.79 | 38.91 | 49 | 0.83 | 38.27 | 40 |
| AC52 | 0.87 | 35.36 | 49 | 0.81 | 34.55 | 47 | 0.84 | 33.82 | 37 |
| AC53 | 0.87 | 39.55 | 46 | 0.88 | 21.82 | 35 | 0.87 | 27.82 | 28 |
| AC54 | 0.98 | 19.82 | 26 | 0.98 | 7.82 | 9 | 0.98 | 10.09 | 5 |
| AC55 | 0.89 | 37.45 | 44 | 0.90 | 16.27 | 28 | 0.89 | 22.82 | 20 |
| AC56 | 0.88 | 37.82 | 45 | 0.88 | 20.64 | 34 | 0.88 | 26.09 | 25 |
| AC57 | 0.82 | 54.64 | 55 | 0.79 | 41.18 | 48 | 0.81 | 49.36 | 54 |
| AC58 | 0.97 | 38.91 | 30 | 0.96 | 15.55 | 13 | 0.96 | 26.55 | 26 |
| AC59 | 0.92 | 37.55 | 38 | 0.82 | 41.27 | 43 | 0.87 | 40.91 | 46 |
| AC60 | 1.08 | 20.82 | 10 | 0.90 | 37.36 | 25 | 0.99 | 28.73 | 30 |
| AC61 | 1.13 | 3.00 | 3 | 1.07 | 3.91 | 2 | 1.10 | 2.73 | 1 |
| AC62 | 1.00 | 20.55 | 24 | 0.88 | 32.64 | 33 | 0.94 | 26.55 | 27 |
| AC63 | 1.11 | 18.82 | 5 | 1.08 | 5.27 | 1 | 1.09 | 7.36 | 3 |
| AC64 | 1.09 | 22.91 | 8 | 0.99 | 24.00 | 7 | 1.04 | 22.45 | 19 |

**Table 4.11 Statistical omics analysis for integrating the responses of tested materials by different biochemical traits in the presence of three different treatments of cadmium.**

| Treatments | Traits | P-value | Significant | Moderate Tolerance | High Tolerance | Low Tolerance |
|---|---|---|---|---|---|---|
| T1 (%7.5 PEG) | PC | 0.00 | Yes | 1158 (a) | 1642 (b) | 821.244 (a) |
| | SSC | 0.00 | Yes | 169.035 (a) | 204.132 (b) | 142.423 (a) |
| | TPC | 0.00 | Yes | 221.242 (a) | 302.840 (b) | 151.414 (a) |
| T2 (%15 PEG) | PC | 0.00 | Yes | 3281 (c) | 1880 (b) | 881.179 (a) |
| | SSC | 0.00 | Yes | 276.489 (c) | 196.863 (b) | 143.448 (a) |
| | TPC | 0.00 | Yes | 482.726 (c) | 353.889 (b) | 241.081 (a) |

## 4.1.2.2 Effects of oak leaf extract, biofertilizer, and soil containing oak leaf powder on tomato growth and biochemical characteristics under water stress conditions

Effect of various treatments on the morpho-physiological and fruit physicochemical traits of tomato under water stress are shown in (Table 4.12). Plant development and growth are essentially the results of cell division, cell enlargement, and differentiation, and they are regulated by a variety of genetic, physiological, ecological, and morphological processes, as well as their interconnections (Ullah *et al.*, 2016). The analysis of variance on morphological characters, relative water content (RWC), and total chlorophyll content (TCC) in the first stress stage (before flowering), the second stress stage (before fruiting), and their combinations revealed that treatments had a significant effect (Appendix 15 and 16). When compared with control plants, all levels of treatments resulted in significant percentage decreases in shoot length (SL), shoot fresh weight (SFW), shoot dry weight (SDW), fruit weight per plant (FWT), relative water content (RWC), and total chlorophyl content (TCC). In comparison with control plants, the stressed plant group (SS) that was not exposed to powdered oak tissue, oak leaf extract, or biofertilizer at any stage had the highest decline percentages for all traits (Table 4.12).

According to the results of the interaction, Braw under SOBS application resulted in the highest increasing percentages of SFW (33.35%), SDW (51.30%), and RFW (145.06%) compared with the irrigated plants (SW) during the first stress stages, while Yadgar under untreated and stressful conditions (SS) resulted in the maximum decreasing values for FWT (50.38%) and RWC (18.72%) (Appendix 17). The interaction results showed that, during the second stress stage, Braw under SOBS application contributed to the greatest increases in SFW (5.03%), SDW (29.64%), and RFW (258.68%) compared with the control conditions, while Yadgar (48.30%) and Sandra (48.11%) under the SOS conditions caused the greatest decreases in FWT and TCC. As indicated in (Appendix 17), the interaction outcomes demonstrated that the Sandra accession under SOBS application contributed to the highest increases in SDW (2.74%), and RDW (255.70%) compared with SW conditions, and Yadgar under the SS condition caused the greatest decreases in SL (26.52%) and FWT (63.89%) during the combination of both stress stages.

**Table 4.12 Effect of oak leaf powder, oak leaf extract, and biofertilizer on the morpho-physiological characteristics of tomato plants at various stress stages. Positive and negative values signify increasing and declining, respectively.**

| Increasing and decreasing percentages compared with irrigated plants in the first stress stage | | | | | | | | | |
|---|---|---|---|---|---|---|---|---|---|
| Treatment | SL (%) | SFW (%) | SDW (%) | RL (%) | RFW (%) | RDW (%) | FWT (%) | RWC (%) | TCC (%) |
| SOBS | −6.70 a ± 5.15 | 1.90 a ± 20.80 | 7.72 a ± 27.31 | 4.94 a ± 16.58 | 74.93 a ± 50.94 | 99.21 a ± 84.92 | −27.30 ab ± 9.53 | −11.94 ab ± 3.26 | −24.30 b ± 12.18 |
| SOES | −6.54 a ± 8.47 | −5.78 b ± 7.59 | −3.24 b ± 8.14 | 0.03 ab ± 12.43 | 44.00 b ± 58.76 | 43.76 ab ± 117.82 | −21.36 a ± 16.97 | −9.44 a ± 8.07 | −9.80 a ± 18.67 |
| SOS | −7.13 a ± 5.52 | −8.93 b ± 6.53 | −11.53 c ± 6.95 | −4.57 b ± 15.55 | 30.05 b ± 26.34 | 22.51 b ± 20.64 | −28.59 b ± 10.07 | −14.24 bc ± 3.89 | −33.34 b ± 8.43 |
| SS | −13.52 b ± 6.56 | −24.76 c ± 15.11 | −22.98 d ± 14.67 | −16.67 c ± 11.37 | 16.16 b ± 29.80 | 26.11 b ± 52.72 | −32.58 b ± 11.58 | −16.75 c ± 4.69 | −31.30 b ± 10.72 |
| Increasing and decreasing percentages compared with irrigated plants in the second stress stage | | | | | | | | | |
| Treatment | SL (%) | SFW (%) | SDW (%) | RL (%) | RFW (%) | RDW (%) | FWT (%) | RWC (%) | TCC (%) |
| SOBS | −9.28 a ± 4.57 | −7.95 a ± 13.31 | 3.02 a ± 18.52 | 15.70 a ± 17.39 | 107.57 a ± 104.78 | 121.80 a ± 91.08 | −28.49 a ± 13.46 | −10.12 a ± 4.59 | −24.18 b ± 12.07 |
| SOES | −9.04 a ± 5.33 | −14.16 b ± 10.09 | −5.18 b ± 11.10 | 5.96 b ± 15.92 | 94.83 a ± 86.72 | 104.54 a ± 82.27 | −30.57 a ± 11.03 | −8.87 a ± 7.42 | −16.85 a ± 12.49 |
| SOS | −13.00 a ± 6.63 | −20.21 c ± 10.32 | −14.03 c ± 8.70 | −11.47 c ± 14.93 | 30.54 b ± 39.65 | 36.60 b ± 26.44 | −35.13 b ± 10.26 | −10.88 a ± 3.99 | −33.29 c ± 10.34 |
| SS | −20.56 b ± 8.62 | −29.05 d ± 15.83 | −25.14 d ± 14.76 | −12.84 c ± 11.43 | 29.59 b ± 39.99 | 31.48 b ± 67.26 | −37.64 b ± 11.47 | −18.14 b ± 6.84 | −31.55 c ± 10.75 |
| Increasing and decreasing percentages compared with irrigated plants in the first and second stress stages | | | | | | | | | |
| Treatment | SL (%) | SFW (%) | SDW (%) | RL (%) | RFW (%) | RDW (%) | FWT (%) | RWC (%) | TCC (%) |
| SOBS | −13.95 a ± 7.95 | −15.22 b ± 12.10 | −9.28 a ± 11.43 | 8.52 a ± 17.42 | 92.02 a ± 83.34 | 101.46 a ± 100.45 | −41.10 a ± 13.87 | −15.59 a ± 6.65 | −26.22 b ± 11.54 |
| SOES | −14.57 a ± 7.57 | −9.14 a ± 12.32 | −8.57 a ± 8.88 | 2.52 ab ± 14.18 | 89.63 a ± 76.02 | 100.71 a ± 106.78 | −39.72 a ± 12.98 | −13.70 a ± 6.65 | −16.24 a ± 14.80 |
| SOS | −17.50 a ± 8.64 | −16.48 b ± 13.13 | −13.05 b ± 10.59 | −3.73 b ± 9.00 | 51.78 b ± 52.23 | 62.56 b ± 117.59 | −40.04 a ± 11.37 | −16.87 a ± 6.45 | −32.91 c ± 8.37 |
| SS | −23.80 b ± 9.89 | −36.00 c ± 23.39 | −27.49 c ± 16.15 | −12.98 c ± 6.97 | 27.08 b ± 35.52 | 37.11 c ± 86.64 | −45.10 b ± 13.80 | −22.06 b ± 5.42 | −35.64 c ± 10.40 |

**SL: shoot length, SFW: shoot fresh weight, SDW: shoot dry weight, RL: root length, RFW: root fresh weight, RDW: root dry weight, FWT: fruits weight per plant, RWC: relative water content, TCC: total chlorophyl content, SS: stressed plants that had not been treated, SOS: stressed plants that had been treated with oak leaf powder, SOES: stressed plants that had been treated with oak leaf powder and oak leaf extract, SOBS: stressed plants that had been treated with oak leaf powder and biofertilizers. Duncan's multiple range test at P ≤ 0.05 indicates that any mean values sharing the same letter in the same column are not statistically significant. The value is represented by trait index ± standard deviation (SD). Each value is the average of eight measurements.**

The analysis of variance (ANOVA) of the data reported significant influences of the treatment on the fruit's physicochemical properties (Appendix 18). As stated in Table 4.13, the titratable acidity (TA). ascorbic acid content (ASC), and total phenolic content (TPC) responded positively to different levels of treatments in all stages of growth. In the first stress stage, the highest increasing percentages of TA, ASC, and TPC were obtained by the treatments SS (11.23%), SOBS (23.50%), and SOES (11.10%), respectively. The TA, ASC, and TPC responded favorably to various treatments during the second stress stage. The highest increasing TA (12.63%), ASC (18.49%), and TPC (12.21%) values were seen in the treatments SS, SOES, and SOBS, respectively. Similarly, when two stress measures were combined, the same results were found. In the SS and SOES applications, the highest percentage increases in TA (19.05%), ASC (13.11%), and TPC (10.42%) were shown. Under all stress conditions, a decreasing amount was also observed in the moisture content (MC), total soluble solids (TSSs), and carotenoid content (CAC). The SS application showed the largest decline in percentage of MC, TSS, and CAC. With the first stress stage, the soluble sugar content (SSC) decreased by 3.27 and 2.78% under SOBS and SOES conditions, respectively. The SSC responded favorably to the SOBS and SOES applications during the second stress stage, increasing by 1.68 and 2.73%, respectively. Under all levels of treatment (SS, SOS, SOES, and SOBS), the SSC values for both stress stages together decreased.

**Table 4.13 Influence of oak leaf powder, oak leaf extract, and biofertilizer on the fruit physicochemical parameters of tomato plants at different stress stages. Increasing and decreasing are labeled by a positive and negative value, respectively.**

| | MC | TA | TSS | ASC | CAC | SSC | TPC |
|---|---|---|---|---|---|---|---|
| Treatment | \multicolumn Increasing and decreasing percentages compared with irrigated plants in the first stress stage | | | | | | |
| SOBS | −0.67 a ± 0.45 | 2.64 bc ± 6.43 | −1.73 a ± 4.35 | 23.50 a ± 12.40 | −3.02 b ± 4.43 | 3.27 a ± 7.73 | 8.55 b ± 12.38 |
| SOES | −0.59 a ± 0.35 | 1.03 c ± 6.81 | −2.08 a ± 3.60 | 22.10 b ± 14.06 | −1.80 a ± 4.25 | 2.78 a ± 8.68 | 11.10 a ± 10.93 |
| SOS | −0.87 b ± 045 | 5.82 b ± 5.12 | −4.43 b ± 4.29 | 15.92 c ± 11.81 | −5.35 c ± 6.19 | −2.34 b ± 7.72 | 9.06 b ± 10.41 |
| SS | −1.28 c ± 0.83 | 11.23 a ± 6.53 | −6.67 c ± 4.37 | 6.41 d ± 10.22 | −7.64 d ± 7.61 | −8.28 c ± 11.84 | 5.06 c ± 10.63 |
| Treatment | Increasing and decreasing percentages compared with irrigated plants in the second stress stage | | | | | | |
| | MC | TA | TSS | ASC | CAC | SSC | TPC |
| SOBS | −0.65 a ± 0.046 | 3.45 b ± 7.54 | −1.59 a ± 3.23 | 17.22 b ± 18.89 | −0.03 a ± 6.67 | 1.68 b ± 10.98 | 12.21 a ± 14.70 |
| SOES | −0.55 a ± 0.039 | 2.12 b ± 7.50 | −2.42 a ± 3.26 | 18.49 a ± 16.69 | −2.40 b ± 7.74 | 2.73 a ± 9.89 | 11.37 a ± 14.15 |
| SOS | −0.87 b ± 0.44 | 6.00 b ± 5.91 | −4.21 b ± 4.64 | 13.40 c ± 17.44 | −4.70 c ± 8.46 | −2.71 c ± 9.85 | 9.37 b ± 12.41 |
| SS | −1.23 c ± 0.78 | 12.63 a ± 11.80 | −6.51 c ± 5.14 | 2.29 d ± 12.86 | −7.80 d ± 10.15 | −8.35 d ± 13.56 | 4.68 c ± 9.02 |
| Treatment | Increasing and decreasing percentages compared with irrigated plants in the first and second stress stages | | | | | | |
| | MC | TA | TSS | ASC | CAC | SSC | TPC |
| SOBS | −0.90 a ± 0.58 | 6.65 c ± 6.14 | −3.86 a ± 4.53 | 12.73 a ± 16.38 | −2.58 a ± 9.90 | −1.49 a ± 10.12 | 9.12 b ± 13.75 |
| SOES | −0.86 a ± 0.59 | 7.19 c ± 5.68 | −4.27 a ± 4.50 | 13.11 a ± 14.22 | −5.12 b ± 9.91 | −1.69 a ± 9.79 | 10.42 a ± 14.22 |
| SOS | −1.21 b ± 0.74 | 11.47 b ± 5.94 | −5.88 b ± 5.32 | 7.04 b ± 17.51 | −6.92 c ± 10.56 | −6.41 b ± 10.34 | 7.65 c ± 12.05 |
| SS | −1.55 c ± 1.01 | 19.05 a ± 11.13 | −8.54 c ± 5.74 | −3.73 c ± 14.17 | −10.10 d ± 11.68 | −12.31 c ± 13.16 | 1.78 d ± 10.46 |

MC: moisture content, TA: titratable acidity, TSS: total soluble solids, ASC: ascorbic acid content, CAC: carotenoid content, SSC: soluble sugar content, TPC: total phenolics content, SS: stressed plants that had not been treated, SOS: stressed plants that had been treated with oak leaf powder, SOES: stressed plants that had been treated with oak leaf powder and oak leaf extract, SOBS: stressed plants that had been treated with oak leaf powder and biofertilizers. Duncan's multiple range test at P ≤ 0.05 indicates that any mean values sharing the same letter in the same column are not statistically significant. The value is represented by trait index ± standard deviation (SD). Each value is the average of eight measurements.

A multivariate analytic technique called principal component analysis (PCA) was used to evaluate the similarity between the levels of treatment. Additionally, it is also used to determine the relationship between attributes. In total, 16 determined variables concerning the morpho-physiological and fruit physicochemical traits under four levels of treatment were subjected to a principal component analysis. Based on an eigenvalue > 1, we extracted a total of two first components with a cumulative distribution of 95.63% (85.05% for the first component and 11.59% for the second component), 96.53% (90.26% for the first component and 6.27% for the second component), and 97.04% (92.95% for the first component and 4.09% for the second component) for the first, second, and their combination stress stages, respectively (Figure 4.9). Different distributions of studied traits and treatments were observed on the PCA plot. Under first stress stage, the most notable contributors to the observed variance along PC1 were SL, SFW, RL, RWC, MC, TA, TSS, ASC, CAC, and SSC. However, the greatest amount of variance along PC2 was caused by SDW, RFW, RDW, FWT, TCC, and TPC (Figure 4.9A). The most noteworthy contributions to the observed variance along PC1 during the second stress stage were SL, SFW, SDW, FWT, MC, TSS, CAC, SSC, and TPC. Nevertheless, RL, RFW, RDW, RWC, TCC, MC, TA, and ASC were responsible for the bulk of the variation along PC2 (Figure 4.9B). Under both stress stages, the SL, SDW, RFW, RDW, RWC, TCC, MC, TA, TSS, ASC, SSC, and TPC were the major contributors to the observed variance along PC1. SFW, RL, FWT, and CAC, on the other hand, were responsible for the majority of the variation along PC2 (Figure 4.9C).

The application of powdered oak leaf, leaf oak extract, and biofertilizers reduced titratable acidity (TA) in fruit in all stress stages compared with the untreated plant under stress conditions and formed the first group in the left of the PCA plot (brown outline). During the first stress stage, the characteristics of the plants treated with SOBS with high percentage values of RL, SFW, SDW, RFW, RDW, TSS, ASC, and SSC were included in the second group on the upper right quadrant (green outline) of the PCA plot. The third group in the lower right quadrant (blue outline) of the PCA plot was made up of attributes in SOES-treated plants with high SL, RWC, TCC, FWT, MC, CAC, and TPC values. Under the second stress stage, the characteristics of the plants treated with SOES that had high values of SL, TSS, MC, SSC, TPC, RWC, and ASC formed the second group in the upper right quadrant (green outline) of the PCA plot. Furthermore, the traits in the plants treated with SOBS with high percentage values of RL, RFW, RDW, SFW, SDW, FWT, TCC, and CAC were included in the third group on the lower right quadrant (blue outline) of the PCA plot. In the combination of both stress stages, traits with high percentage values of SFW, SDW, FWT, RWC, TCC, TPC, and ASC in plants treated with SOES comprised the second group in the upper right quadrant (green outline) of the PCA plot. The third group was in the lower right quadrant (blue outline) of the PCA plot. It was made up of plants treated with SOBS and having high values of SL, RL, RFW, RDW, MC, CAC, TSS, and SSC.

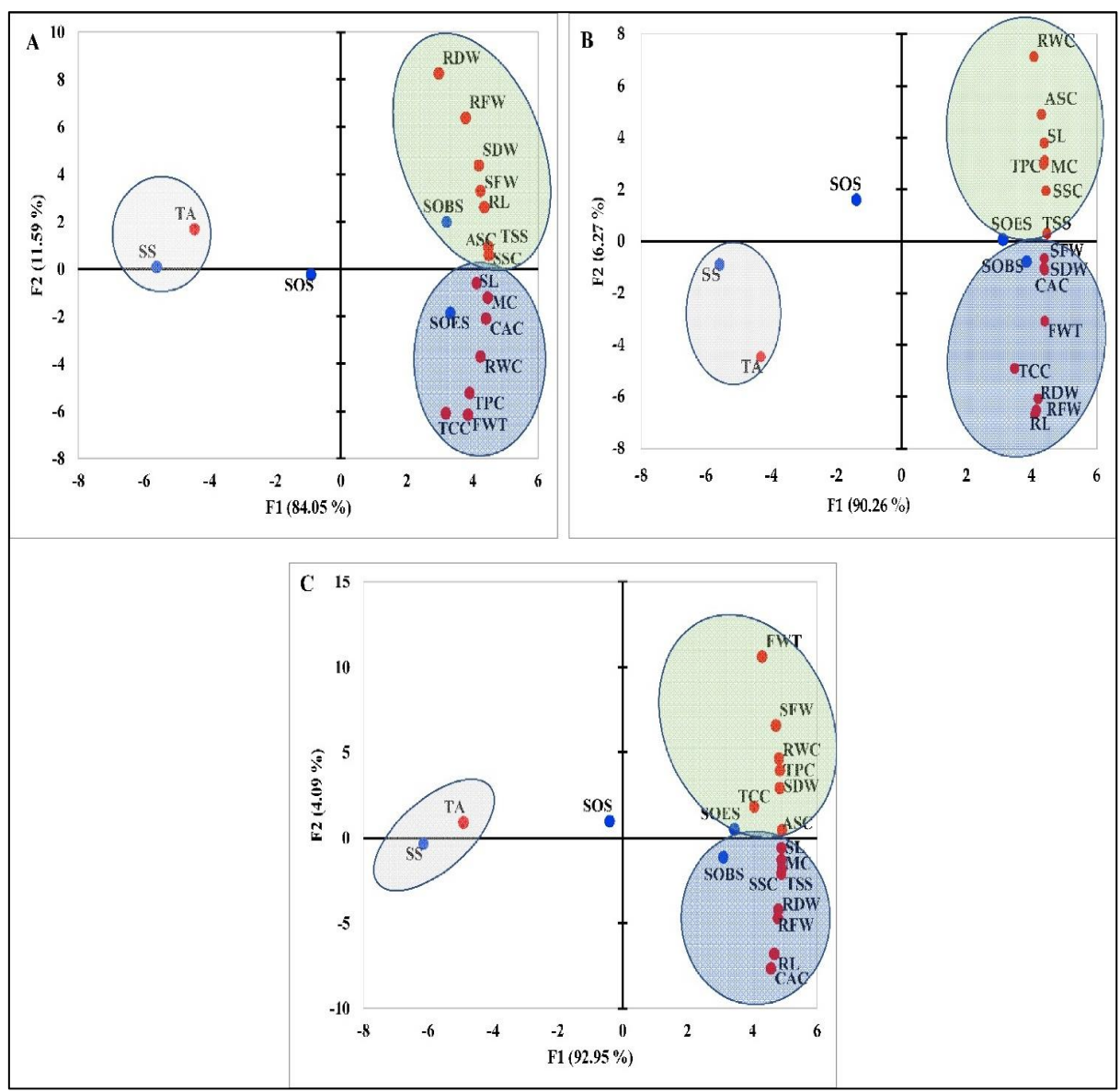

**Figure 4.9 PCA plot showing the distribution of various morpho-physiological and fruit physicochemical traits and treatments under stress conditions. SL: shoot length, SFW: shoot fresh weight, SDW: shoot dry weight, RL: root length, RFW: root fresh weight, RDW: root dry weight, FWT: fruits weight per plant, RWC: relative water content, TCC: total chlorophyl content, MC: moisture content, TA: titratable acidity, TSS: total soluble solids, ASC: ascorbic acid content, CAC: carotenoid content, SSC: soluble sugar content, TPC: total phenolic content, SS: stressed plants that had not been treated, SOS: stressed plants that had been treated with oak leaf powder, SOES: stressed plants that had been treated with oak leaf powder and oak leaf extract, SOBS: stressed plants that had been treated with oak leaf powder and biofertilizers. F1 and F2 represent the first and second components, respectively.**

*Influence of accessions on the morpho-physiological and physicochemical characteristics of tomato fruit under application of SS, SOS, SOES, and SOBS*

Under conditions of water stress, analysis of variance (ANOVA) revealed highly significant accession effects on the morpho-physiological traits of the first stress stage (before blooming), the second stress stage (before fruiting), and their combinations (Appendix 15). Shoot length (SL), shoot fresh weight (SFW), shoot dry weight (SDW), fruit weight per plant (FWT), relative water content (RWC), and total chlorophyll content (TCC) were all significantly lower in all accessions as compared with control plants. SL (13.13%), SFW (17.96%), SDW (17.83%), FWT (42.63%), and RWC (15.68%) exhibited the largest decreasing percentages in the stressed Yadgar accession. In all stress stages, the tolerant accessions (Raza Pashayi and Sandra) had lower decreasing amounts of SL and FWT than the sensitive accessions (Braw and Yadgar) (Table 4.14). Root fresh weight (RFW) and root dry weight (RDW) demonstrated high increasing percentages in all four accessions for all stress levels under water stress circumstances.

The analysis of variance (ANOVA) of the data obtained in the fruit physicochemical traits found a significant accession effect (Appendix 17). According to Table 4.15, all stress stages contributed to a reduction in the four accessions' moisture content (MC). The fruit of tolerant accessions showed higher increasing values in the ASC, CAC, and TPC characteristics than sensitive accessions under all stress stages. Under all stages of stress, the TA was higher in sensitive accessions than in tolerant accessions. Additionally, the Sandra accession showed an increase in CAC of 2.22, 5.01, and 6.95% for the first, second, and both of them together, respectively. In accordance with (Appendix 18), the mean pairwise comparison for the interaction of accessions and different treatments showed that Sandra had the highest increasing percentages in CAC (3.18%) and SSC (16.32%) in the presence of the SOES application, followed by Raza Pashayi with the highest increasing percentages in ASC (36.58%) and TPC (22.43%). With the exception of TA with the treatment of SS and SOS, the Braw accession reported the highest declining values in all physicochemical parameters under the first stress stage. The Sandra accession registered the largest percentage increases in TSS (2.63%), ASC (38.21%), CAC (6.02%), and SSC (16.67%) compared with irrigated plants (Appendix 19), while the Braw accession showed declining trends in all physicochemical measures except TA under the second stress stage. Sandra had the largest increasing percentages in TSS (1.75%), CAC (9.47%), and SSC (11.81%) with SOBS application, followed by Raza Pashayi in ASC (28.71%) and TPC (27.20%) in the presence of SOES application during both stress stages (Appendix 19).

**Table 4.14 Impact of tomato accessions treated with oak leaf powder, oak leaf extract, and biofertilizer at different stress stages on the morpho-physiological traits. Increasing and declining percentages are represented by positive and negative values, respectively.**

| Accessions | SL (%) | SFW (%) | SDW (%) | RL (%) | RFW (%) | RDW (%) | FWT (%) | RWC (%) | TCC (%) |
|---|---|---|---|---|---|---|---|---|---|
| **Increasing and decreasing percentages compared with irrigated plants during the first stress stage** | | | | | | | | | |
| Raza Pashayi | −4.98 a ± 5.24 | −5.75 a ± 2.20 | −3.79 b ± 1.95 | −6.52 bc ± 11.01 | 34.19 b ± 17.88 | 53.76 ab ± 16.65 | −19.47 a ± 6.44 | −11.75 ab ± 4.63 | −31.53 b ± 8.38 |
| Sandra | −5.14 a ± 5.81 | −8.03 a ± 7.99 | −9.21 c ± 9.88 | −5.18 b ± 7.93 | 36.34 b ± 60.49 | 98.55 a ± 146.60 | −21.26 ab ± 14.64 | −10.13 a ± 8.03 | −32.62 b ± 23.06 |
| Braw | −10.65 b ± 6.62 | −5.83 a ± 30.20 | 0.79 a ± 35.77 | −14.79 c ± 13.11 | 66.29 a ± 64.37 | 24.59 b ± 41.84 | −26.48 b ± 5.15 | −14.81 b ± 5.28 | −17.60 a ± 8.38 |
| Yadgar | −13.12 b ± 7.06 | −17.96 b ± 9.03 | −17.83 d ± 9.88 | 10.23 a ± 19.20 | 28.30 b ± 25.01 | 14.68 b ± 16.65 | −42.63 c ± 6.24 | −15.68 b ± 3.01 | −16.98 a ± 11.40 |
| **Increasing and decreasing percentages compared with irrigated plants during the second stress stage** | | | | | | | | | |
| Raza Pashayi | −8.71 a ± 4.44 | −6.32 a ± 1.83 | −4.32 a ± 1.17 | −2.97 bc ± 17.02 | 43.51 b ± 27.94 | 76.21 b ± 24.00 | −18.68 a ± 8.55 | −12.57 ± 4.76 | −31.19 c ± 7.80 |
| Sandra | −9.35 a ± 5.40 | −13.32 b ± 8.57 | −7.98 a ± 10.95 | −11.49 c ± 10.65 | 53.32 b ± 51.17 | 158.73 a ± 94.25 | −29.87 b ± 4.25 | −8.93 a ± 8.25 | −39.90 d ± 8.22 |
| Braw | −18.44 b ± 9.70 | −21.72 c ± 20.98 | −6.05 a ± 28.75 | −0.54 b ± 20.29 | 150.13 a ± 110.13 | 52.48 b ± 66.73 | −36.66 c ± 3.49 | −14.55 b ± 4.13 | −14.13 a ± 9.99 |
| Yadgar | −15.39 b ± 6.74 | −30.00 d ± 6.22 | −22.97 b ± 7.07 | 12.36 a ± 20.13 | 15.57 b ± 22.11 | 6.99 c ± 11.82 | −46.60 d ± 6.52 | −11.96 ab ± 8.40 | −20.65 b ± 7.58 |
| **Increasing and decreasing percentages compared with irrigated plants during the first and second stress stages** | | | | | | | | | |
| Raza Pashayi | −9.97 a ± 3.90 | −6.23 a ± 1.68 | −3.28 a ± 1.61 | 0.14 b ± 11.89 | 36.62 c ± 23.30 | 53.07 b ± 26.86 | −25.90 a ± 5.29 | −22.04 c ± 2.75 | −33.06 c ± 7.58 |
| Sandra | −12.01 a ± 6.33 | −13.64 b ± 9.47 | −10.01 b ± 10.39 | −9.82 c ± 8.32 | 73.29 b ± 45.59 | 238.51 a ± 47.09 | −36.77 b ± 3.01 | −15.54 ab ± 6.41 | −40.04 d ± 6.58 |
| Braw | −27.55 c ± 7.28 | −24.46 c ± 29.40 | −18.78 c ± 18.75 | −4.75 bc ± 14.41 | 143.07 a ± 75.82 | 12.30 c ± 48.49 | −44.83 c ± 4.28 | −18.42 bc ± 5.54 | −15.80 a ± 10.58 |
| Yadgar | −20.29 b ± 6.15 | −32.50 d ± 8.00 | −26.31 d ± 6.58 | 8.77 a ± 16.95 | 7.52 d ± 18.88 | −2.04 c ± 17.34 | −58.46 d ± 5.43 | −12.23 a ± 4.80 | −22.11 b ± 13.26 |

SL: shoot length, SFW: shoot fresh weight, SDW: shoot dry weight, RL: root length, RFW: root fresh weight, RDW: root dry weight, FWT: fruits weight per plant, RWC: relative water content, TCC: total chlorophyl content. Any mean values sharing the same letter in the same column are not statistically significant, according to Duncan's multiple range test at P ≤ 0.05. The values are represented by the standard deviation of the trait index. Each value is the average of eight measurements. The value is represented by trait index ± standard deviation (SD). Each value is the average of eight measurements.

## Impact of various treatments on the biochemical responses of the leaves of tomato plants under conditions of water stress

To gain a better understanding of the mechanism of tolerance in plants treated with SS, SOS, SOES, and SOBS under water deficit stress, a number of biochemical measurements were performed on the leaves of tomato plants. As shown in Appendix 20, significant variations were detected among different levels of treatments for all biochemical characters of the leaves of the tomato under all stress stages. The maximum values of proline content (PC), soluble sugar content (SSC), guaiacol peroxidase (GPA), and catalase (CAT) were recorded by the tomato plants treated with SOES, while the highest values of total phenolic content (TPC) and antioxidant activity (AC) were observed by the plants treated with SOBS under the first and second stress stages. Moreover, under the combination of first and second stress stages, the plants treated with SOBES displayed the greatest values of all biochemical traits, with the

**Table 4.15 Effect of tomato accessions treated at various stress stages with oak leaf powder, oak leaf extract, and biofertilizer on the fruit physicochemical traits. Increasing and decreasing percentages are indicated by positive and negative values, respectively.**

| | Increasing and Decreasing Percentages Compared with Irrigated Plants during the First Stress Stage | | | | | | |
|---|---|---|---|---|---|---|---|
| **Accessions** | **MC (%)** | **TA (%)** | **TSS (%)** | **ASC (%)** | **CAC (%)** | **SSC (%)** | **TPC (%)** |
| Raza Pashayi | −0.45 a ± 0.14 | −1.49 b ± 9.61 | −2.06 b ± 2.35 | 29.29 a ± 6.52 | −0.84 b ± 1.37 | 5.73 b ± 3.59 | 19.82 a ± 2.46 |
| Sandra | −0.35 a ± 0.18 | 5.97 a ± 5.27 | 1.32 a ± 2.84 | 25.46 b ± 10.82 | 2.22 a ± 0.73 | 9.07 a ± 3.84 | 14.18 b ± 2.72 |
| Braw | −1.46 c ± 0.68 | 7.46 a ± 6.65 | −6.68 c ± 2.81 | 0.26 d ± 5.12 | −8.24 c ± 3.86 | −9.45 c ± 8.47 | −8.43 d ± 3.12 |
| Yadgar | −1.14 b ± 0.25 | 8.77 a ± 5.84 | −7.49 c ± 3.07 | 12.93 c ± 6.40 | −10.95 d ± 3.73 | −9.94 c ± 4.69 | 8.20 c ± 1.97 |
| | Increasing and decreasing percentages compared with irrigated plants during the second stress stage | | | | | | |
| **Accessions** | MC (%) | TA (%) | TSS (%) | ASC (%) | CAC (%) | SSC (%) | TPC (%) |
| Raza Pashayi | −0.39 a ± 0.20 | −3.71 c ± 5.59 | −1.93 b ± 1.54 | 24.72 b ± 4.60 | 2.04 b ± 1.51 | 2.73 b ± 2.79 | 24.29 a ± 6.83 |
| Sandra | −0.33 a ± 0.17 | 6.01 b ± 6.63 | 1.32 a ± 2.09 | 27.38 a ± 12.61 | 5.01 a ± 1.03 | 12.67 a ± 4.54 | 15.19 b ± 4.13 |
| Braw | −1.44 c ± 0.57 | 14.87 a ± 7.93 | −6.15 c ± 3.28 | −11.16 d ± 2.62 | −13.05 d ± 2.80 | −15.41 d ± 7.74 | −8.02 d ± 1.74 |
| Yadgar | −1.14 b ± 0.29 | 7.03 b ± 5.53 | −7.97 d ± 3.12 | 10.46 c ± 8.75 | −8.92 c ± 7.82 | −6.64 c ± 4.02 | 6.18 c ± 1.03 |
| | Increasing and decreasing percentages compared with irrigated plants during the first and second stress stages | | | | | | |
| **Accessions** | MC (%) | TA (%) | TSS (%) | ASC (%) | CAC (%) | SSC (%) | TPC (%) |
| Raza Pashayi | −0.42 a ± 0.20 | 3.12 c ± 5.58 | −2.90 b ± 3.32 | 20.78 a ± 7.76 | 0.19 b ± 2.75 | 0.33 b ± 3.46 | 21.28 a ± 6.70 |
| Sandra | −0.65 b ± 0.14 | 9.65 b ± 5.28 | 0.24 a ± 2.71 | 20.57 a ± 9.39 | 6.95 a ± 1.89 | 7.81 a ± 4.09 | 12.68 b ± 3.67 |
| Braw | −2.14 d ± 0.70 | 17.57 a ± 10.89 | −8.79 c ± 1.59 | −17.15 c ± 6.30 | −16.17 c ± 4.31 | −18.87 d ± 7.90 | −11.39 d ± 2.23 |
| Yadgar | −1.31 c ± 0.32 | 14.01 a ± 5.66 | −11.09 d ± 2.59 | 4.96 b ± 6.22 | −15.68 c ± 2.75 | −11.17 c ± 3.38 | 6.39 c ± 1.82 |

MC: moisture content, TA: titratable acidity, TSS: total soluble solids, ASC: ascorbic acid content, CAC: carotenoid content, SSC: soluble sugar content, TPC: total phenolics content. Any mean values sharing the same letter in the same column are not statistically significant, as determined by the Duncan's multiple range test at P ≤ 0.05. The value is represented by trait index ± standard deviation (SD). Each value is the average of eight measurements.

exception of the LP trait. Furthermore, the control plants (SW) exhibited the minimum values of all chemical characters of the leaves of tomato under all stress stages. Low amounts of lipid peroxidation were observed by SW (5.24 nmol g$^{-1}$ FLW), followed by SOES (7.15 nmol g$^{-1}$ FLW) and SOBS (8.46 nmol g$^{-1}$ FLW), under the first, second, and their combination stress stages (Table 4.16). Seven different variables relating to the biochemical parameters of leaves treated with SW, SS, SOS, SOES, and SOBS were subjected to a principal component analysis (PCA). Based on an eigenvalue greater than one, the first two components displayed cumulative distributions of 93.53, 93.35, and 98.11% for the first, second, and their combined stress stages, respectively (Figure 4.10 A, B and C). The biochemical characteristics and treatments were dispersed in various ways across the PCA plot throughout the first, second, and combined stages of stress. The characteristics that had the most significance in affecting the observed variance along PC1 were PC, SSC, TPC, AC, GPA, and CAT. However, the LP characteristic was the primary driver of variance along PC2. In comparison with untreated plants (SS), the application of SOBS, SOES, and SOS reduced the amount of lipid

**Table 4.16 Impact of oak leaf powder, oak leaf extract, and biofertilizer on the biochemical characteristics of the leaves of tomato plants under various stress stages.**

| Treatment | PC (µg g⁻¹) | SSC (µg g⁻¹) | TPC (µg g⁻¹) | AC (µg g⁻¹) | LP (nmol g⁻¹) | GPA (units min⁻¹ g⁻¹) | CAT (units min⁻¹ g⁻¹) |
|---|---|---|---|---|---|---|---|
| **First stress stage** | | | | | | | |
| SOBS | 1546.37 b ± 503.08 | 569.04 b ± 99.21 | 433.90 a ± 98.38 | 1010.20 a ± 173.44 | 8.46 c ± 1.13 | 0.26 b ± 0.06 | 139.61 b ± 42.49 |
| SOES | 1956.50 a ± 489.76 | 612.64 a ± 109.34 | 399.21 b ± 90.59 | 1006.99 b ± 175.26 | 7.15 d ± 0.98 | 0.34 a ± 0.06 | 160.71 a ± 56.00 |
| SOS | 1322.91 c ± 619.10 | 524.14 c ± 96.68 | 344.91 c ± 57.07 | 966.79 c ± 171.26 | 11.05 b ± 2.24 | 0.25 b ± 0.08 | 118.51 c ± 65.65 |
| SS | 1307.65 d ± 578.09 | 417.19 d ± 108.74 | 325.57 d ± 56.09 | 892.80 d ± 94.45 | 13.10 a ± 2.26 | 0.16 c ± 0.09 | 87.66 d ± 45.71 |
| SW | 1054.58 e ± 425.20 | 374.14 e ± 91.47 | 312.23 e ± 63.52 | 893.31 d ± 129.46 | 5.24 e ± 0.78 | 0.13 d ± 0.06 | 64.94 e ± 42.26 |
| **Second stress stage** | | | | | | | |
| SOBS | 2058.81 b ± 426.81 | 742.65 b ± 110.77 | 428.24 a ± 20.91 | 986.05 a ± 120.82 | 9.92 c ± 1.40 | 0.24 b ± 0.08 | 126.62 b ± 30.89 |
| SOES | 2534.00 a ± 433.44 | 782.93 a ± 89.71 | 402.68 b ± 29.48 | 974.43 b ± 159.49 | 8.17 d ± 1.68 | 0.33 a ± 0.08 | 159.09 a ± 40.27 |
| SOS | 1813.81 c ± 396.27 | 627.76 c ± 146.44 | 378.99 c ± 35.48 | 909.19 c ± 163.83 | 10.59 b ± 1.32 | 0.23 c ± 0.06 | 113.64 c ± 24.69 |
| SS | 1616.82 d ± 444.00 | 529.00 d ± 122.78 | 322.51 e ± 70.59 | 902.40 d ± 139.83 | 12.83 a ± 2.85 | 0.17 d ± 0.07 | 81.17 d ± 15.04 |
| SW | 1126.37 e ± 533.06 | 501.76 e ± 145.17 | 334.84 d ± 32.40 | 895.51 e ± 109.70 | 7.20 e ± 0.51 | 0.15 e ± 0.08 | 64.94 e ± 40.50 |
| **Combination of both stress stages** | | | | | | | |
| SOBS | 2057.01 b ± 391.73 | 764.63 b ± 121.99 | 453.15 b ± 58.23 | 1029.29 b ± 96.55 | 10.67 c ± 1.11 | 0.30 b ± 0.11 | 155.84 b ± 63.92 |
| SOES | 2217.65 a ± 330.37 | 856.54 a ± 96.24 | 493.69 a ± 122.67 | 1092.47 a ± 120.92 | 8.94 d ± 1.21 | 0.44 a ± 0.19 | 217.53 a ± 90.38 |
| SOS | 1689.96 c ± 485.67 | 670.35 c ± 109.86 | 407.97 c ± 59.26 | 938.58 c ± 96.75 | 12.53 b ± 2.66 | 0.27 c ± 0.10 | 137.99 c ± 20.56 |
| SS | 1661.24 d ± 268.52 | 580.69 d ± 12.76 | 316.07 e ± 76.85 | 907.34 d ± 118.09 | 13.95 a ± 2.80 | 0.17 d ± 0.09 | 95.78 d ± 17.32 |
| SW | 1126.37 e ± 533.06 | 501.76 e ± 145.17 | 334.84 d ± 32.40 | 895.51 e ± 109.70 | 7.20 e ± 0.51 | 0.15 e ± 0.08 | 64.94 e ± 40.50 |

PC: proline content, SSC: soluble sugar content, TPC: total phenolic content, AC: antioxidant activity, LP: lipid peroxidation, GPA: peroxidase, CAT: catalase, SS: stressed plants that had not been treated, SOS: stressed plants that had been treated with oak leaf powder, SOES: stressed plants that had been treated with oak leaf powder and oak leaf extract, SOBS: stressed plants that had been treated with oak leaf powder and biofertilizers. Duncan's multiple range test at P ≤ 0.05 indicates that any mean values sharing the same letter in the same column are not statistically significant. The value is represented by mean ± standard deviation (SD). Each value is the average of eight measurements.

peroxidation in the leaves during all stages of stress. On the right side (blue outline) of the PCA plot, characteristics of SOBS- and SOES-treated plants with high PC, TPC, AC, SSC, GPA, and CAT values were noted throughout the first, second, and their combined stress stages. Meanwhile, the plants that received SS treatment produced more LP (brown outline).

*Impact of different accessions treated with SW, SS, SOS, SOES, and SOBS on the biochemical responses of the leaves of tomato plants under circumstances of water stress*

Tomato plant leaves were analyzed chemically in order to acquire a better knowledge of the mechanism of tolerance in accessions treated with SS, SOS, SOES, and SOBS.

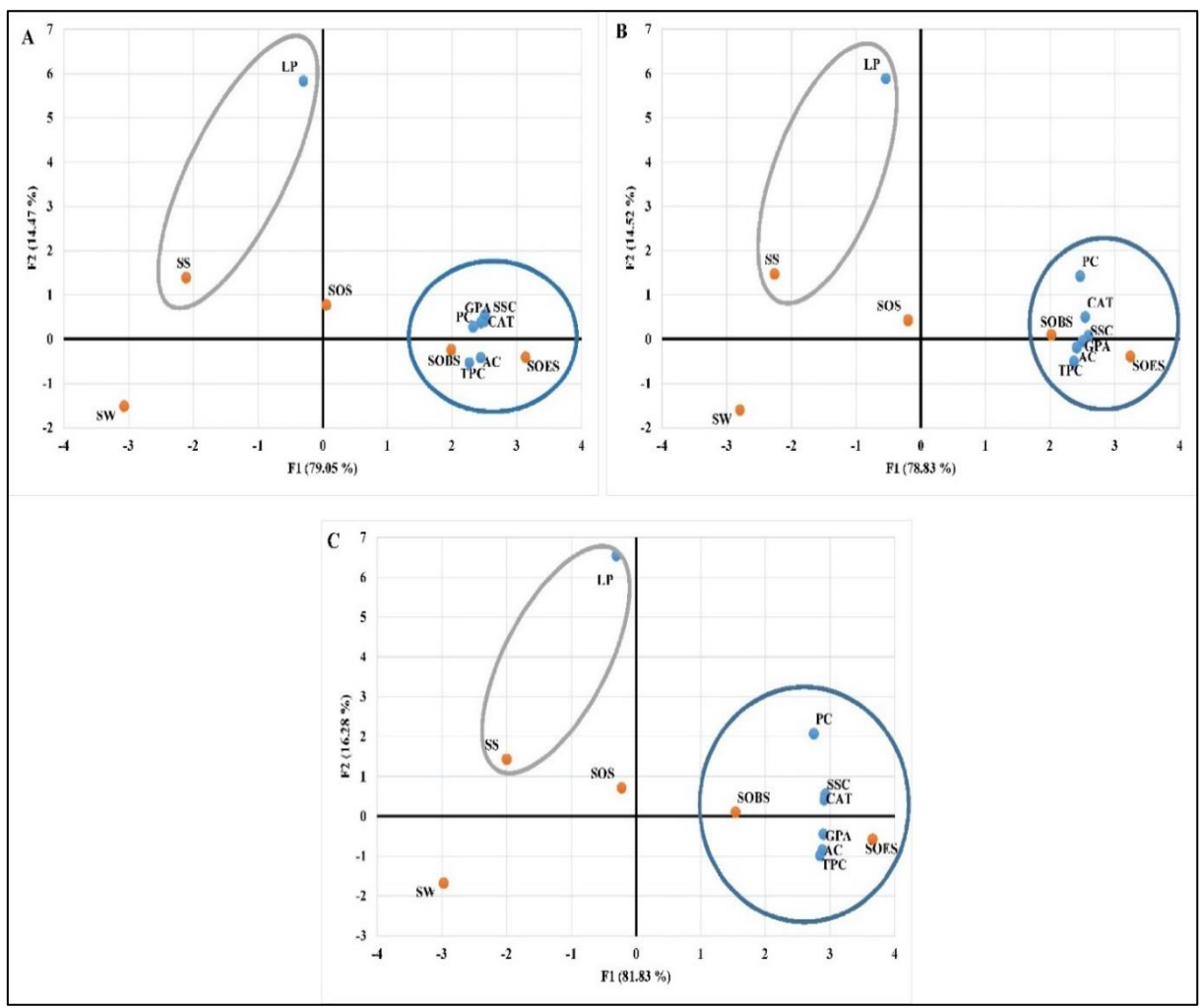

**Figure 4.10 PCA plot illustrating the distribution of leaf biochemical characteristics and treatments under different stress circumstances. PC: proline content, SSC: soluble sugar content, TPC: total phenolic content, AC: antioxidant activity, LP: lipid peroxidation, GPA: peroxidase, CAT: catalase, SS: stressed plants that had not been treated, SOS: stressed plants that had been treated with oak leaf powder, SOES: stressed plants that had been treated with oak leaf powder and oak leaf extract, SOBS: stressed plants that had been treated with oak leaf powder and biofertilizers. F1 and F2 represent the first and second components, respectively.**

As demonstrated in Appendix 20, substantial differences were identified between different accessions for all biochemical characteristics of tomato leaves under all stress stages. The tolerant accessions Sandra had the highest values of PC, SSC, and AC during the first stress stage, whereas the tolerant accession Raza Pashayi had the highest scores of TPC, GPA, and CAT traits. The sensitive genotype Yadgar showed the minimum values of all chemical characteristics with the exception of the LP trait. As a comparison between tolerant and sensitive accessions, the mean values of SSC, GPA, and CAT in tolerant accessions were higher than those obtained in sensitive plants. The highest scores of LP were found in sensitive plants (Table 4.17). Under the second stress stage, the tolerant accession Sandra had the highest values of PC, TPC, AC, and CAT, while the tolerant accession Raza Pashayi had the highest value of GPA. Except for the PC and LP features, the sensitive accession Yadgar displayed the lowest values for all biochemical parameters (Table 4.17). Comparing tolerant and sensitive accessions, the mean TPC, AC, and GPA values of tolerant accessions were greater than those of sensitive plants. The susceptible plants (Braw and Yadgar) had the highest levels of LP. Sandra accession exhibited the greatest PC, AC, and CAT scores in response to both stress periods. With the exception of the LP trait, Yadgar accessions had the lowest values for all leaf biochemical parameters (Table 4.17).

**Table 4.17 Effects of oak leaf powder, oak leaf extract, and biofertilizer on the biochemical traits of the leaves of tomato plants under different levels of stress.**

| | | | First stress stage | | | | |
|---|---|---|---|---|---|---|---|
| Accessions | PC (µg g$^{-1}$) | SSC (µg g$^{-1}$) | TPC (µg g$^{-1}$) | AC (µg g$^{-1}$) | LP (nmol g$^{-1}$) | GPA (units min$^{-1}$ g$^{-1}$) | CAT (units min$^{-1}$ g$^{-1}$) |
| Raza Pashayi | 1093.08 d ± 146.59 | 517.72 b ± 160.20 | 433.82 a ± 65.06 | 987.30 c ± 34.97 | 6.86 d ± 2.03 | 0.30 a ± 0.06 | 172.73 a ± 40.01 |
| Sandra | 2220.97 a ± 257.08 | 610.25 a ± 88.53 | 371.72 c ± 70.52 | 1080.95 a ± 106.77 | 9.39 b ± 2.34 | 0.26 b ± 0.12 | 132.47 b ± 46.10 |
| Braw | 1252.82 b ± 497.38 | 499.26 c ± 75.14 | 393.63 b ± 48.06 | 1020.41 b ± 103.02 | 9.28 c ± 3.50 | 0.19 c ± 0.07 | 93.51 c ± 59.81 |
| Yadgar | 1183.54 c ± 526.99 | 370.49 d ± 69.11 | 253.48 d ± 21.77 | 727.43 d ± 34.97 | 10.46 a ± 3.77 | 0.16 d ± 0.08 | 58.44 d ± 17.35 |
| | | | Second stress stage | | | | |
| Accessions | PC (µg g$^{-1}$) | SSC (µg g$^{-1}$) | TPC (µg g$^{-1}$) | AC (µg g$^{-1}$) | LP (nmol g$^{-1}$) | GPA (units min$^{-1}$ g$^{-1}$) | CAT (units min$^{-1}$ g$^{-1}$) |
| Raza Pashayi | 1620.41 c ± 526.17 | 606.11 c ± 181.90 | 378.73 b ± 51.11 | 1010.00 b ± 59.28 | 8.93 c ± 1.88 | 0.31 a ± 0.06 | 103.90 bc ± 27.77 |
| Sandra | 2278.41 a ± 491.93 | 658.77 b ± 100.11 | 414.37 a ± 27.83 | 1102.97 a ± 47.37 | 7.93 d ± 1.19 | 0.23 b ± 0.12 | 128.57 a ± 68.90 |
| Braw | 1621.13 c ± 938.94 | 795.56 a ± 82.71 | 373.60 c ± 23.28 | 858.92 c ± 27.47 | 10.77 b ± 2.48 | 0.19 c ± 0.07 | 106.49 b ± 47.70 |
| Yadgar | 1799.90 b ± 121.84 | 486.85 d ± 110.53 | 327.12 d ± 74.40 | 762.16 d ± 54.98 | 11.34 a ± 3.01 | 0.17 d ± 0.04 | 97.40 c ± 18.69 |
| | | | Combination of both stress stages | | | | |
| Accessions | PC (µg g$^{-1}$) | SSC (µg g$^{-1}$) | TPC (µg g$^{-1}$) | AC (µg g$^{-1}$) | LP (nmol g$^{-1}$) | GPA (units min$^{-1}$ g$^{-1}$) | CAT (units min$^{-1}$ g$^{-1}$) |
| Raza Pashayi | 1903.64 b ± 567.18 | 658.83 c ± 190.89 | 432.28 b ± 97.33 | 1003.38 b ± 57.89 | 9.30 c ± 2.13 | 0.40 a ± 0.13 | 132.47 b ± 47.07 |
| Sandra | 2114.72 a ± 403.89 | 724.07 b ± 157.46 | 405.43 c ± 18.45 | 1082.30 a ± 52.67 | 9.01 d ± 1.28 | 0.29 b ± 0.21 | 176.62 a ± 125.44 |
| Braw | 1382.77 d ± 117.24 | 793.94 a ± 72.50 | 459.33 a ± 118.99 | 955.36 c ± 168.75 | 12.48 a ± 3.63 | 0.17 d ± 0.06 | 103.90 c ± 36.73 |
| Yadgar | 1600.67 c ± 669.52 | 522.35 d ± 126.03 | 307.53 d ± 45.65 | 849.50 d ± 168.75 | 11.85 b ± 3.14 | 0.20 c ± 0.06 | 124.68 b ± 59.02 |

PC: proline content, SSC: soluble sugar content, TPC: total phenolic content, AC: antioxidant activity, LP: lipid peroxidation, GPA: peroxidase, CAT: catalase. Duncan's multiple range test at $p \leq 0.05$ reveals that any mean values in the same column that share the same letter are not statistically significant. The value is represented by mean ± standard deviation (SD). Each value is the average of eight measurements.

Raza Pashayi had the highest values in SSC (711.79 µg g$^{-1}$), TPC (518.13 µg g$^{-1}$), and CAT (220.78 units min$^{-1}$ g$^{-1}$) in the availability of the SOES treatment, while Sandra had the highest values in PC (2446.05 µg g$^{-1}$) and GPA (0.42 units min$^{-1}$ g$^{-1}$) under the first stress stage, as shown in Appendix 21. In comparison with irrigated plants during the second stress stage, the Sandra accession recorded the highest values for AC (1151.89 µg g$^{-1}$), GPA (0.40 units min$^{-1}$ g$^{-1}$), and CAT (214.29 units min$^{-1}$ g$^{-1}$). When SOES was applied, Sandra had the highest SSC (976.60 µg g$^{-1}$), GPA (0.62 units min$^{-1}$ g$^{-1}$), and CAT (363.64 units min$^{-1}$ g$^{-1}$) scores, while Raza Pashayi had the highest PC (2571.69 µg g$^{-1}$) score during both stress stages.

*GC/MS analysis of oak leaf extract*

Table 4.18 and Figure 4.11 displays the phytochemical composition of the extracts as determined by GC/MS analysis. The extract contained twenty-four components. The major compounds were heptasiloxane, 1,1,3,3,5,5,7,7,9,9,11,11,13,13-tetradecamethyl-(32.50%), silane, dimethoxydimethyl-(11.67), octasiloxane, 1,1,3,3,5,5,7,7,9,9,11,11,13,13,15,15-hexadecamethyl-(10.88%), 1-hexadecanol (9.37%), behenic alcohol (8.86%), 2,4-di-tert-butylphenol (7.02%), 1-octadecene (6.73%), acetic acid, chloro-, octadecyl ester (1.55%), dichloroacetic acid, 4-hexadecyl ester (1.52%), coumatetralyl isomer-2 ME (1.49%), 1-dodecanol (1.29%), chloroacetic acid, and pentadecyl ester (1.01%).

**Table 4.18 Compounds detected by GC/MS analysis in the leaf water extract of *Quercus aegilops*.**

| Name of Compound | Retention Time (min) | Peak Area | Concentration (%) | Biological Activity of Major Compounds |
|---|---|---|---|---|
| Silane, dimethoxydimethyl- | 5.17 | 7,202,705.00 | 11.67 | Antibacterial |
| Cyclotrisiloxane, hexamethyl- | 6.93 | 327,535.00 | 0.53 | |
| Silane, methyldimethoxyethoxy- | 8.30 | 268,638.00 | 0.44 | |
| Oxime-, methoxy-phenyl- | 9.38 | 231,616.00 | 0.38 | |
| Tetraethyl silicate | 10.54 | 314,412.00 | 0.51 | |
| 1-Dodecanol | 13.90 | 796,766.00 | 1.29 | Antibacterial |
| 1-Hexadecanol | 16.73 | 1,941,423.00 | 9.37 | Reduction of evaporation |
| Carbonic acid, decyl undecyl ester | 16.84 | 397,832.00 | 0.64 | |
| 7-Tetradecene | 16.90 | 307,446.00 | 0.50 | |
| Chloroacetic acid, tetradecyl ester | 17.04 | 250,824.00 | 0.41 | |
| 2,4-Di-tert-butylphenol | 18.29 | 4,329,760.00 | 7.02 | Antioxidant |
| Carbonic acid, eicosyl vinyl ester | 19.31 | 422,435.00 | 0.68 | |
| Dichloroacetic acid, 4-hexadecyl ester | 19.36 | 536,815.00 | 1.52 | Antimicrobial |
| 1-Octadecene | 21.45 | 4,150,402.00 | 6.73 | Antioxidant and antimicrobial |
| Acetic acid, chloro-, octadecyl ester | 21.58 | 542,636.00 | 1.55 | No activity was reported |
| 1,2-Benzenedicarboxylic acid, bis(2-methylpropyl) ester | 22.32 | 229,532.00 | 0.37 | |
| 18-Norabietane | 23.10 | 242,753.00 | 0.39 | |
| Behenic alcohol | 23.48 | 3,132,644.00 | 8.86 | Antifungal |
| Chloroacetic acid, pentadecyl ester | 23.58 | 273,527.00 | 1.01 | No activity was reported |
| Coumatetralyl isomer-2 ME | 23.67 | 918,610.00 | 1.49 | No activity was reported |
| Acetic acid, chloro-, octadecyl ester | 24.34 | 507,902.00 | 0.82 | |
| Cyclotetrasiloxane, octamethyl- | 27.02 | 268,576.00 | 0.44 | |
| Heptasiloxane, 1,1,3,3,5,5,7,7,9,9,11,11,13,13-tetradecamethyl- | 32.59 | 21,294,993.00 | 32.50 | Insecticidal and antibacterial |
| Octasiloxane, 1,1,3,3,5,5,7,7,9,9,11,11,13,13,15,15-hexadecamethyl- | 38.79 | 6,719,421.00 | 10.88 | |

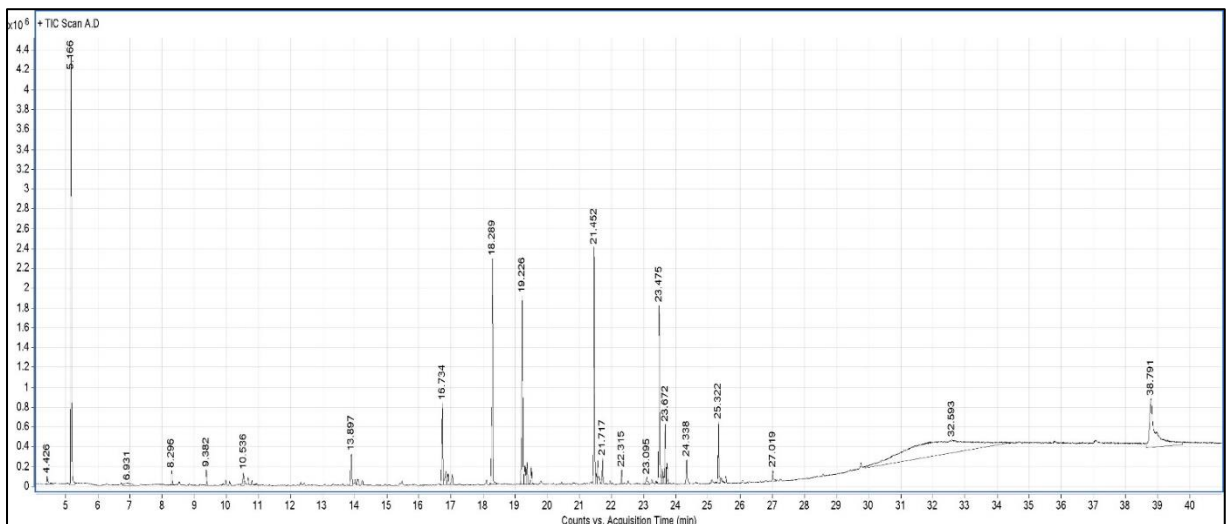

**Figure 4.11 Chromatograms of the oak leaf extract by GC/MS.**

### 4.1.3 Assessment of tomato plants to heavy metal

### 4.1.2.1 *In vitro* responses of tomato accessions to heavy metal stress using cadmium (Cd)

#### *The effect of Cd on seedling morphological traits*

One of the heavy metals known to be the most stressful to plants is cadmium (Cd). Most crop species experience physiological and morphological changes as a result of excessive Cd exposure. Cd causes reduction in nutrient uptake, biomass production, and crop yield (Chtouki *et al.*, 2021)

In this study, four treatments (0, 150, 300, and 450 µM) of cadmium (Cd) were conducted to reveal the response of 64 tomato accessions under *in vitro* conditions. According to the ANOVA analysis, radar charts, and box charts, there were significant differences among the accessions under both control and Cd-induced (Cd treatments) conditions for tomato seedling morphological characteristics: germination percentage (GP), root length (RL), shoot length (SL), seedling fresh weight (FW), and seedling dry weight (DW) (Table 4.19 and Figures 4.12 and 4.13). Table 4.20 showed that there were highly significant differences among accessions, Cd concentrations, and interactions between them for all morphological traits (GP, RL, SL, FW, and DW) ($P \le 0.001$). As Cd concentration increased, morphological traits significantly declined for all accessions in all morphological characteristics (Table 4.19). GP recorded data ranged from 80% to 100% with an average of 92.38% and RL from 5.23 to 10.95 cm, with an average of 8.37 cm; while SL from 5.23 to 9.1 cm, with an average of 7.25 cm; also, FW from 31.25 to 65.58 mg, with an average of 47.36 mg, and DW from 1.02 to 4.80 mg, with an average of 2.03 mg, these data were revealed under the control conditions (0 µM Cd concentration) (Table 4.19 and Appendix 22).

At T1 (150 µM) Cd induced, the values of GP ranged from 72.00% to 100.00% with an average of 89.52%; and RL from 0.85 to 4.53 cm with an average of 2.77 cm, SL from 4.75 to 8.00 cm with an average of 6.08 cm, FW from 23.04 to 61.16 mg with a mean of 39.68 mg, and DW from 0.70 to 2.60 mg with an average of 1.68 mg (Table 4.19 and Appendix 23). Under the effect Cd (300 μM) T2, the results show that the GP values varied between 60.00% to 100.00% with an average of 86.54%, and the RL value from 0.55 to 2.73 cm with a mean of 1.21 cm, the SL value 2.25 to 6.55 cm with a mean of 4.71 cm, while the FW value from 20.50 to 49.96 mg with an average of 33.96 mg and the DW varied from 1.00 to 2.50 mg with a mean of 1.66 mg (Table 4.19 and Appendix 24). Under induced Cd stress T3 (450 μM) (Table 4.19 and appendix 25), the GP value ranged from 60.00% to 100.00% with the mean of 84.48%, while the values of RL and SL from 0.25 to 1.15 and 0.48 to 5.15 cm with average values of 0.55 and 2.57 cm, respectively. The FW value also varied between 3.65 to 43.12 mg with a mean of 27.03 mg, DW ranged from 0.23 to 2.46 mg with an average of 1.69 mg.

The radar and box charts of all traits illustrated significant variations among T0 (0 μM), T1 (150 μM), T2 (300 μM), and T3 (450 μM), as shown for each characteristic by the lower and upper box plot limits (Figures 4.12 and 4.13). The morphological traits values of all tomato accessions under normal conditions, when compared to plants that had been exposed to Cd stress, showed significantly higher trait values. These results indicated that all accessions were affected by all Cd concentrations, particularly root length (RL) which had been more significantly decreased by Cd-induced (Figures 4.12 B and 4.13 B).

In Appendix 26, the mean values of all morphological traits of all accessions under all Cd concentrations were revealed, and the results showed that there was a lot of variation among accessions. The highest and lowest values of GP indicated for AC7 and AC13 which were 98.33% and 75.00%, respectively, while AC25 and AC6 showed the longest and shortest root length (RL) by 4.07 cm and 2.31 cm, respectively. The longest and shortest shoot length (SL) were recorded by AC12 (6.48 cm) and AC37 (4.07 cm), respectively. The highest values of FW and DW were indicted by AC20 and AC60 with 50.88 mg and 2.25 mg, respectively, while the lowest values of these traits recorded by AC52 with 28.54 mg and AC11 by 1.01 mg.

In Appendix 27, the interaction value between the tomato accessions and the Cd treatments was displayed. There were highly significant differences among them. The highest germination percentage (GP) (98.67%) was recorded by the combinations of AC5*C, AC7*C, AC25*C, AC60*C, AC63*C, AC5*T1, and AC7*T1, while the lowest values were shown under AC32*T3 with 60.00 %. The interactions of AC29*C and AC12*C recorded the longest root length (RL) and shoot length (SL), which were 10.59 cm and 8.97 cm, respectively, whereas the combination of AC61*T3 and AC21*T3 revealed the shortest values of these traits, 0.28 cm and 0.51 cm, respectively. The interactions of AC15*C and AC11*T3 recorded the highest and lowest values of FW, which were 62.66 mg and 10.42 mg, respectively. The combination of AC8*C recorded the greatest value of DW, 3.18 mg, whereas the interaction of AC11*T1 showed the lowest value by 0.89 mg.

**Table 4.19 Descriptive statistics of morphological and biochemical traits under different Cd concentration.**

| Traits | T0 | | | | | T1 | | | | |
|--------|---------|---------|---------|---------|---------|---------|----------|---------|---------|---------|
| | Minimum | Maximum | Mean | F | Pr > F | Minimum | Maximum | Mean | F | Pr > F |
| GP | 80.00 | 100.00 | 92.38 | 4.56 | < 0.0001 | 72.00 | 100.00 | 89.52 | 12.69 | < 0.0001 |
| RL | 5.23 | 10.95 | 8.37 | 22.51 | < 0.0001 | 0.85 | 4.53 | 2.77 | 27.66 | < 0.0001 |
| SL | 5.23 | 9.10 | 7.25 | 11.74 | < 0.0001 | 4.75 | 8.00 | 6.08 | 18.39 | < 0.0001 |
| FW | 31.25 | 65.58 | 47.36 | 17.99 | < 0.0001 | 23.04 | 61.16 | 39.68 | 20.43 | < 0.0001 |
| DW | 1.02 | 4.80 | 2.03 | 6.56 | < 0.0001 | 0.70 | 2.60 | 1.68 | 9.54 | < 0.0001 |
| PC | 394.51 | 1478.62 | 888.81 | 542.37 | < 0.0001 | 915.03 | 2684.77 | 1631.77 | 1566.66 | < 0.0001 |
| SSC | 63.95 | 204.69 | 124.53 | 163.42 | < 0.0001 | 106.54 | 321.36 | 204.56 | 472.21 | < 0.0001 |
| TPC | 50.15 | 152.02 | 88.49 | 317.52 | < 0.0001 | 81.24 | 235.17 | 151.10 | 699.26 | < 0.0001 |
| AC | 497.70 | 692.30 | 607.35 | 45.46 | < 0.0001 | 684.19 | 1073.38 | 932.72 | 2614.75 | < 0.0001 |
| GPA | 0.21 | 13.92 | 5.47 | 107.78 | < 0.0001 | 0.38 | 30.58 | 9.13 | 326.44 | < 0.0001 |
| CAT | 25.97 | 168.83 | 87.05 | 18.36 | < 0.0001 | 77.92 | 285.71 | 148.84 | 42.68 | < 0.0001 |
| LP | 3.45 | 7.48 | 5.25 | 722.72 | < 0.0001 | 2.74 | 8.61 | 5.89 | 1583.75 | < 0.0001 |
| | T2 | | | | | T3 | | | | |
| | Minimum | Maximum | Mean | F | Pr > F | Minimum | Maximum | Mean | F | Pr > F |
| GP | 60.00 | 100.00 | 86.54 | 10.29 | < 0.0001 | 60.00 | 100.00 | 84.48 | 18.86 | < 0.0001 |
| RL | 0.55 | 2.73 | 1.21 | 21.28 | < 0.0001 | 0.25 | 1.15 | 0.55 | 11.86 | < 0.0001 |
| SL | 2.25 | 6.55 | 4.71 | 29.80 | < 0.0001 | 0.48 | 5.15 | 2.57 | 72.77 | < 0.0001 |
| FW | 20.50 | 49.96 | 33.96 | 26.30 | < 0.0001 | 3.65 | 43.12 | 27.03 | 17.33 | < 0.0001 |
| DW | 1.00 | 2.50 | 1.66 | 9.21 | < 0.0001 | 0.23 | 2.46 | 1.69 | 8.26 | < 0.0001 |
| PC | 950.92 | 3919.64 | 2242.51 | 4034.05 | < 0.0001 | 685.28 | 3445.79 | 1750.51 | 1788.65 | < 0.0001 |
| SSC | 157.78 | 457.16 | 262.27 | 1028.39 | < 0.0001 | 130.62 | 452.84 | 243.73 | 159.32 | < 0.0001 |
| TPC | 124.68 | 289.48 | 189.99 | 766.26 | < 0.0001 | 93.97 | 287.60 | 166.80 | 209.37 | < 0.0001 |
| AC | 782.84 | 1127.43 | 966.54 | 2294.38 | < 0.0001 | 647.70 | 1112.57 | 886.22 | 447.22 | < 0.0001 |
| GPA | 0.54 | 26.87 | 9.77 | 266.70 | < 0.0001 | 0.49 | 25.89 | 8.51 | 282.96 | < 0.0001 |
| CAT | 90.91 | 324.68 | 176.31 | 78.75 | < 0.0001 | 90.91 | 324.68 | 153.79 | 57.02 | < 0.0001 |
| LP | 0.55 | 9.45 | 6.73 | 41.52 | < 0.0001 | 4.68 | 11.06 | 7.64 | 124.98 | < 0.0001 |

GP: germination percentage (%), RL: root length (cm), SL: shoot lenght (cm), FW: fresh weight (g), DW: dry weight (g), PC: proline content (µg/g FW), SSC: soluble sugar content (µg/g FW), TPC: total phenolic content (µg/g FW), AC: antioxidant capacity (µg/g FW), GPA: guaiacol peroxidase activity (units/min/g FW), CAT: catalase (units/min/g FW), and LP: lipid peroxidation (nmol/g FW).

**Table 4.20 Statistics that describe morpho-chemical parameters of accessions, treatments and combination of them under different Cd-indiced.**

| Traits | Minimum | Maximum | Mean | Accessions | Pr > F | Treatment | Pr > F | Accessions* Treatment | Pr > F |
|--------|---------|---------|--------|------------|----------|-----------|----------|-----------|----------|
| GP | 60.00 | 100.00 | 88.23 | 35.65 | < 0.0001 | 220.19 | < 0.0001 | 2.51 | < 0.0001 |
| RL | 0.25 | 10.95 | 3.22 | 33.64 | < 0.0001 | 44911.69 | < 0.0001 | 19.84 | < 0.0001 |
| SL | 0.48 | 9.10 | 5.15 | 48.15 | < 0.0001 | 12923.07 | < 0.0001 | 21.03 | < 0.0001 |
| FW | 3.65 | 65.58 | 37.00 | 56.98 | < 0.0001 | 1958.56 | < 0.0001 | 7.33 | < 0.0001 |
| DW | 0.23 | 4.80 | 1.76 | 19.71 | < 0.0001 | 156.89 | < 0.0001 | 4.06 | < 0.0001 |
| PC | 394.51 | 3919.64 | 1628.40 | 4876.36 | < 0.0001 | 131101.58 | < 0.0001 | 989.93 | < 0.0001 |
| SSC | 63.95 | 457.16 | 208.78 | 575.72 | < 0.0001 | 23485.39 | < 0.0001 | 155.89 | < 0.0001 |
| TPC | 50.15 | 289.48 | 149.10 | 743.56 | < 0.0001 | 39259.90 | < 0.0001 | 220.73 | < 0.0001 |
| AC | 497.70 | 1127.43 | 848.21 | 887.80 | < 0.0001 | 108833.83 | < 0.0001 | 310.08 | < 0.0001 |
| GPA | 0.21 | 30.58 | 8.22 | 890.04 | < 0.0001 | 1971.01 | < 0.0001 | 38.19 | < 0.0001 |
| CAT | 25.97 | 324.68 | 141.50 | 114.15 | < 0.0001 | 3051.41 | < 0.0001 | 28.04 | < 0.0001 |
| LP | 0.55 | 11.06 | 6.38 | 314.04 | < 0.0001 | 5103.38 | < 0.0001 | 33.00 | < 0.0001 |

GP: germination percentage (%), RL: root length (cm), SL: shoot lenght (cm), FW: fresh weight (g), DW: dry weight (g), PC: proline content (µg/g FW), SSC: soluble sugar content (µg/g FW), TPC: total phenolic content (µg/g FW), AC: antioxidant capacity (µg/g FW), GPA: guaiacol peroxidase activity (units/min/g FW), CAT: catalase (units/min/g FW), and LP: lipid peroxidation (nmol/g FW).

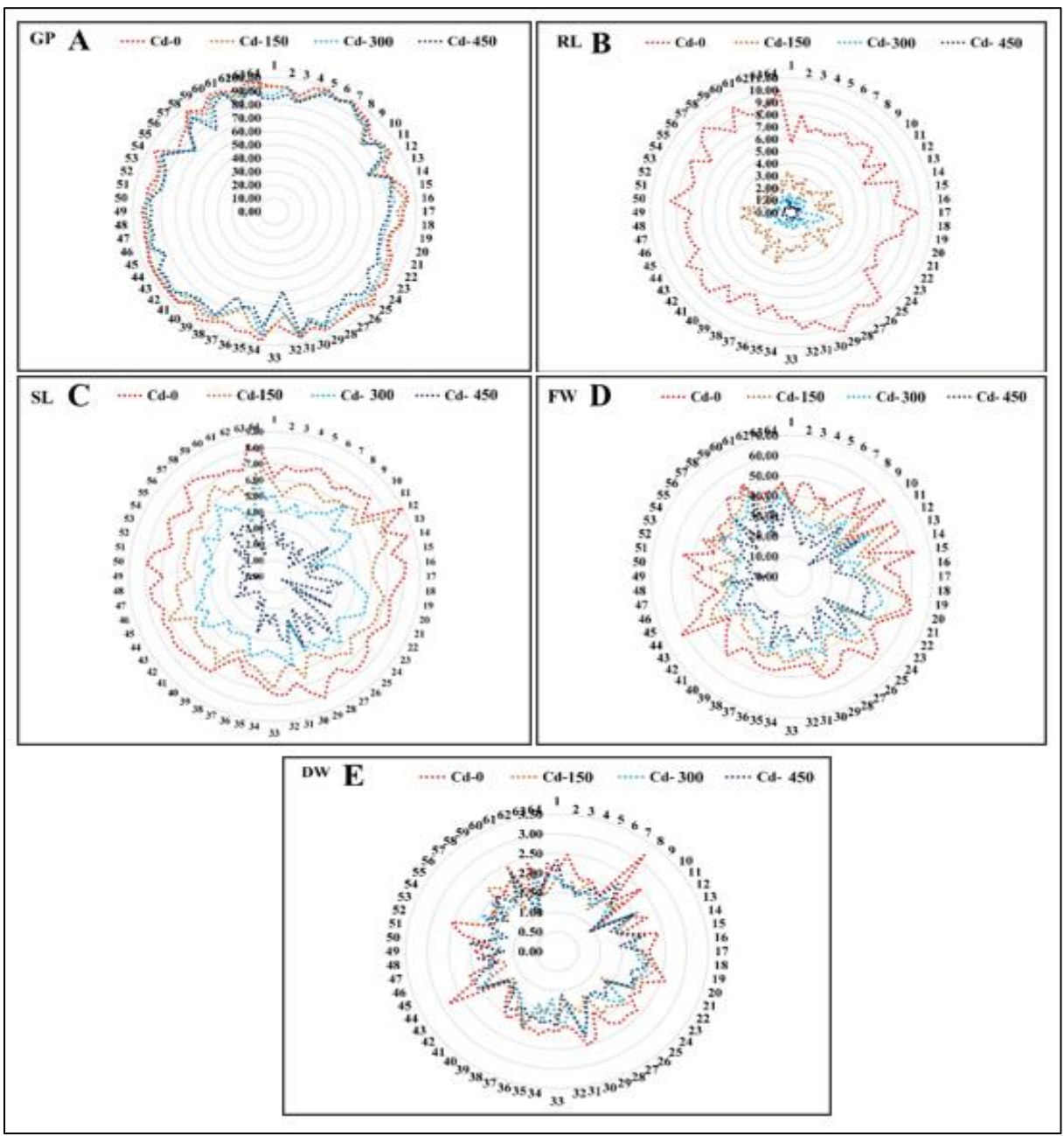

**Figure 4.12 Radar box illustrating the variation in phenotypic traits under control and Cd stress conditions. (A) Germination percentage (GP), (B) Root length (RL), (C) Shoot length (SL), (D) Seedling fresh weight (FW), and (E) Seedling dry weight (DW).**

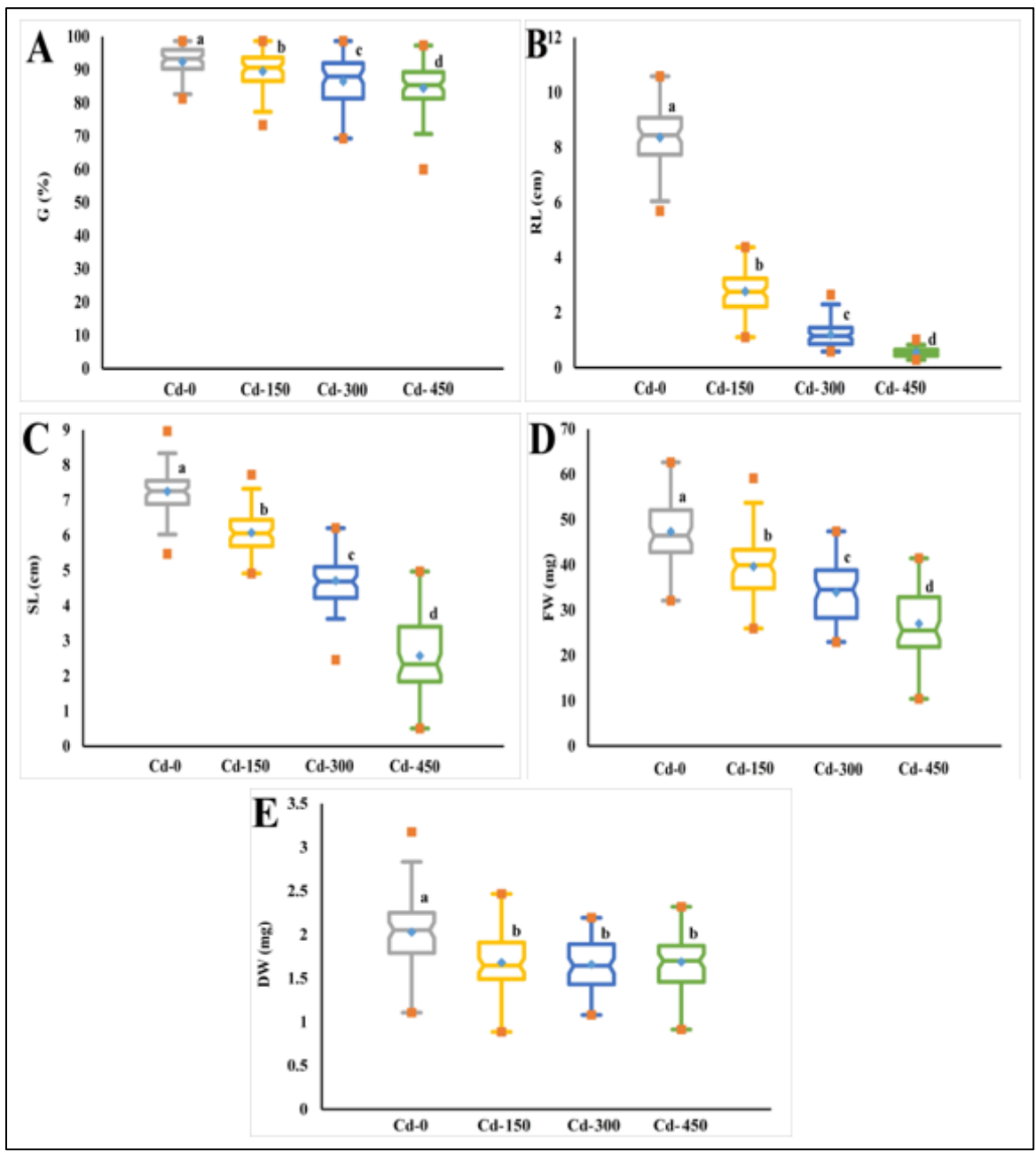

**Figure 4.13 Box chart illustrating the variation in phenotypic traits under control and Cd stress conditions. (A) Germination percentage (GP), (B) Root length (RL), (C) Shoot length (SL), (D) Seedling fresh weight (FW), and (E) Seedling dry weight (DW). The values specified are the mean values determined for the four measurements collected for control (T0) (0 µM Cd), T1 (150 µM Cd), T2 (300 µM Cd), and T3 (450 µM Cd). Different letters represent a significant difference between the mean values according to Duncan's Multiple-Range Test (P ≤ 0.01). A blue dot in the box indicates the mean, while orang dots represent the minimum and maximum values.**

### Relationship between morphological parameters and tomato accessions

Principal component analysis (PCA) was carried out to clarify the relationships among the morphological traits and the tomato accessions under control and Cd-induced conditions. The relationship between each variable and the main component is used to calculate the differential influence of the variables in each principal component.

Under the control conditions, the first component (PC1) and the second component (PC2) jointly explained 65.65% of the observed variation, and the first component (PC1) described 36.19% of the variation and positively influenced by the traits (FW and DW), which were positively correlated with the first component and these traits recorded the highest values in the accessions of AC08, AC10 AC15, AC18, AC20, and AC44. The second component (PC2) clarified 29.46% of the variance, which was positively correlated with GP, RL and SL traits (Figure 4.14 A). PC1 and PC2 were able to hold together 64.59% (45.94% by PC1 and 18.64% by PC2) of the initial variation at T1 (150 µM Cd-induced). The PC1 correlated positively with traits of RL, SL, FW and DW and negatively with GP. While RL, SL, FW and DW were responsible for the differentiation of AC12, AC18, AC19, AC20 AC21, AC22, AC24 and AC25), these accessions were considered as the highest tolerant accessions to 150 µM Cd-induced, however AC7, AC11, and AC42 were considered the lowest tolerant accessions under this treatment (Figure 4.14 B). Under the T2 treatment (300 µM Cd-induced), the first two components together explained 69.82% of the observed variation, PC1 accounted for the largest proportion of variance by 48.38%, while PC2 represented about 21.44% of the variance (Figure 4.14 C). The PC1 was positively associated with FW and RL, and these traits revealed the highest values in the accessions AC19, AC20, AC21, AC22 and AC60. The PC2 was positively correlated with GP, SL and DW. Figure 4.14 D, shows the distribution of morphological traits with the accessions under the T3 treatment (450 µM Cd-induced). The two components (PC1 and PC2) explained with a total variation of 61.98% (40.74% for PC1 and 21.23% for PC2), SL, FW, and DW were positively associated with the PC1, and the highest values of these traits revealed by the accessions AC12, AC18, AC19, AC22, AC25, AC57, AC60, and AC64, whereas the second component (PC2) positively related with GP and RL.

*Seedling biochemical estimation under Cd stress*

According to the ANOVA analysis and bar chart, there was a significant difference between the tomato accessions under both control (T0) and Cd-induced (Cd treatments) conditions for phytochemical parameters: proline content (PC), soluble sugar content (SSC), total phenolic content (TPC), antioxidant capacity (AC), guaiacol peroxidase activity (GPA), catalase (CAT), and lipid peroxidation (LP) (Table 4.19 and Figure 4.15). Table 4.20 shows that there were highly significant differences between accessions, Cd concentrations, and interactions between them for all biochemical traits (PC, SSC, TPC, AC, GPA, CAT, and LP) (P ≤ 0.001).

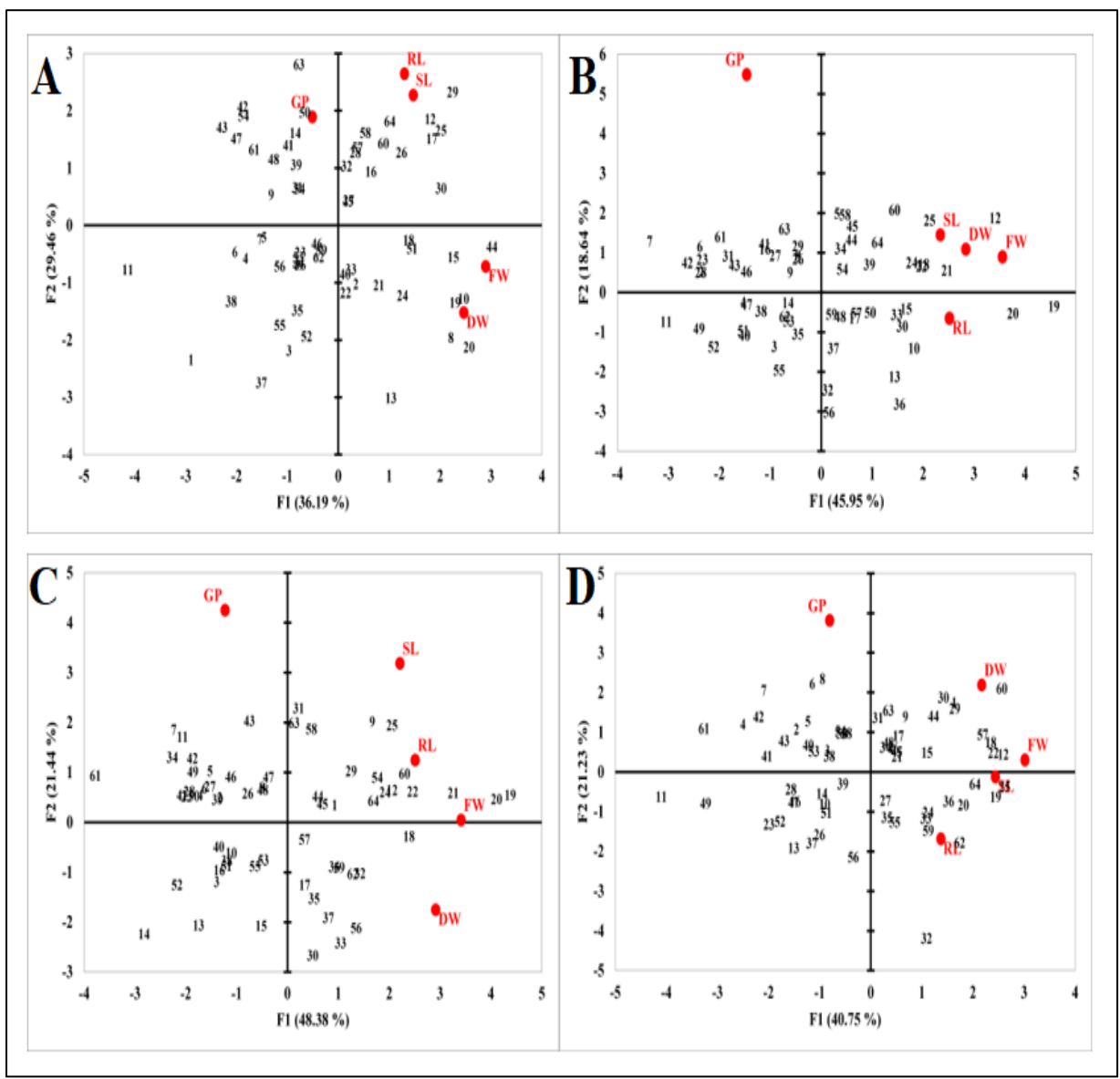

**Figure 4.14 Principal component analysis (PCA) indicates the rlationship between the morphological traits and accessions under control and cadmium stress conditions. A: control (0 µM), B: T1 (150 µM), C: T2 (300 µM), and D: (450 µM). GP: germination percentage, RL: root lenght, SL: shoot lenght, FW: fresh weight, and DW: dry weight.**

Under the control conditions (T0: 0 µM Cd-induced) (Table 4.19 and Appendix 28), PC and SSC values ranged from 394.51 to 1478.62 µg/g FW with a mean of 888.81 µg/g FW and 63.95 to 204.69 µg/g FW with an average 124.53 µg/g FW, respectively. The values of TPC and AC varied from 50.15 to 152.02 µg/g FW and 497.70 to 692.30 µg/g FW with a mean of 88.49 µg/g FW and 607.35 µg/g FW respectively. The GPA and CAT values were ranged from 0.21 to 13.92 units/min/g FW and 25.97 to 168.83 units/min/g FW with the average 5.47 units/min/g FW and 87.05 units/min/g FW, respectively. The maximum LP value was 7.48 nmol/g FW and the minimum value was 3.45 nmol/g FW with a mean of 5.25 nmol/g FW.

The highest and the lowest values of PC were 2684.77 and 915.03 µg/g FW, respectively, with an average 1631.77 µg/g FW, while SSC recorded the maximum value with 321.36 µg/g FW and the minimum value by 106.54 µg/g FW, with the mean 204.56 µg/g FW. The maximum values of TPC and AC were 235.17 µg/g FW and 1073.38 µg/g FW, respectively, whereas the minimum values of these traits were 81.24 µg/g FW and 684.19 µg/g FW, respectively and also the average of traits were 151.10 µg/g FW and 932.72 µg/g FW, respectively. The GPA and CAT values ranged from 0.38 to 30.58 units/min/g FW, and 77.92 to 285.71 units/min/g FW, with the average of 9.13 and 148.84 units/min/g FW, respectively. The LP value varied from 2.74 to 8.61 nmol/g FW with a mean of 5.89 nmol/g FW. these data were revealed under the T1 treatment (150 µM Cd-induced) (Table 4.19 and Appendix 29).

At T2 (300 µM Cd induced), the value of PC ranged from 950.92 to 3919.64 µg/g FW with an average of 2242.51 µg/g FW, and SSC from 157.78 to 457.16 µg/g FW with an average of 262.27 µg/g FW, TPC from 124.68 to 289.48 µg/g FW with an average of 189.99 µg/g FW, AC from 782.84 to 1127.43 µg/g FW with a mean of 966.54 µg/g FW, GPA from 0.54 to 26.87 units/min/g FW with an average of 9.77 units/min/g FW, CAT from 90.91 to 324.68 units/min/g FW, with a mean of 176.31 units/min/g FW, and LP from 0.55 to 9.45 nmol/g FW with an average of 6.73 nmol/g FW (Table 4.19 and Appendix 30).

Under Cd-induced T3 (450 µM), the PC values varied between 685.28 to 3445.79 µg/g FW with an average of 1750.51 µg/g FW, while the values of SSC and TPC were ranged from 130.62 to 452.84 µg/g FW and 93.97 to 287.60 µg/g FW with the average values of 243.73 and 166.80 µg/g FW, respectively. The AC values ranged from 657.70 to 1112.57 µg/g FW with a mean of 886.22 µg/g FW. The values of GPA and CAT ranged from 0.49 to 25.89 units/min/g FW and 90.91 to 324.68 units/min/g FW, with the average values of 8.51 and 153.79 units/min/g FW, respectively. The LP value also varied from 4.68 to 11.06 nmol/g FW, with a mean 7.64 nmol/g FW (Table 4.19 and Appendix 31).

The bar chart of all biochemical traits revealed significant differences between T0 (0 µM), T1 (150 µM), T2 (300 µM), and T3 (450 µM) (Figure 4.15). The phytochemical traits values of all tomato accessions under normal conditions revealed the lowest values, compared to plants that had been exposed to Cd stress. The T2 treatment (300 µM Cd-induced) recorded the highest value of all biochemical traits (Figure 4.17 A, B, C, D, E, and F), except the LP trait which recorded the maximum value under T3 treatment (450 µM Cd-induced) (Figure 4.15 G).

The mean value of all biochemical traits for all tomato accession under all Cd treatments were shown in Appendix 32. The findings revealed significant differences among accessions. The highest PC and SSC values were found in AC44 and AC8, which were 2771.82 µg/g FW and 328.92 µg/g FW, respectively, and the lowest values were found in AC51 and AC56, which were 906.14 µg/g

FW and 132.01 μg/g FW, respectively. AC7 had the highest TPC and AC values, which were 225.43 μg/g FW and 957.62 μg/g FW, respectively, while the lowest values showed by AC22 and AC 56 with 97.15 μg/g FW and 703.61 μg/g FW, respectively. The maximum values of GPA and CAT were revealed in AC63 and AC7 with 22.08 units/min/g FW and 219.16 units/min/g FW, respectively, while the minimum values were recorded by AC1 and AC32 by 0.45 units/min/g FW and 92.53 units/min/g FW, respectively. The highest and the lowest value of LP recorded by AC36 (8.61 nmol/g FW) and AC34 (4.45 nmol/g FW).

The interaction values between the tomato accessions and the Cd treatments are indicated in Appendix 33, highly significant differences revealed between them. The maximum and minimum values of PC were recorded by the interaction of AC44*Cd-300    and  AC1*Cd-0  and  AC  with 3894.51 μg/g FW and 419.64 μg/g FW, respectively. The combination of AC7*Cd-450 and AC43*Cd-0 recorded the highest and the lowest values of SCC with 449.75 μg/g FW and 68.27 μg/g FW, respectively. The maximum values of TPC and AC were shown by the interaction of AC7*Cd-300 and AC5*Cd-300 with 278.42 μg/g FW and 1124.05 μg/g FW, respectively, while the minimum values indicated by the combinations of AC51*Cd-0 (50.90 μg/g FW) and AC27*Cd-0 (501.76 μg/g FW), respectively. The interactions of AC63*Cd-150 and AC25*Cd-300, revealed the maximum values of GPA and CAT by 29.97 units/min/g FW and 311.69 units/min/g FW, respectively, whereas the minimum values recorded by AC1*Cd-0 (0.23 units/min/g FW), and AC25*Cd-0 (38.96 units/min/g FW), respectively. The highest and the lowest LP values were indicated by the interactions of AC19*Cd-450 and AC32*Cd-150, with 10.97 nmol/g FW and 2.79 nmol/g FW, respectively.

### *Multivariance analysis between phytochemical traits and tomato accessions under Cd-induced conditions*

The relationship between the biochemical traits and tomato accessions was estimated by using principal component analysis (PCA), under control and Cd-induced conditions.

The first component (PC1) and the second competent (PC2) together described 46.22% (26.85% for PC1 and 19.37% for PC2) of the total variance under the control condition. The first component positively correlated with the TPC, CAT, GPA, and LP. AC2, AC5, AC7, AC8, AC9, AC12, AC14, AC15, AC16, and AC50, situated the right side, near the TPC and CAT, which recorded the highest value of these traits. The second component (PC2) positively associated with PC, SSC, and AC, these traits revealed the highest value in the accessions AC22, AC26, AC32, AC48, and AC52 (Figure 4.16 A).

Under the T1 conditions, the first component (PC1) and the second component (PC2) jointly explained 45.20% of the observed variation, and the first component (PC1) described 27.64% of the variation and positively associated by the traits AC, CAT, PC, and GPA, these traits recorded the highest values in the accessions AC31, AC42, AC44, AC46, AC60, AC61, and AC63. The second component (PC2) clarified 17.56% of the variance, which was positively correlated with SSC, TPC, and LP traits (Figure 4.17 AB). PC1 and PC2 together described 60.25% (43.91% by PC1 and 16.34% by PC2) of the total variation under T2. The PC1 correlated positively with the traits LP, CAT, AC, SSC, TPC, and AC negatively with GPA. While SSC, TPC, AC, and CAT were responsible for the differentiation of AC5, AC7, and AC9, these accessions were considered as the highest tolerant to 300 µM Cd-induced (Figure 4.17 C). Under the T3 treatment, the first two components together described 71.44% of the observed variation, PC1 accounted for 56.17%, the of variance. while PC2 represented about 15.27% of the variance (Figure 4.15 D). The PC1 was positively associated with PC, LP, SSC, TPC, CAT, and AC. Whereas the SSC, TPC, AC, and CAT were responsible for the differentiation of AC5 and AC7 and these accessions were selected as the highest resistant to 450 µM Cd-induced.

### *Ranking of tomato accessions under Cd stress condition using seedling morphological traits*

All tomato accessions were ranked based on the germination percentage, root and shoot length, fresh and dry weights of seedlings exposed to Cd concentrations, and compared with the measured data of the seedlings under control conditions, which were measured at the seedling stage, while grown in the *in vitro* condition. In the case of determination of the best tomato accessions to Cd stress, the accessions which had the lowest average rank (AR) and the highest stress tolerance index (STI), were considered as the highest tolerant accession to Cd stress.

Under the T1 (150 µM Cd-induced) (Table 4.21), the accessions AC25, AC63, and AC5 were considered as the highest tolerant to Cd stress, while these accessions recorded the lowest AR value and the highest STI. AC13, AC56, and AC55 were selected as more susceptible to cadmium under the same condition. AC7, AC63, and AC60 were recommended as the Cd resistance accessions and AC13, AC33, and AC56 were considered as the lowest tolerant to Cd exposure under T2 (300 µM Cd-induced) (Table 4.21).

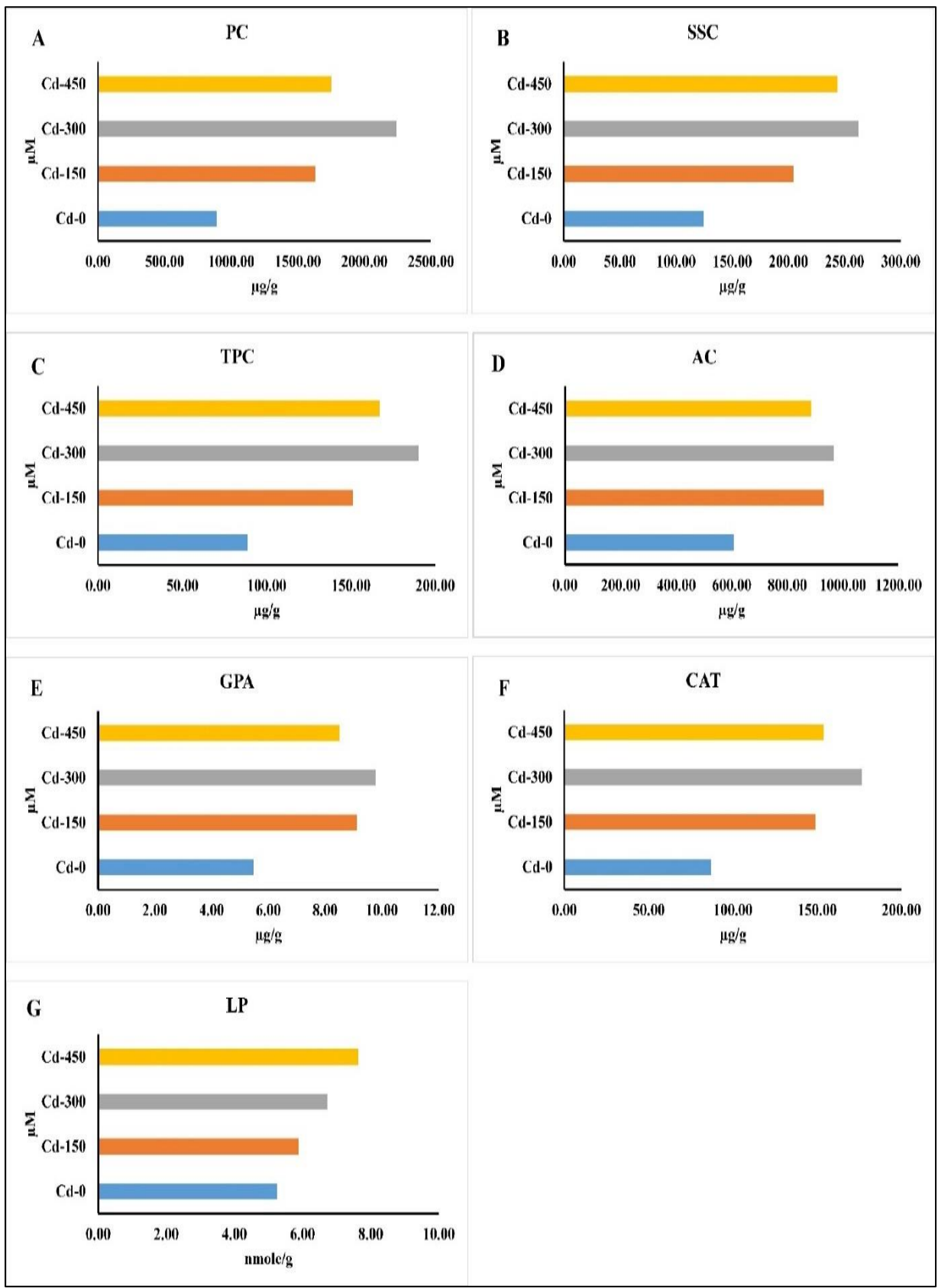

**Figure 4.15 The bar charts illustrates the effect of control and Cd stress conditions on the biochemical traits, A: proline conten, B: soluble sugar content, C: total phenolic content, D: antioxidant capacity, E: gouaucol peroxidase activity, F: catalase activity, G: lipid peroxidation.**

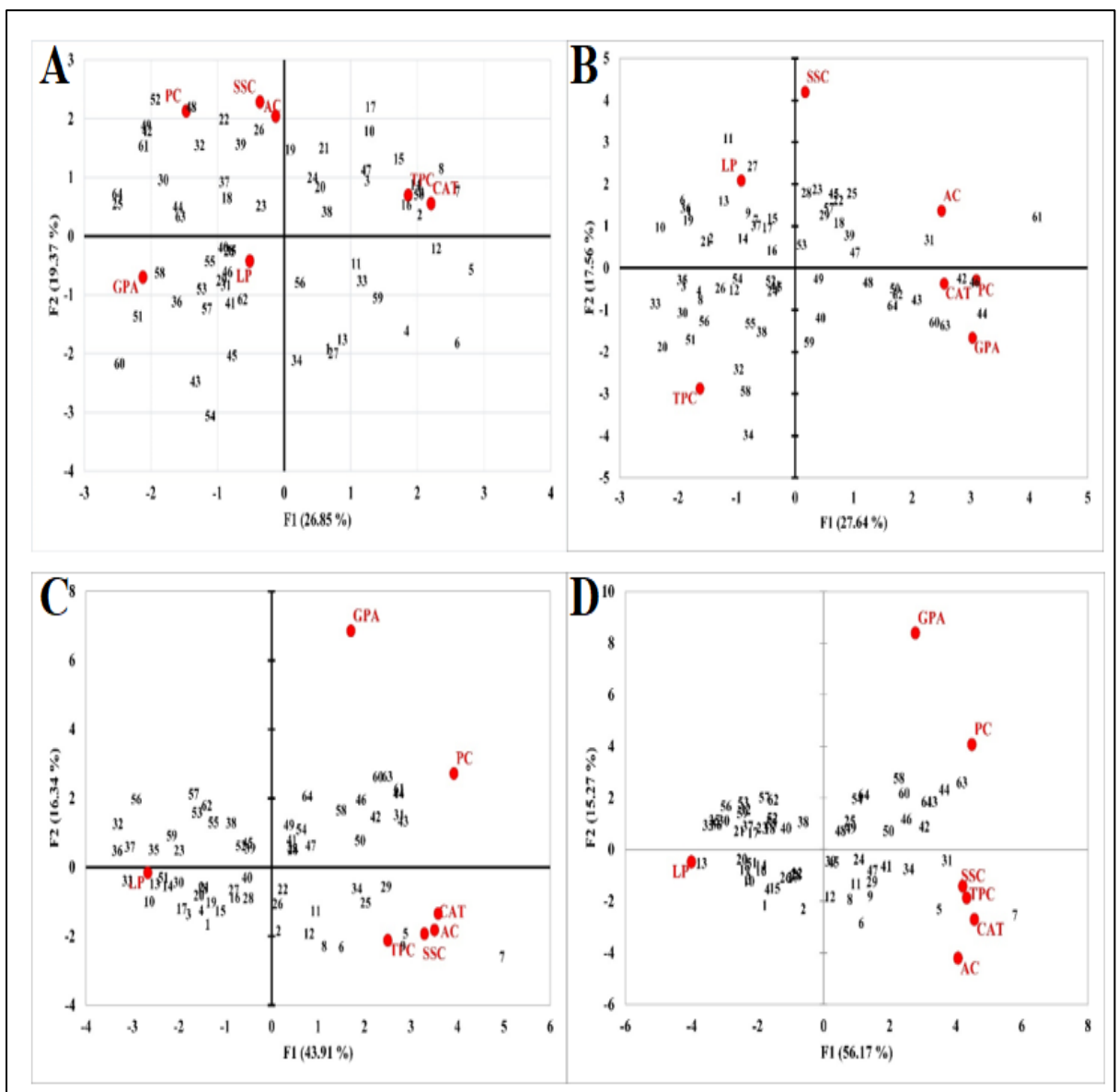

**Figure 4.16 PCA showed the distribution and relationship of all biochemical traits among all accessions. PC: proline content, SSC: soluble sugar content, TPC: total phenolic content, AC: antioxidant activity, LP: lipid peroxidation, GPA: guaicol peroxidase activity, CAT: catalase activity. A: control (0 µM), B: T1 (150 µM), C: T2 (300 µM), and D:T3: (450 µM).**

The highest tolerant accessions to Cd, under T3 (450 µM Cd-induced) were AC63, AC29, and AC7, whereas the susceptible accessions were AC32, AC13, and AC37 (Table 4.22).

The important ranking which can be considered for selection of the best accessions to Cd stress, in which the mean values of the traits from all three Cd stress conditions were compared with data from control condition (Table 4.22). It was revealed that AC7, AC5 and AC31 were considered as the best-performed tomato accessions resistant to Cd stress, while the AC56, AC32, and AC13 were chosen as sensitive accessions to Cd stress.

**Table 4.21 The ranking of tomato accessions, using the stress tolerance index (STI) and average number of ranks (AR). According to the germination percentage and growth traits of seedlings, grown under T1 and T2.**

| Accessions | T1 | | | T2 | | |
|---|---|---|---|---|---|---|
| | STI | AR | Rank | STI | AR | Rank |
| AC1 | 0.92 | 20.36 | 32 | 0.84 | 31.00 | 40 |
| AC2 | 0.93 | 26.64 | 29 | 0.91 | 16.09 | 22 |
| AC3 | 0.80 | 49.64 | 56 | 0.77 | 38.45 | 52 |
| AC4 | 0.93 | 36.45 | 28 | 0.88 | 33.27 | 28 |
| AC5 | 1.04 | 3.64 | 3 | 0.99 | 5.09 | 4 |
| AC6 | 0.95 | 21.45 | 26 | 0.92 | 15.09 | 20 |
| AC7 | 1.03 | 6.73 | 6 | 1.01 | 2.18 | 1 |
| AC8 | 0.95 | 29.55 | 24 | 0.91 | 20.91 | 21 |
| AC9 | 0.92 | 40.55 | 33 | 0.93 | 14.18 | 15 |
| AC10 | 0.77 | 60.09 | 61 | 0.76 | 51.27 | 55 |
| AC11 | 0.90 | 25.09 | 41 | 0.86 | 20.91 | 33 |
| AC12 | 0.98 | 16.00 | 18 | 0.91 | 25.73 | 23 |
| AC13 | 0.69 | 61.18 | 64 | 0.64 | 60.27 | 64 |
| AC14 | 0.84 | 35.00 | 50 | 0.78 | 42.55 | 50 |
| AC15 | 0.90 | 44.36 | 38 | 0.82 | 48.09 | 42 |
| AC16 | 0.97 | 24.55 | 20 | 0.86 | 40.82 | 35 |
| AC17 | 0.89 | 47.09 | 42 | 0.81 | 49.45 | 44 |
| AC18 | 0.91 | 29.00 | 36 | 0.81 | 45.64 | 43 |
| AC19 | 0.80 | 38.64 | 55 | 0.75 | 43.00 | 56 |
| AC20 | 0.79 | 39.00 | 58 | 0.74 | 43.82 | 60 |
| AC21 | 0.87 | 36.00 | 47 | 0.77 | 50.18 | 51 |
| AC22 | 0.88 | 32.64 | 43 | 0.84 | 28.73 | 39 |
| AC23 | 0.97 | 13.91 | 19 | 0.88 | 34.36 | 31 |
| AC24 | 0.90 | 34.18 | 40 | 0.88 | 23.73 | 32 |
| AC25 | 1.06 | 11.55 | 1 | 0.99 | 16.82 | 5 |
| AC26 | 0.93 | 35.82 | 30 | 0.88 | 32.45 | 30 |
| AC27 | 0.94 | 28.18 | 27 | 0.85 | 42.27 | 38 |
| AC28 | 0.95 | 26.91 | 23 | 0.91 | 24.91 | 24 |
| AC29 | 1.03 | 18.91 | 5 | 0.97 | 20.09 | 8 |
| AC30 | 0.87 | 51.36 | 46 | 0.80 | 50.27 | 45 |
| AC31 | 1.00 | 13.45 | 12 | 0.99 | 5.27 | 6 |
| AC32 | 0.78 | 59.91 | 60 | 0.76 | 55.82 | 53 |
| AC33 | 0.82 | 34.64 | 53 | 0.64 | 63.00 | 63 |
| AC34 | 0.99 | 25.27 | 16 | 0.96 | 14.82 | 10 |
| AC35 | 0.87 | 45.82 | 45 | 0.76 | 57.09 | 54 |
| AC36 | 0.80 | 58.00 | 57 | 0.74 | 58.73 | 59 |
| AC37 | 0.79 | 57.82 | 59 | 0.75 | 53.73 | 58 |
| AC38 | 0.91 | 31.91 | 37 | 0.85 | 33.09 | 36 |
| AC39 | 1.00 | 25.18 | 13 | 0.92 | 28.91 | 18 |
| AC40 | 0.85 | 45.09 | 49 | 0.79 | 45.55 | 48 |
| AC41 | 1.01 | 9.64 | 9 | 0.92 | 26.36 | 16 |
| AC42 | 0.99 | 24.09 | 17 | 0.94 | 18.82 | 13 |
| AC43 | 0.99 | 17.00 | 15 | 0.95 | 13.09 | 12 |
| AC44 | 1.01 | 11.36 | 8 | 0.96 | 13.55 | 11 |
| AC45 | 1.00 | 11.45 | 14 | 0.86 | 40.82 | 34 |
| AC46 | 0.95 | 16.82 | 22 | 0.92 | 12.73 | 17 |
| AC47 | 0.95 | 24.09 | 25 | 0.89 | 29.27 | 27 |
| AC48 | 0.92 | 29.82 | 31 | 0.88 | 25.27 | 29 |
| AC49 | 0.91 | 35.73 | 35 | 0.90 | 19.09 | 25 |
| AC50 | 0.92 | 34.45 | 34 | 0.89 | 21.73 | 26 |
| AC51 | 0.86 | 50.18 | 48 | 0.79 | 49.27 | 46 |
| AC52 | 0.81 | 44.27 | 54 | 0.75 | 50.36 | 57 |
| AC53 | 0.84 | 44.64 | 52 | 0.79 | 39.73 | 47 |
| AC54 | 0.96 | 34.18 | 21 | 0.94 | 22.09 | 14 |
| AC55 | 0.72 | 50.45 | 62 | 0.71 | 39.09 | 61 |
| AC56 | 0.71 | 62.45 | 63 | 0.68 | 61.64 | 62 |
| AC57 | 0.84 | 53.18 | 51 | 0.82 | 38.09 | 41 |
| AC58 | 1.02 | 15.64 | 7 | 0.97 | 16.55 | 9 |
| AC59 | 0.87 | 44.36 | 44 | 0.78 | 52.18 | 49 |
| AC60 | 1.03 | 21.64 | 4 | 1.00 | 11.09 | 3 |
| AC61 | 1.01 | 17.00 | 10 | 0.97 | 9.45 | 7 |
| AC62 | 0.90 | 43.09 | 39 | 0.85 | 40.18 | 37 |
| AC63 | 1.04 | 19.64 | 2 | 1.01 | 10.91 | 2 |
| AC64 | 1.01 | 23.27 | 11 | 0.92 | 31.00 | 19 |

**Table 4.22 The ranking of tomato accessions, using the stress tolerance index (STI) and average number of ranks (AR). According to the germination percentage and growth traits of seedlings, grown under T3 and all treatments.**

| Accessions | T3 | | | All | | |
|---|---|---|---|---|---|---|
| | STI | AR | Rank | AR | STI | Rank |
| AC1 | 0.79 | 33.55 | 40 | 27.18 | 0.85 | 27 |

| AC2 | 0.84 | 25.73 | 26 | 19.36 | 0.89 | 17 |
|---|---|---|---|---|---|---|
| AC3 | 0.75 | 36.45 | 50 | 40.45 | 0.77 | 42 |
| AC4 | 0.86 | 22.64 | 20 | 29.45 | 0.89 | 31 |
| AC5 | 0.94 | 6.91 | 4 | 4.45 | 0.99 | 2 |
| AC6 | 0.91 | 8.18 | 12 | 11.91 | 0.93 | 7 |
| AC7 | 0.97 | 3.00 | 3 | 2.45 | 1.00 | 1 |
| AC8 | 0.90 | 12.45 | 14 | 17.36 | 0.92 | 16 |
| AC9 | 0.91 | 9.73 | 13 | 16.82 | 0.92 | 15 |
| AC10 | 0.74 | 44.91 | 51 | 53.64 | 0.75 | 56 |
| AC11 | 0.81 | 22.82 | 35 | 22.64 | 0.86 | 19 |
| AC12 | 0.85 | 28.55 | 21 | 23.64 | 0.92 | 21 |
| AC13 | 0.61 | 58.27 | 63 | 60.55 | 0.64 | 62 |
| AC14 | 0.79 | 27.64 | 42 | 33.36 | 0.80 | 35 |
| AC15 | 0.82 | 35.82 | 32 | 45.00 | 0.85 | 49 |
| AC16 | 0.80 | 45.18 | 39 | 40.18 | 0.87 | 40 |
| AC17 | 0.78 | 46.27 | 45 | 48.91 | 0.83 | 54 |
| AC18 | 0.76 | 47.64 | 47 | 43.09 | 0.83 | 47 |
| AC19 | 0.69 | 52.00 | 59 | 46.18 | 0.75 | 52 |
| AC20 | 0.69 | 46.36 | 56 | 43.45 | 0.74 | 48 |
| AC21 | 0.71 | 53.73 | 54 | 48.91 | 0.78 | 55 |
| AC22 | 0.79 | 36.91 | 41 | 31.45 | 0.84 | 33 |
| AC23 | 0.76 | 51.36 | 48 | 37.55 | 0.87 | 38 |
| AC24 | 0.84 | 21.82 | 28 | 23.45 | 0.87 | 20 |
| AC25 | 0.93 | 20.27 | 7 | 15.82 | 1.00 | 14 |
| AC26 | 0.79 | 46.91 | 43 | 39.45 | 0.87 | 39 |
| AC27 | 0.83 | 33.09 | 31 | 35.73 | 0.87 | 37 |
| AC28 | 0.81 | 41.09 | 36 | 32.64 | 0.89 | 34 |
| AC29 | 0.97 | 5.82 | 2 | 13.00 | 0.99 | 8 |
| AC30 | 0.83 | 28.27 | 30 | 45.73 | 0.84 | 51 |
| AC31 | 0.93 | 8.91 | 9 | 7.73 | 0.97 | 3 |
| AC32 | 0.59 | 62.09 | 64 | 60.73 | 0.71 | 63 |
| AC33 | 0.64 | 59.00 | 61 | 59.18 | 0.70 | 60 |
| AC34 | 0.87 | 27.82 | 18 | 21.64 | 0.94 | 18 |
| AC35 | 0.69 | 58.18 | 57 | 56.00 | 0.77 | 58 |
| AC36 | 0.73 | 54.55 | 52 | 57.18 | 0.76 | 59 |
| AC37 | 0.63 | 62.00 | 62 | 59.18 | 0.72 | 61 |
| AC38 | 0.83 | 23.36 | 29 | 28.55 | 0.86 | 30 |
| AC39 | 0.85 | 34.09 | 23 | 30.45 | 0.92 | 32 |
| AC40 | 0.78 | 35.18 | 46 | 41.64 | 0.81 | 44 |
| AC41 | 0.85 | 32.09 | 22 | 24.27 | 0.93 | 22 |
| AC42 | 0.93 | 9.55 | 8 | 14.82 | 0.96 | 9 |
| AC43 | 0.89 | 19.82 | 15 | 15.27 | 0.95 | 11 |
| AC44 | 0.92 | 12.82 | 11 | 11.45 | 0.96 | 6 |
| AC45 | 0.81 | 41.55 | 34 | 35.45 | 0.89 | 36 |
| AC46 | 0.87 | 14.64 | 17 | 15.09 | 0.92 | 10 |
| AC47 | 0.84 | 29.36 | 25 | 27.00 | 0.89 | 26 |
| AC48 | 0.84 | 25.64 | 27 | 26.45 | 0.88 | 25 |
| AC49 | 0.82 | 34.64 | 33 | 28.18 | 0.88 | 29 |
| AC50 | 0.84 | 23.27 | 24 | 25.64 | 0.88 | 23 |
| AC51 | 0.78 | 42.73 | 44 | 48.55 | 0.81 | 53 |
| AC52 | 0.73 | 42.18 | 53 | 45.09 | 0.76 | 50 |
| AC53 | 0.75 | 42.45 | 49 | 42.36 | 0.79 | 46 |
| AC54 | 0.88 | 23.73 | 16 | 25.64 | 0.93 | 24 |
| AC55 | 0.69 | 37.73 | 58 | 40.36 | 0.71 | 41 |
| AC56 | 0.65 | 59.27 | 60 | 62.27 | 0.68 | 64 |
| AC57 | 0.80 | 31.36 | 38 | 41.64 | 0.82 | 45 |
| AC58 | 0.92 | 17.64 | 10 | 15.45 | 0.97 | 12 |
| AC59 | 0.71 | 56.45 | 55 | 53.73 | 0.79 | 57 |
| AC60 | 0.94 | 16.45 | 5 | 15.73 | 0.99 | 13 |
| AC61 | 0.93 | 9.91 | 6 | 10.00 | 0.97 | 4 |
| AC62 | 0.80 | 39.55 | 37 | 40.73 | 0.85 | 43 |
| AC63 | 0.97 | 8.55 | 1 | 11.27 | 1.01 | 5 |
| AC64 | 0.87 | 30.09 | 19 | 27.45 | 0.93 | 28 |

**4.1.2.1 Assessment of cadmium resistance under greenhouse conditions**

*Analysis of variance (ANOVA), increasing and decreasing in percentages in morphological traits of four tomato accessions under Cd stress*

The results of a two-way analysis of variance (ANOVA) (Table 4.23) on agro-morphological traits showed highly significant effects of treatments for all traits except RL, RFW, and RDW, while TCC showed the only significant difference. Also, highly significant differences were revealed for the effect of accessions on all traits except the TCC, and the interaction of treatments and accessions effects was highly significant for the traits of SFW and SDW. Table (4.24) indicates the increasing and decreasing percentages of all morphological parameters for the treatments under (Cd+Soil) and (Cd+Soil+Oak) compared with control. According to the results, all statuses have significant differences compared with the control for all traits. The highest decreases were recorded under Cd+Soil stress for the RL, SL, SFW, and TFW by 14.35%, 18.00%, 6.53%, 5.88%, and 13.85%, respectively, while the lowest decrease for the RL, SL, SFW, and TFW traits were recorded under Cd+Soil+Oak by 10.68%, 10.22%, 0.46%, and 10.10%, with a slight increase recorded for the SDW trait at 0.42%. The highest increases were recorded under Cd+Soil for the RFW, RDW, and TCC traits, with 20.77%, 27.33%, and 23.54%, respectively, and the lowest increases were indicated under Cd+Soil+Oak for these traits, with 17.47%, 24.21%, and 8.47%, respectively.

The increasing and decreasing percentages of morphological traits for all accessions are indicated in Table 4.25. The results revealed that all accessions have significant differences compared with control for all traits. The RL trait was significantly reduced in AC56, AC07, and AC05 by 21.24%, 17.36%, and 13.25%, respectively, and in AC32, it was slightly increased by 1.79%. The highest and lowest reductions of SL were indicated by AC32 (29.88%) and AC07 (2.55%), respectively. The RFW parameter was significantly increased in AC05, AC07, and AC32 by 22.73%, 43.89%, and 24.71%, respectively, while reduced in AC56 by 14.85%. AC56 recorded the highest reduction in SFW by 16.09%, and AC05 showed a significant increase of 5.36%. RDW was significantly increased in AC32, AC05, and AC07 by 44.41%, 37.88%, and 34.06%, respectively, and decreased in AC65 by 12.46%. AC56 recorded the highest reduction in SDW with 9.34%, and AC05 showed an increase of 3.11%. The highest increase of TCC was recorded by AC56 (23.24%) and the lowest increase of this trait was shown by AC05, which was 3.77%. The TFW parameter was significantly reduced for all accessions, the highest and the lowest reductions were found by AC56 (16.00%) and AC07 (4.63%), respectively.

**Table 4.23 Two-way ANOVA for measured agro-morphological traits was tested for four tomato accessions under Cd stress.**

| | | RL | SL | RFW | SFW | RDW | SDW | TFW |
|---|---|---|---|---|---|---|---|---|
| Treatments | F | $2.43^{NS}$ | $12.77^{**}$ | $0.62^{NS}$ | $33.92^{**}$ | $0.44^{NS}$ | $41.93^{**}$ | $9.16^{**}$ |

| | | | | | | | |
|---|---|---|---|---|---|---|---|
| | Pr > F | 0.14 | 0.00 | 0.44 | < 0.0001 | 0.51 | < 0.0001 | 0.01 |
| Accessions | F | 18.34** | 27.54** | 34.14** | 74.73** | 24.10** | 27.98** | 17.15** |
| | Pr > F | < 0.0001 | < 0.0001 | < 0.0001 | < 0.0001 | < 0.0001 | < 0.0001 | < 0.0001 |
| Treatments*Accessions | F | 2.41$^{NS}$ | 2.38$^{NS}$ | 2.22$^{NS}$ | 18.28** | 1.20$^{NS}$ | 7.43** | 2.78$^{NS}$ |
| | Pr > F | 0.11 | 0.11 | 0.13 | < 0.0001 | 0.34 | 0.00 | 0.07 |

**RL: root length (cm), SL: shoot length (cm), RFW: root fresh weight (g), SFW: shoot fresh weight (g), RDW: root dry weight (g), SDW: shoot dry weight (g), TFW: total fruit weight per plant.**

**Table 4.24 The increasing and decreasing of morphological traits under Cd stress compared to control.**

| Treatments | RL | SL | RFW | SFW | RDW | SDW | TFW |
|---|---|---|---|---|---|---|---|
| Cd+Soil | 14.35 a | 18.00 a | -20.77 a | 6.53 a | -27.73 a | 5.88 a | 13.85 a |
| Cd+Soil+Oak | 10.68 a | 10.22 b | -17.47 a | 0.46 b | -24.21 a | -0.42 b | 10.10 b |

**Increasing and decreasing in percentage of all morphological traits under Cd stress compared to control. RL: root length (cm), SL: shoot length (cm), RFW: root fresh weight (g), SFW: shoot fresh weight (g), RDW: root dry weight (g), SDW: shoot dry weight (g), TFW: total fruit weight per plant (g). (-= increasing, += decreasing).**

**Table 4.25  The increasing and decreasing morphological traits of accessions compared to control.**

| Accessions | RL | SL | RFW | SFW | RDW | SDW | TFW |
|---|---|---|---|---|---|---|---|
| AC05 | 13.25 b | 11.71 b | -22.73 b | -5.36 c | -37.88 b | -3.11 c | 12.31 a |
| AC07 | 17.36 ab | 2.55 c | -43.89 c | 1.41 b | -34.06 b | 1.56 b | 4.63 b |
| AC32 | -1.79 c | 29.88 a | -24.71 b | 1.85 b | -44.41 b | 3.13 b | 14.96 a |
| AC56 | 21.24 a | 12.31 b | 14.85 a | 16.09 a | 12.46 a | 9.34 a | 16.00 a |

**RL: root length (cm), SL: shoot length (cm), RFW: root fresh weight (g), SFW: shoot fresh weight (g), RDW: root dry weight (g), SDW: shoot dry weight (g), TFW: total fruit weight per plant (g).**

### The principal component analysis (PCA) of morphological traits of four tomato accessions under Cd stress

The principal component analysis (PCA) divides the input and response variables into many clusters based on their similarity and differences in correlation and variance. A multivariate analysis (PCA) was carried out on morphological traits for accessions under Cd stress (Figure 4.17). PCA (Figure 4.17) data revealed that the first two dimensions (F1 and F2) explained 86.43% of the total variance. The first axis (F1) describes 56.69% of the variation, and the second axis (F2) explains 29.74%. The first principal component (F1) was positively associated with RL, RFW, SFW, RDW, SDW, and TFW traits, and these traits also correlated positively with AC56 and negatively with AC05, AC07, and AC32.

### The effect of the interaction of treatments and accessions on morphological traits

Table 4.26 shows the increasing and decreasing of morphological traits from the interaction of treatments and accessions compared to control. In the case of the RL trait, the highest reductions were revealed in the interaction (Cd+Soil*AC56) by 20.93%, and the interaction (Cd+Soil+Oak*AC32) recorded an increase of 8.30%. The interaction Cd+Soil+Oak*AC07 showed a significant increase in SL by 4.71%, while Cd+Soil*AC32 indicated the highest reduction with 31.95%. The highest increase and decrease in RFW traits were revealed in the interaction of Cd+Soil*AC07 (45.41%) and Cd+Soil*AC56 (20.03%). The interaction Cd+Soil*AC56 recorded a significant increase in SFW with 14.83%, while Cd+Soil*AC56 recorded the highest reduction with 18.74%. With the exception of the Cd+Soil*AC56 and Cd+Soil+Oak*AC56, which decreased by 14.43% and 10.49%, respectively, all interactions increased significantly in the RDW trait. The highest and the lowest decreases in SDW were recorded at the interaction of Cd+Soil*AC56 (12.23%) and Cd+Soil+Oak*AC32 (2.90%), respectively, whereas the interaction of Cd+Soil+OaK*AC05 showed an increase of 9.78%. All interactions in the TCC trait were significantly increased except for the interaction Cd+Soil+OaK*AC05, which was decreased by 7.96%. All interactions in TFW were significantly decreased, and the highest reduction was recorded at Cd+Soil*AC56, and the lowest reduction was shown at Cd+Soil+Oak*AC07.

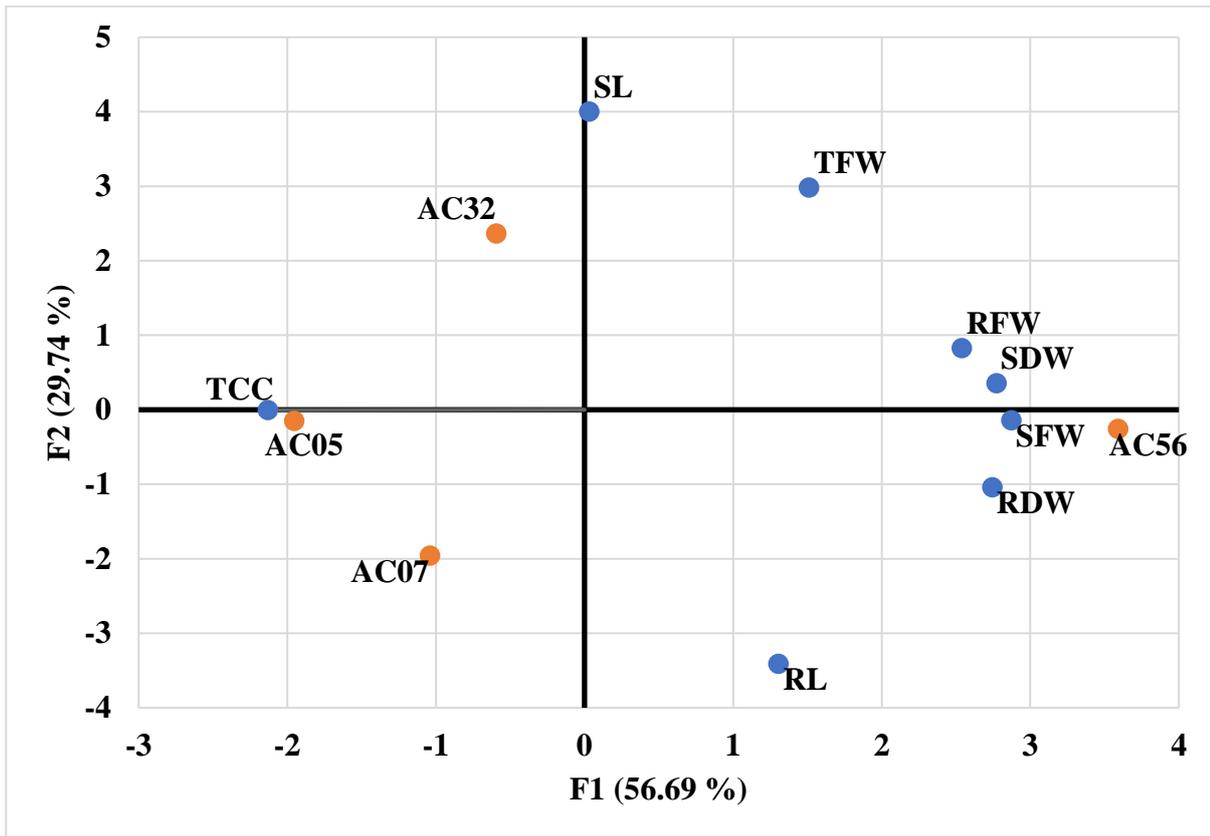

**Figure 4.17 Principal component analysis (PCA) plot presenting the distribution and relationship between accessions (AC05, AC07, AC32 and AC56) and studied traits. Studied traits:( RL: root length, SL: shoot length, RFW: root fresh weight, SFW: shoot fresh weight, RDW: root dry weight, SDW: shoot dry weight, TCC: total chlorophyll content, TFW: total fruit weight per plant).**

**Table 4.26 The increasing and decreasing of morphological traits of the interaction treatments and accessions.**

|  | RL | SL | RFW | SFW | RDW | SDW | TCC | TFW |
|---|---|---|---|---|---|---|---|---|
| Cd+Soil*AC05 | 11.54 ab | 17.81 b | -23.02 bc | 4.10 c | -37.02 b | 3.55 b | -15.50 ab | 15.30 ab |

| | | | | | | | |
|---|---|---|---|---|---|---|---|
| Cd+Soil+OaK*AC05 | 14.95 ab | 5.60 c | -22.43 bc | -14.83 d | -38.74 b | -9.78 d | 7.96 a | 9.32 cd |
| Cd+Soil*AC07 | 20.20 a | 9.81 bc | -45.41 d | 1.30 c | -33.53 b | 4.36 b | -24.97 ab | 5.38 d |
| Cd+Soil+Oak*AC07 | 14.51 ab | -4.71 d | -42.37 d | 1.52 c | -34.59 b | -1.25 c | -12.48 ab | 3.88 d |
| Cd+Soil*AC32 | 4.71 b | 31.95 a | -34.69 cd | 2.00 c | -54.82 b | 3.36 b | -28.96 b | 14.57 abc |
| Cd+Soil+Oak*AC32 | -8.30 c | 27.80 a | -14.73 b | 1.70 c | -34.00 b | 2.90 b | -7.61 ab | 15.35 ab |
| Cd+Soil*AC56 | 20.93 a | 12.43 bc | 20.03 a | 18.74 a | 14.43 a | 12.23 a | -24.74 ab | 20.16 a |
| Cd+Soil+Oak*AC56 | 21.54 a | 12.20 bc | 9.66 a | 13.44 b | 10.49 a | 6.45 b | -21.74 ab | 11.84 bc |

**RL: root length (cm), SL: shoot length (cm), RFW: root fresh weight (g), SFW: shoot fresh weight (g), RDW: root dry weight (g), SDW: shoot dry weight (g), TFW: total fruit weight per plant (g).**

### *Analysis of variance, and means comparison of chemical characteristics*

The results of the two-way analysis of variance (ANOVA) of phytochemical characters (PC, SSC, TPC, AC, GPA, and CAT) of the fresh leaves of four tomato accessions grown under Cd stress and control are shown in (Table 4.27), which indicates to significant differences in all characteristics for treatments, accessions, and the interaction of treatments and accessions. Table 4.28 shows the effects of treatments and controls on the chemical characteristics of the leaves of four tomato accessions. The results showed a significant difference between the treatments of Cd+Soil, Cd+Soil+Oak and control for chemical characteristics. The highest values of proline content (PC), soluble sugar content (SSC), antioxidant capacity (AC), and guaiacol peroxidase activity (GPA) were recorded under the Cd+Soil+Oak treatment, which were 1772.46 µg/g FLW, 687.18 µg/g FLW, 1025.74 µg/g FLM, and 0.43 units/min/g FW, respectively, while the treatment of Cd+Soil showed the highest values of total phenolic compound (TPC) and catalase activity (CAT) with 400.43 g/g FLW and 158.72 units/min/g FW, respectively. The lowest values of all characters were indicated under the control condition. Table 4.29 indicates the mean comparison of each accession for the chemical characteristics. AC07 recorded the highest values of proline content (PC), guaiacol peroxidase activity (GPA) and catalase activity (CAT) of 1767.42 µg/g FLW, 0.32 units/min/g FW and 150.37 units/min/g FW, respectively. AC05 revealed the highest values of soluble sugar content (SSC), total phenolic content (TPC) and antioxidant capacity, with 822.18 µg/g FLW, 420.37 µg/g FLW and 1016.52 µg/g FLM, respectively. The lowest values of proline content (PC), antioxidant activity (AC), guaiacol peroxidase activity (GPA) and catalase activity (CAT) were recorded by AC32, which were 1317.08 µg/g FLW, 857.84 µg/g FLM, 0.22 units/min/g FW and 104.52 units/min/g FW, while the lowest values of soluble sugar content (SSC) and total phenolic content (TPC) were indicted by AC56 with 496.56 µg/g FLW and 363.45 µg/g FLW, respectively.

**Table 4.27 Two-way analysis of variance (ANOVA) for chemical characteristics of four tomato accessions under the Cd stress.**

| | | PC | SSC | TPC | AC | GPA | CAT |
|---|---|---|---|---|---|---|---|
| Accessions | F | 1541.90** | 94132.54** | 7769.14** | 6909.2** | 124.26** | 144.77** |
| | Pr > F | < 0.0001 | < 0.0001 | < 0.0001 | < 0.0001 | < 0.0001 | < 0.0001 |
| Treatment | F | 2883.27** | 20349.79** | 11225.64** | 8786.97** | 1604.36** | 1013.97** |

| | Pr > F | < 0.0001 | < 0.0001 | < 0.0001 | < 0.0001 | < 0.0001 | < 0.0001 |
|---|---|---|---|---|---|---|---|
| Accessions*Treatment | F | 252.16** | 3162.41** | 4757.11** | 1759.82** | 54.54** | 127.41** |
| | Pr > F | < 0.0001 | < 0.0001 | < 0.0001 | < 0.0001 | < 0.0001 | < 0.0001 |

PC: proline content, SSC: soluble sugar content, TPC: total phenolic content, AC: antioxidant activity, LP: lipid peroxidation, GPA: guaiacol peroxidase activity, CAT: catalase activity.

**Table 4.28 Comparison for chemical characteristics means of treatments.**

| Treatments | PC (µg/g FLW) | SSC (µg/g FLW) | TPC (µg/g FLW) | AC (µg/g FLM) | GPA (units/min/g FW) | CAT (units/min/g FW) |
|---|---|---|---|---|---|---|
| Control | 1274.38 c | 542.58 c | 342.47 c | 894.29 c | 0.12 c | 79.65 c |
| Cd+Soil | 1481.31 b | 633.63 b | 400.43 a | 919.45 b | 0.29 b | 158.72 a |
| Cd+Soil+Oak | 1772.46 a | 687.18 a | 393.78 b | 1025.74 a | 0.43 a | 153.80 b |

PC: proline content, SSC: soluble sugar content, TPC: total phenolic content, AC: antioxidant activity, LP: lipid peroxidation, GPA: guaiacol peroxidase activity, CAT: catalase activity.

**Table 4.29 Comparison for chemical characteristics means of accessions.**

| Accessions | PC (µg/g FLW) | SSC (µg/g FLW) | TPC (µg/g FLW) | AC (µg/g FLM) | GPA (units/min/g FW) | CAT (units/min/g FW) |
|---|---|---|---|---|---|---|
| AC05 | 1595.28 b | 822.18 a | 420.37 a | 1016.52 a | 0.31 a | 137.64 b |
| AC07 | 1767.42 a | 724.34 b | 381.49 b | 990.27 b | 0.32 a | 150.37 a |
| AC32 | 1317.08 d | 496.56 c | 363.45 c | 857.84 d | 0.22 c | 104.52 d |
| AC56 | 1357.76 c | 441.42 d | 350.27 d | 921.35 c | 0.26 b | 130.36 c |

PC: proline content, SSC: soluble sugar content, TPC: total phenolic content, AC: antioxidant activity, LP: lipid peroxidation, GPA: guaiacol peroxidase activity, CAT: catalase activity.

*Relationship between the accessions with chemical characteristics under Cd treatments*

To determine the relationship between chemical characteristics with accessions, a multivariate analysis (PCA) was carried out (Figure 4.18). PCA of the first two dimensions (F1 and F2) accounted for 97.13% of total variation. The first component (F1) described 84.28% of the variation and it was jointed positively with the chemical characteristics of proline content (PC), soluble sugar content (SSC), total phenolic content (TPC), antioxidant capacity (AC), Guaiacol peroxidase activity (GPA), and Catalase activity (CAT) and also positively correlated with the AC05 and AC07, but negatively correlated with the AC32 and AC56. The second component (F2) clarified 12.85% of the variance, which was positively associated with AC32 and AC56.

*The effect of the interaction of treatments and accessions on phytochemical characteristics*

Table 4.30 shows the mean comparison of chemical characteristics from the interaction of treatments and accessions. The height value of PC was recorded for the interaction of AC07*Cd+Soil+Oak, which was 2038.62 µg/g FLW, and the lowest value was revealed at

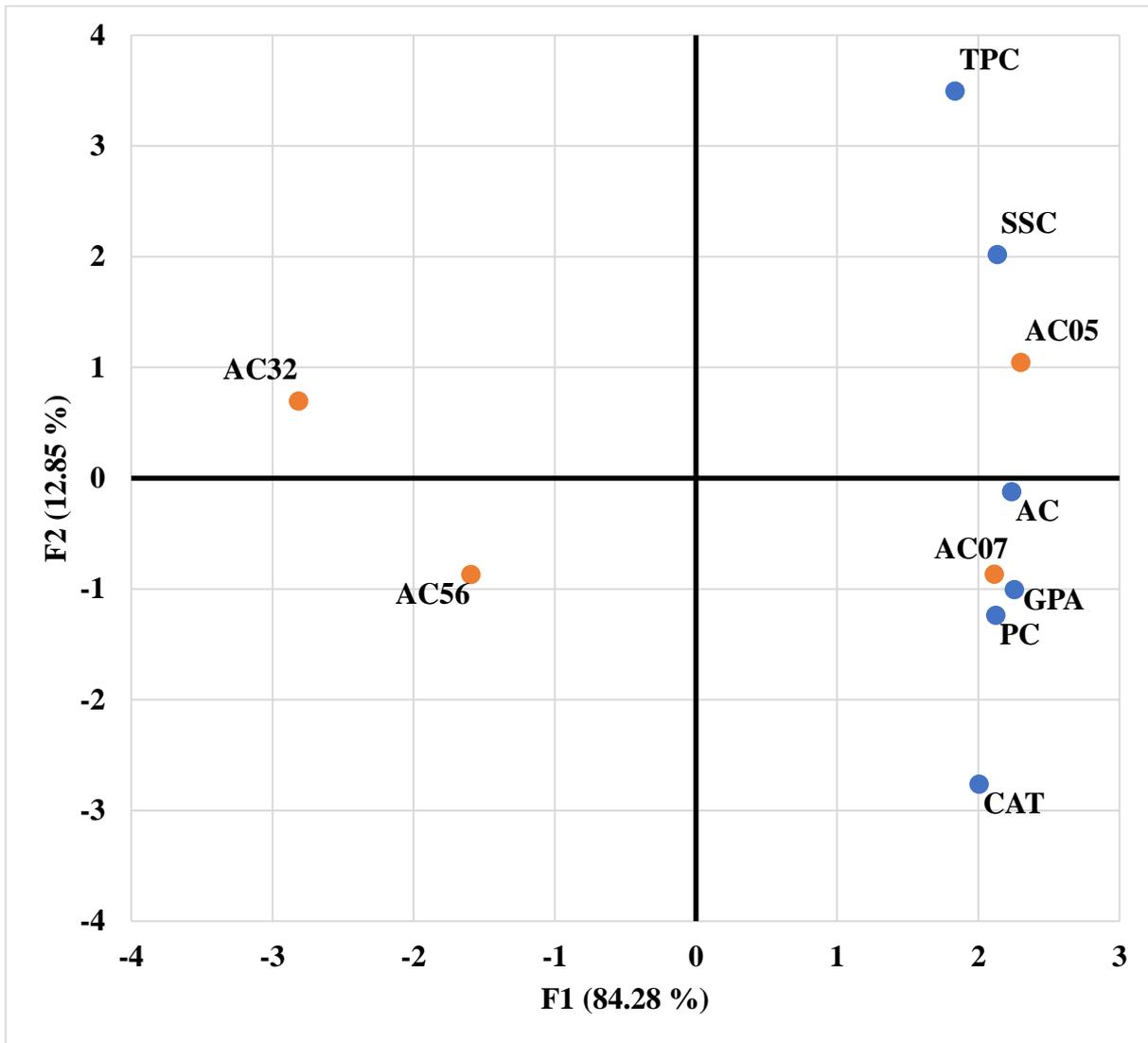

**Figure 4.18 Principal component analysis (PCA) plot revealing the distribution and relationship between tomato accessions (AC05, AC07, AC32, and AC56) with the chemical characteristics. (PC: proline content, SSC: soluble sugar content, TPC: total phenolic content, AC: antioxidant capacity, GPA: guaiacol peroxidase activity, CAT: Catalase activity).**

AC32*Cd+Soil, with 1207.59 µg/g FLW. AC05*Cd+Soil+Oak had the highest SSC value of 950.68 µg/g FLW, while AC56*Control had the lowest value of 390.19 µg/g FLW. The highest and the lowest values of TPC were revealed by AC05*Cd+Soil (495.47 µg/g FLW) and AC32*Control (320.00 µg/g FLW), respectively. The highest value of AC was recorded by AC05*Cd+Soil+Oak, which was 1168.65 µg/g FLM, and the lowest value was indicated by AC32*Cd+Soil, with 836.22 µg/g FLM. AC05*Cd+Soil+Oak (0.48 units/min/g FW) and AC07*Cd+Soil (203.78 units/min/g FW) had the highest GPA and CAT values, respectively, while AC07*Control and AC32*Control had the lowest GPA (0.10 units/min/g FW) and only AC07*Control had the lowest CAT (58.44 units/min/g FW).

**Table 4.30 The comparison mean of morphological traits of the interaction of treatments and accessions.**

| Treatment*Accession | PC (µg/g FLW) | SSC (µg/g FLW) | TPC (µg/g FLW) | AC (µg/g FLM) | GPA (units/min/g FW) | CAT (units/min/g FW) |
|---|---|---|---|---|---|---|
| AC05*Control | 1243.23 gh | 649.75 f | 370.19 e | 918.51 g | 0.15 f | 93.92 f |
| AC05*Cd+Soil | 1562.72 d | 866.11 b | 495.47 a | 962.39 c | 0.34 d | 187.62 b |
| AC05*Cd+Soil+Oak | 1979.90 b | 950.68 a | 395.47 d | 1168.65 a | 0.45 b | 131.38 e |
| AC07*Control | 1378.36 f | 656.23 e | 336.48 i | 904.46 h | 0.10 g | 58.44 h |
| AC07*Cd+Soil | 1885.28 c | 734.32 d | 410.82 c | 950.41 d | 0.39 c | 203.78 a |
| AC07*Cd+Soil+Oak | 2038.62 a | 782.47 c | 397.15 d | 1115.93 b | 0.48 a | 188.87 b |
| AC32*Control | 1247.08 gh | 474.14 i | 320.00 j | 857.84 k | 0.10 g | 71.99 g |
| AC32*Cd+Soil | 1207.59 i | 490.80 h | 343.60 h | 836.22 l | 0.16 f | 101.30 f |
| AC32*Cd+Soil+Oak | 1496.56 e | 524.75 g | 426.74 b | 879.46 j | 0.40 c | 140.26 d |
| AC56*Control | 1228.87 hi | 390.19 k | 343.22 h | 896.35 i | 0.15 f | 94.24 f |
| AC56*Cd+Soil | 1269.64 g | 443.27 j | 351.84 g | 928.78 f | 0.25 e | 142.16 d |
| AC56*Cd+Soil+Oak | 1574.77 d | 490.80 h | 355.77 f | 938.92 e | 0.38 c | 154.67 c |

**PC: proline content, SSC: soluble sugar content, TPC: total phenolic content, AC: antioxidant capacity, GPA: guaiacol peroxidase activity, CAT: Catalase activity.**

*Determination of Cd concentration in tomato roots, stems, leaves, and fruits*

The concentrations of Cd in tomato roots, stems, leaves, and fruits are shown in Table 4.31. In the case of the Cd concentrations in the tomato roots, the highest level of Cd was accumulated in the roots of the treatment combination AC07*Cd+Soil with 93.80 ppm, while using oak leaf powder in the treatment combination AC05*Cd+Soil+Oak recorded the highest reduction in Cd absorption in the roots with 13.58%. The maximum Cd accumulation in stem revealed in the treatment AC56*Cd+Soil by 6.50 ppm, the highest reduction of Cd absorption was indicated by the treatment AC32*Cd+Soil+Oak with 35%. AC32*Cd+Soil in the tomato leaf recorded the highest Cd accumulation which was 18.30 ppm, and the same accession treated with oak leaf powder revealed the maximum Cd absorption reduction which was 34.43%.

The concentration of Cd in tomato fruit which is more important to human health. The results indicated that in all treatments with AC05 and AC07, Cd did not detect in the tomato fruits, while the treatments Cd+Soil and Cd+Soil+Oak in the AC32 showed the highest concentration of Cd, which were 0.527 ppm and 0.436 ppm, respectively. The treatments of Cd+Soil and Cd+Soil+Oak of the AC56 also revealed the Cd concentration of 0.181 ppm and 0.134 ppm, respectively.

**4.1.4 Evaluation of the tomato accessions for nematode infection**

Four tomato accessions (AC14, AC43, AC53, and AC63) were tested for nematode resistance. These accessions were chosen based on morphological and molecular diversity. There were three treatments used: control, nematode, and nematode with oak leaf powder. Oak leaf powder was used to enhance the accessions to resist the nematode infection. The eggs of root-knot nematode (*Meloidogyne spp.*) were used and two species were confirmed

**Table 4.31 The concertation of Cd in different parts of tomato plant.**

| Accession*Treatment | Cd concentration (ppm) | Reduction in Cd absorption (%) | Accession*Treatment | Cd concentration (ppm) | Reduction in Cd absorption (%) |
|---|---|---|---|---|---|
| Tomato root | | | | | |
| AC05*Control | 4.30 | | AC32*Control | 5.20 | |
| AC05*Cd+Soil | 85.40 | | AC32*Cd+Soil | 76.50 | |
| AC05*Cd+Soil+Oak | 73.80 | 13.58 | AC32*Cd+Soil+Oak | 68.00 | 11.11 |
| AC07*Control | 3.70 | | AC56*Control | 4.50 | |
| AC07*Cd+Soil | 93.80 | | AC56*Cd+Soil | 74.40 | |
| AC07*Cd+Soil+Oak | 81.70 | 12.90 | AC56*Cd+Soil+Oak | 67.80 | 8.87 |
| Tomato stem | | | | | |
| AC05*Control | 0.82 | | AC32*Control | 0.75 | |
| AC05*Cd+Soil | 4.00 | | AC32*Cd+Soil | 6.00 | |
| AC05*Cd+Soil+Oak | 3.30 | 17.50 | AC32*Cd+Soil+Oak | 3.90 | 35.00 |
| AC07*Control | 0.78 | | AC56*Control | 0.63 | |
| AC07*Cd+Soil | 3.60 | | AC56*Cd+Soil | 6.50 | |
| AC07*Cd+Soil+Oak | 2.80 | 22.22 | AC56*Cd+Soil+Oak | 4.50 | 30.77 |
| Tomato leaf | | | | | |
| AC05*Control | 0.47 | | AC32*Control | 0.94 | |
| AC05*Cd+Soil | 5.60 | | AC32*Cd+Soil | 18.30 | |
| AC05*Cd+Soil+Oak | 5.20 | 7.14 | AC32*Cd+Soil+Oak | 12.00 | 34.43 |
| AC07*Control | 0.63 | | AC56*Control | 0.73 | |
| AC07*Cd+Soil | 6.90 | | AC56*Cd+Soil | 9.40 | |
| AC07*Cd+Soil+Oak | 6.60 | 4.35 | AC56*Cd+Soil+Oak | 9.60 | -2.13 |
| Tomato fruit | | | | | |
| AC05*Control | 0 | | AC32*Control | 0 | |
| AC05*Cd+Soil | 0 | | AC32*Cd+Soil | 0.527 | |
| AC05*Cd+Soil+Oak | 0 | 0.00 | AC32*Cd+Soil+Oak | 0.436 | 17.27 |
| AC07*Control | 0 | | AC56*Control | 0 | |
| AC07*Cd+Soil | 0 | | AC56*Cd+Soil | 0.181 | |
| AC07*Cd+Soil+Oak | 0 | 0.00 | AC56*Cd+Soil+Oak | 0.134 | 25.97 |

(*M. javanica and M. incognita*) in a previous study carried out in Sulaymaniyah governorate by (Mahmood, 2017). According to the ANOVA analysis, there was a significant difference between the tomato accessions under both treatment conditions for the traits shoot length (SL), shoot dry weight (SDW), root length (RL), root fresh weight (RFW), total fruit weight per plant (TFW), and disease severity index (DS) (Table 4.32). Table 4.33 shows that for all traits studied (SL, SDW, RL, RFW, TFW, and DSI), there were highly significant differences among accessions, treatments, and interactions between them (P ≤ 0.001). The maximum and minimum values of SL were 209.33 cm and 51.66 cm, respectively, with an average of 121.72 cm. The mean SDW was 108.09 g, with a range of 59.22 to 204.39 g. The lowest and highest values of RL and RFW varied from 12.00 to 39.00 cm and 10.78 to 55.66 g, with averages of 22.38 cm and 25.01 g, respectively. The highest value of TFW was 1118.74 g and the lowest value was 102.36 g, with an average of 504.17 g. DSI value ranged from 0.00 to 100.00% with a mean of 52.59% (Table 4.32). The effect of nematode and nematode+oak on the tomato root was shown in Figure 4.19.

The mean values of all traits for tomato accession under the three used treatments are shown in Table 4.34. Significant differences were recorded among all accessions for all traits. The highest values of SL and SDW were recorded by the AC43 and AC53 by 148.74 cm and 149.97 g, respectively, while the lowest values were revealed in the AC63 with 65.44 cm and

**Table 4.32 The morphological traits mean values and DS%.**

| Traits | Minimum | Maximum | Mean | F | Pr > F |
|---|---|---|---|---|---|
| SL | 51.66 | 209.33 | 121.72 | 151.16*** | < 0.0001 |
| SDW | 59.22 | 204.39 | 108.09 | 127.90*** | < 0.0001 |
| RL | 12.00 | 39.00 | 22.38 | 60.77*** | < 0.0001 |
| RFW | 10.78 | 55.66 | 25.01 | 61.08*** | < 0.0001 |
| TFW | 102.36 | 1118.74 | 504.17 | 472.00*** | < 0.0001 |
| DS | 0.00 | 100.00 | 52.59 | 532.61*** | < 0.0001 |

SL: shoot length (cm), SDW: shoot dry weight (g), RL: root length (cm), RFW: root fresh weight (g), TFW: total fruit weight per plant (g), and DS: disease severity (%).

**Table 4.33 Statistics that describe all morphological characterstics of accessions, treatments and thier combinations.**

| Traits | Accessions | | Treatment | | Accessions*Treatment | |
|---|---|---|---|---|---|---|
| | F | Pr > F | F | Pr > F | F | Pr > F |
| SL | 282.72*** | < 0.0001 | 348.86*** | < 0.0001 | 19.48*** | < 0.0001 |
| SDW | 296.61*** | < 0.0001 | 225.20*** | < 0.0001 | 11.11*** | < 0.0001 |
| RL | 32.40*** | < 0.0001 | 250.93*** | < 0.0001 | 11.57*** | < 0.0001 |
| RFW | 7.06*** | 0.00 | 285.81*** | < 0.0001 | 13.18*** | < 0.0001 |
| TFW | 51.34*** | < 0.0001 | 2205.16*** | < 0.0001 | 104.61*** | < 0.0001 |
| DS% | 106.58*** | < 0.0001 | 2688.93*** | < 0.0001 | 26.84*** | < 0.0001 |

SL: shoot length (cm), SDW: shoot dry weight (g), RL: root length (cm), RFW: root fresh weight (g), TFW: total fruit weight per plant (g), and DS: disease severity (%).

76.19 g, respectively. AC63 and AC43 recorded the maximum values of RL and RFW with 25.11 cm and 26.91 g, respectively, whereas AC14 recorded the lowest values of these traits by 18.07 cm and 21.36 g, respectively. The highest and lowest TFW values recorded by the AC53 and AC14 with 584.67 g, and 437.14 g respectively. AC14 indicated the highest value of DSI by 64.07%, while AC63 recorded the lowest value which was 38.89%.

**Table 4.34 The mean values of all traits for tomato accessions.**

| Accessions | SL | SDW | RL | RFW | TFW | DS% |
|---|---|---|---|---|---|---|
| AC14 | 148.33 a | 78.37 c | 18.07 c | 21.36 b | 437.14 d | 64.07 a |
| AC43 | 148.74 a | 127.85 b | 24.59 a | 26.91 a | 466.20 c | 51.85 c |
| AC53 | 124.40 b | 149.97 a | 21.74 b | 25.90 a | 584.67 a | 55.56 b |
| AC63 | 65.44 c | 76.19 c | 25.11 a | 25.84 a | 528.66 b | 38.89 d |
| Pr > F | < 0.0001 | < 0.0001 | < 0.0001 | 0.00 | < 0.0001 | < 0.0001 |
| Significant | Yes | Yes | Yes | Yes | Yes | Yes |

SL: shoot length (cm), SDW: shoot dry weight (g), RL: root length (cm), RFW: root fresh weight (g), TFW: total fruit weight per plant (g), and DS: disease severity (%).

Table 4.35 shows the treatment's mean values for all traits as well as significant differences between treatments. The control treatment recorded the highest values for the traits SL, SDW, RL, RFW, and TFW with 165.25 cm, 139.96 g, 31.36 cm, 40.75 g, and 935.55 g, respectively, while the nematode treatment indicated the lowest values for these traits by 98.39 cm, 89.71 g, 17.58 cm, 16.62 g, and 274.78 g, respectively. The highest and lowest DSI values were revealed by the nematode and control treatments, which were 79.44% and 0.00%, respectively.

**Table 4.35 The mean values of all traits for the treatments.**

| Treatments | SL | SDW | RL | RFW | TFW | DS% |
|---|---|---|---|---|---|---|
| Control | 165.25 a | 139.96 a | 31.36 a | 40.75 a | 935.55 a | 0.00 b |
| Nematode+Oak | 101.55 b | 94.61 b | 18.19 b | 17.64 b | 302.17 b | 78.33 a |
| Nematode | 98.39 b | 89.71 b | 17.58 b | 16.62 b | 274.78 c | 79.44 a |
| Pr > F | < 0.0001 | < 0.0001 | < 0.0001 | < 0.0001 | < 0.0001 | < 0.0001 |
| Significant | Yes | Yes | Yes | Yes | Yes | Yes |

**SL: shoot length (cm), SDW: shoot dry weight (g), RL: root length (cm), RFW: root fresh weight (g), TFW: total fruit weight per plant (g), and DS: disease severity (%).**

The interactions between the accessions and the treatments are shown in Table 4.36, and significant differences were revealed in these interactions. The highest and lowest values of SL were observed in the interactions AC43*Control and AC63*Nematode with 203.00 cm and 53.22 cm, respectively, whereas the maximum and minimum values of SDW were shown by the interactions AC53*Control and AC63*Nematode with 187.44 g and 66.77 g, respectively. The interaction of AC14*Nematode+Oak recorded the lowest RL with 12.55 cm, while the highest value of this trait was revealed by the interaction of AC43*Control. The highest and lowest values of RFW were indicated by the interactions of AC43*Control and AC14*Nematode which were 49.07 g and 11.78 g, respectively. The interaction of AC53*Control recorded the maximum value of TFW with 1077.85 g and the minimum value (128.84 g) of this trait was shown by the interaction of AC14*Nematode. The highest DSI ratio recorded by the interaction of AC14*Nematode with 96.67%, while the lowest ratio of DSI recorded by the interaction of all accessions with control condition, which was 0.00%.

**Table 4.36 The mean values of the interactions among the accessions and the treatments**

| Accessions*Treatment | SL | SDW | RL | RFW | TFW | DS% |
|---|---|---|---|---|---|---|
| AC14*Control | 178.22 b | 105.39 e | 28.44 c | 39.15 b | 1039.65 a | 0.00 e |
| AC14*Nematode+ Oak | 135.22 c | 65.02 g | 12.55 g | 13.17 fg | 142.92 g | 95.56 a |
| AC14*Nematode | 131.55 cd | 64.69 g | 13.22 g | 11.78 g | 128.84 g | 96.67 a |
| AC43*Control | 203.00 a | 175.19 b | 36.22 a | 49.07 a | 932.36 b | 0.00 e |
| AC43*Nematode+ Oak | 123.44 cd | 104.65 e | 19.22 e | 16.43 fg | 249.77 f | 77.78 c |
| AC43*Nematode | 119.77 d | 103.69 e | 18.33 ef | 15.24 fg | 216.46 f | 77.78 c |
| AC53*Control | 191.33 a | 187.44 a | 32.33 b | 42.72 b | 1077.85 a | 0.00 e |
| AC53*Nematode+ Oak | 92.89 e | 138.78 c | 17.44 ef | 17.59 ef | 357.01 e | 83.33 b |
| AC53*Nematode | 89.00 e | 123.67 d | 15.44 fg | 17.40 ef | 319.15 e | 83.33 b |
| AC63*Control | 88.44 e | 91.82 f | 28.44 c | 32.07 c | 692.36 c | 0.00 e |
| AC63*Nematode+ Oak | 54.66 f | 69.99 g | 23.55 d | 23.39 d | 458.98 d | 56.67 d |
| AC63*Nematode | 53.22 f | 66.77 g | 23.33 d | 22.08 de | 434.65 d | 60.00 d |
| Pr >F | < 0.0001 | < 0.0001 | < 0.0001 | < 0.0001 | < 0.0001 | < 0.0001 |
| Significant | Yes | Yes | Yes | Yes | Yes | Yes |

**SL: shoot length (cm), SDW: shoot dry weight (g), RL: root length (cm), RFW: root fresh weight (g), TFW: total fruit weight per plant (g), and DS: disease severity (%).**

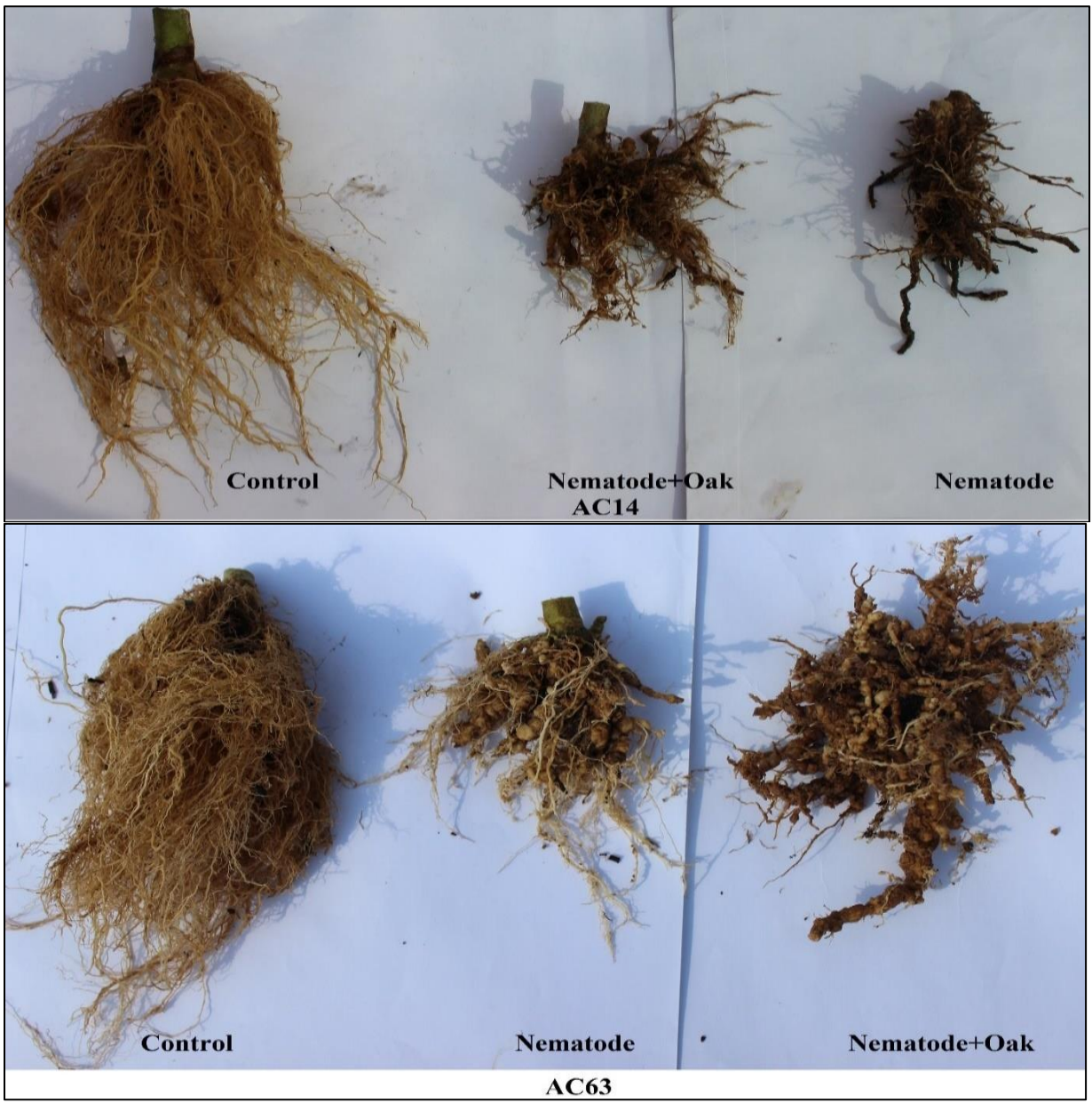

**Figure 4.19 The tomato root, control, using nematode and nematode+oak, for the AC14 and AC63.**

## 4.2 Discussion

Genetic variability has been analyzed for two reasons: first, it is necessary for long-term sustainability, and second, it is a crucial source of plant conservation and breeding programs. As a result, gathering sufficient knowledge on the levels of genetic variation patterns is essential in order to develop methods for effective conservation and sustainable use of genetic resources. Genetic drift is not a significant contributor to genetic variation in managed populations. Maintaining genetic diversity is one of the primary goals of plant accession conservation and breeding. Understanding genetic variation between and within populations is crucial for designing appropriate management strategies aimed at maintaining them. Thus, in the current work, morphological and molecular descriptions were used to emphasize tomato germplasm diversification and to reveal the genetic similarities between 64 tomato accessions taken from various locations of Kurdistan region-Iraq.

## 4.2.1 Fruit traits evaluation

Tomato productivity has declined significantly in comparison to other plants due to various biotic and abiotic factors, and the establishment of adequate varieties based on genetic diversity research to conquer these stresses defines the major aim, among several approaches. Awareness about genetic variability can be assessed using many approaches that use genetic markers to offer relevant data for the development of breeding programs, including marker-assisted selection and genetic management systems. It is worth noting, as, with all other plants, the tomato breeding program is primarily dependent on local populations that contain the crucial genetic makeup for unique attributes. In Iraq, agro-ecosystem diversifying has resulted in the creation of specialized accessions tailored to local climatic and edaphic conditions, while retaining high genetic diversity. The goal of this part of the research was to determine the fruit morphological potential of Iraqi tomato accessions using quantitative parameters.

There were significant differences among the morphological characteristics. According to the results of morphological traits, it can be seen that the maximum and the minimum values of plant height were revealed in AC39 (334.33 cm) and AC8 (84.33 cm), respectively, while in a study, Singh *et al.* (2015) showed that the highest tomato plant was 197 cm. The longest root recorded in AC60 which was 51 cm, and the shortest length of root with 23.33 cm was recorded in AC21. The maximum and minimum plant dry weights were recorded in AC51 (454.80 g) and AC8 (44.29 g), respectively. The highest and the lowest total chlorophyll contents were recorded in AC13 (93.4 SPAD) and AC6 (11.03 SPAD), respectively. Similar findings previously reported by other researchers (Heuvelink *et al.*, 2005; Jiang *et al.*, 2017).

The analysis of variance found significant differences between accessions for all fruit characteristics, demonstrating a high level of fruit phenotypic variability among the Iraqi Kurdistan region tomato accessions investigated in this study. This conclusion is in line with the past studies showing morphological attributes in tomato germplasm from Pakistan (Hussain *et al.*, 2018), El Salvador (Chávez-Servia *et al.*, 2018), Bangladesh (Salim *et al.*, 2020), Mexican Republic (Marín-Montes *et al.*, 2016), and Colombia (Vargas *et al.*, 2020). In our study the average fruit weight (106.05 g) was lower than the fruit weight described by Rodríguez and Servia (2010) (120.77 g), but higher than those of Hernández-Bautista *et al.* (2015) (8.30 g) and Chávez-Servia *et al*. (2018) (35.78 g). The mean diameter of the fruit in this study (57.38 mm) was slightly lower than that found by Hernández-Bautista *et al.* ((2015) (60.90 mm), but larger than that discovered by Rodríguez and Servia (2010) (20.30 mm). In this study, the average tomato production per plant (1718.62 g) was larger than that obtained by Hernández-Bautista *et al.* (2015) (1252.40 g). As witnessed in statistical phenotypic analysis, our collection had a wide variation in FS, FW, NSS, and FWP traits. These morphological characteristics are critical markers for developing tomato breeding strategies. According to the PCA results, FS, FW, FT, FD, and FF were the most distinguishing attributes, followed by NSS. Furthermore, the PCA results could aid parents to develop an efficient segregating population for identifying specific QTLs. Cluster analysis and PCA were unable to differentiate tomato accessions depending on geographic sources. The number of clusters in our findings (7 clusters) were more than that reported (5 clusters) by Ayenan *et al.* (2019). The genetic makeup and number of accessions used were accounted for the variations between our findings and those of other studies.

### 4.2.2 Molecular characterization

As anticipated, adopting morphological and agronomic characteristics for the generation of gene pools is insufficient due to environmental influence, plant growth stage, or limited variability. To overcome the aforementioned disadvantages, molecular markers are being employed as novel methods for detecting genetic diversity and evaluating germplasm in tomato populations. To the best of our knowledge, no studies on the use of molecular markers for the evaluation of genetic variability in tomato have been published in Iraq, and this work could serve as a starting point for future research on genetic diversity and breeding in the Iraqi tomato germplasm. The allelic richness of plant accessions is a parameter of genetic diversity enrichment, which is frequently exploited by informative molecular markers that designate populations for selection, breeding, and conservation. Diverse molecular markers and primers resulted in different amplification products, representing the polymorphism of the genomic areas; consequently, diverse marker design approaches will yield different results. The data reliability should potentially improve as the number of markers and genome coverage increases. When two or more categories of molecular marker systems are being used to assess genetic diversity in populations with the same accessions, the relationship between marker systems may be of relevant for research. Different authors noticed varying responses, extending from consonance to no correlation in the relationships (Archak *et al.*, 2003; Maccaferri *et al.*, 2007).

As a consequence, 585 bands were formed by three different types of molecular markers. The high rates of polymorphism exhibited by various markers proved their effectiveness in genetic variability investigations in Iraqi tomato accessions. SCoTs are gene-targeted markers that target the small conserved region enclosing a gene's start codon, whereas CDDP markers target conserved sections of functioning genes. Unlike arbitrary ISSR markers, which mainly target non-coding areas of the genome, SCoT and CDDP can deal with specific genes and attributes (Amom *et al.*, 2020). A large number of polymorphisms, coupled with the high number of polymorphic alleles derived per primer in our study, could be clarified by both the broad range of genetic diversity and geographical collection area, as well as the performance of ISSR, SCoT, and CDDP markers in reaching adequate polymorphism in targeted regions of the tomato genome accessions. The average number of ISSR polymorphic alleles per locus (9.31) was higher than the mean number of bands (2.57) for six tomato lines by El-Mansy *et al.* (2021), the average number of alleles (4.33) for eight tomato genotypes by Abdein *et al.* (2018), and the average number of alleles (7.66) for eleven tomato genotypes by Kiani and Siahchehreh (2018), but lower than the number obtained (14.29) by Vargas *et al.* (2020). The disparity in the number of alleles per locus between our results and previous researches can be related to differences in the diversity of the population analyzed, the number of accessions tested, and the primers used. Compared with our results, fewer SCoT polymorphic products were detected by other authors Shahlaei *et al.* (2014), Abdein *et al.* (2018), and Abdeldym *et al.* (2020). However, Abdeldym *et al.* (2020), found fewer CDDP polymorphism products per primer (5.6 bands). The findings demonstrated that both the SCoT and CDDP markers are comparably effective at detecting polymorphism and also have a higher detection capacity than ISSR.

GD and PIC values are both indicators of genetic variety among accessions in breeding populations, which give light on the evolutionary impact on alleles and the mutation rate of a locus may have encountered over time. The increased levels of GD and PIC observed after ISSR, SCoT and CDDP markers analysis suggest that the primers examined were highly informative. According to Botstein *et al.* (1980), a PIC value greater than 0.5 represents a highly informative marker. The PIC averages of ISSR obtained by Abdein *et al.* (2018) (0.687) and El-Mansy *et al.* (2021) (0.36) were lower than that found in this investigation. Furthermore, the PIC of the investigated SCoT marker in this work was higher than that found by El-Mansy *et al.* (2021), who stated that the maximum (0.31) and minimum (0.18) PIC values were found for primers SCoT2 and SCoT12, respectively.

ISSR, SCoT, and CDDP markers did not distinguish tomato accessions based on provenance location. For all 3 methods, UPGMA clustering provided a comparable grouping distribution. Some discordance between dendrograms acquired by different marker types could be explained by the distinct nature of each marker, the different sections of the genome covered by different marker technologies, the level of polymorphism found, and the number of loci studied. Several researchers have explored the grouping of diverse tomato accessions and reported that accessions collected from distinct clusters due to their great genetic diversity (Marín-Montes *et al.*, 2016; Kiani and Siahchehreh 2018; Abdeldym *et al.*, 2020; Vargas *et al.*, 2020). The use of UPGMA dissimilarity dendrograms based on ISSR, SCoT, and CDDP facilitated the grouping of 64 tomato accessions into two main clusters with several sub-clusters. El-Mansy *et al.* (2021) observed similar results, who studied the genetic grouping of six tomato genotypes using ISSR and SCoT markers. Vargas *et al.* (2020) stated four clusters in their analysis of various *Solanum* species. The difference in the number of clusters between our results and Vargas *et al*. (2020) research can be attributed to variances in the number of species and accessions assessed, and the primers used.

The cophenetic correlation coefficient between the dendrogram and the original distance matrix obtained from the ISSR, SCoT, CDDP and their combination markers in this study were 0.93, 0.92, 0.87, and 0.93, indicating a good fit of molecular data for genetic diversity. The strong agreement between the genetic matrices of ISSR, SCoT, and CDDP revealed that each marker technique was successful in determining the genetic relationship between the diverse tomato accessions.

Population differentiation analysis can help to understand genetic diversity and makes genome-wide association studies (GWAS) results more accurate. The presence of genetic structure within a population might lead to false positives in mapping studies. Therefore, considerable focus is placed on thoroughly analyzing the underlying population structure of any population to be used for marker-trait correlations. As a result, analyzing population structure is regarded as the initial step in conducting GWAS for real marker-trait relationships. The structure analysis results of three markers type mostly conformed to the groups found in the cluster analysis, in which the 64 tomato accessions were divided into two genetic groups based on the delta K value. Similarly, there were no grouping trends among the accessions based on their geographical origin. This could be attributed to some degree of gene flow across the tomato accessions studied, as a result of interspecific crosses, processes of allele introgression from wild species to cultivated species, the type of reproduction distinctive of each species, anthropogenic events, and other factors. The two clusters were substantially admixed, implying that the majority of variability exists among the accessions within the groups. Based on the population size and the difference in the number of accessions representing the six provinces from which they were gathered, two populations may be appropriate for our panel.

The genetic differentiation of a population represents the interactions of various evolutionary processes, including dispersal shifts, habitat change, population separation, mutation, genetic drift, mating system, gene flow, and natural selection. Geographic isolation, community fragmentation, breeding systems, and genetic drifts are all potential sources of significant population variation. The fixation index of population differentiation (Fst) assesses the degree of diversification on a scale of 0 to 1, with 0 representing complete genetic material sharing and 1 suggesting no sharing. In discriminating populations, a Fst value greater than 0.15 is considered significant (Frankham, 1995). Fst measures genetic divergence caused by population structure. When the three types of molecular markers based on Fst values were compared, the results revealed the highest value of Fst (0.55) was found in the second population, confirming the existence of significant genetic variation within the individuals of the second population, whereas the CDDP-Fst exhibited the lowest value (0.30) in the second population, confirming the existence of low genetic variation within the individuals of the second population. As shown in our findings of three markers, the expected heterozygosity values of tomato within-population were high, indicating that they contain relatively high levels of genetic variation. Hence, STRUCTURE analysis found considerable differentiation in both populations. This is consistent with the AMOVA results of three markers, in which the majority of the variation was explained for within populations.

Another approach for evaluating evolutionary divergence in a population is gene flow, which can similarly change the genetic structure. It allows individuals from one gene pool to mate with individuals from another, allowing allele frequencies to shift and the degree of population divergence to diminish. As a result, the greater the degree of genetic differentiation, the more evident the gene flow. In population genetics, gene flow estimates are classified as minimum (gene flow 1), moderate (gene flow > 1), and substantial (gene flow > 4) degree (Slatkin, 1987; Kumar *et al.*, 2014). The high observed gene flow value among populations in our study, together with considerable levels of genetic variation within geographical areas, suggest a reasonable migrant-pool migratory model (Wade and McCauley, 1988). The competency displayed by each marker system revealed that they can be used in combination with one another to successfully investigate the genetic diversity of various tomato accessions. Consequently, our work provides preliminary data on intraspecific and interspecific diversity in tomato, as well as a baseline database to aid biologists and breeders in accession delimitation.

### 4.2.3 Drought stress

Plant growth results from cell division, cell enlargement, and differentiation and is regulated by a wide range of genetic, physiological, ecological, and morphological processes, as well as the interaction between these factors (Ullah *et al.*, 2016). Damage to physiological and biochemical processes, such as a delay in stomatal conductance, a decrease in nutrient uptake, a breakdown of leaf pigments, a decrease in photosynthesis, a stop in the rate of net assimilation and photosystem photochemical efficiency parameters, an increase in ROS, and oxidative damage caused by water stress, reduced the morphological features (Ibrahim *et al.*, 2020b). Our results, under *in vitro* conditions by using different PEG concentration, revealed that all tomato accessions have different response to all PEG concentrations, for all morphological and biochemical traits. The *in vitro* experiment showed natural variations among the studied accessions, which revealed different growth characteristics in the control and stress conditions of the sixty-four tomato accessions, and significant decreases were noted as the concentration of PEG increased. These results are similar with previous studies and provide further evidence of the suitability of PEG as a molecule for simulating droughts under *in vitro* conditions. These findings are corresponding with previous research and show the response to droughts depends on both accessions and levels of stress (George *et al.,* 2013). The morphological parameters GP, RL, SL and FW were significantly decreased as the PEG concentration increased, GP value was 90.1% under T0, while decreased to 86.34% and 80.04% under T1 and T2, respectively. RL values were 8.21, 6.14 and 4.65 cm under T0, T1, and T2, respectively. These results are similar with the finding by (Zhou *et al.,* 2017), while the DW was increased, this result may relate to the accumulation of sugar and PEG molecule inside the plant cell, which exposed to the drought via PEG (Brdar-Jokanović *et al.,* 2014). The biochemical traits PC, SSC, TPC, AC, GPA, CAT, and LP were gradually increased, while the PEG concentration increased, particularly in the tolerant accessions. For example, PC values were 854.45, 1447.35 and 2295.87 µg/g FW, and SSC values were 129.97, 189.86 and 220.88 µg/g FW, under T0, T1 and T2 respectively. SSC has several roles, including osmotic modification, carbon preservation, detoxification of ROS, defense of membrane integrity, safety of DNA structures, and protein stabilization. In serious dehydrated conditions, sugars become an important water substitute, perhaps more than proline, for the hydration of proteins (Bowne *et al.,* 2012). Proline is osmotic, plays a significant role in the stabilization of the membrane. It also works by scavenging free radicals and syncing the redox ability of the cells, which allows the plants to fight abiotic stress (Khan *et al.,* 2015). Prolonged stress, however, may induce ROS aggregation at the plasma membrane, and resultant damage to cells. Therefore, to minimize ROS generation, the plant needs the up-regulation of antioxidant/detoxifying systems including APX, SOD, CAT, and POD (Barna *et al.,* 2003; Zhang *et al.,* 2020).

Under the greenhouse conditions, The fresh weight, plant height, and productivity of the stressed tomato plants were all lower than those of the control plant (watered plants), as was found by previous studies (Eziz *et al.*, 2017; Wang and Xing, 2017; Zhou *et al.*, 2017; Janni *et al.*, 2019). Relative water content and total chlorophyll content also decreased under SS conditions. The same results were also found in tomato plants studied by Khan *et al.* (2015), Ullah *et al.* (2016), and Ibrahim *et al.* (2020b).

Root fresh weight (RFW) and root dry weight (RDW) under situations of water stress have shown significant increased percentages for all degrees of treatment under all stress stages. The plant treated with SOBS and SOES had significantly higher RFW and RDW trait values than those of the control group (SW) during all stress stages. As a comparison among the three stress stages, the plants treated with SOBS showed the greatest increases in RWF (107%) and RDW (127.80%) in the second stress stage. The increased root surface area and root volume in plants during the search for water in the soil is mostly responsible for the higher RFW and RDW observed across all stress stages in comparison to untreated and unstressed plants. Additionally, a large number of prominent compounds found in leaf extract, including silane, heptasiloxane, and octasiloxane, are thought to be silicon (Si) sources and are responsible for the increasing RL, RFW, and RDW in plants exposed to SOES at all stress levels. The leaf extract also had the compounds 2,4-Di-tert-butylphenol and 1-octadecene, which have antioxidant properties that reduce the synthesis of ROS products and membrane lipid peroxidation (Varsha *et al.*, 2015; Kolyada *et al.*, 2018; Tonisi *et al.*, 2020; Zhao *et al.*, 2020). In addition, the leaf extract contained a 1-hexadecanol compound, which is used to reduce water evaporation in reservoirs. Si-enhanced cell-wall extensibility in the root's growth zone likely contributes to root elongation. Root density and length were both increased by Si in  purslane (Kafi *et al.*, 2011). Sorghum's root length was found to be increased by Si, according to the research of Sonobe *et al.* (2010). It's also likely that the higher RFW and RDW in SOES-treated plants are due to the ability of Si and 2,4-Di-tert-butylphenols to minimize ROS overproduction, which reduces membrane lipid peroxidation. On the other hand, our research showed that both SFW and SDW were lower in the SOES-treated plants. This may be due to the fact that Si controls the levels of polyamine and 1-aminocyclopropane-1-carboxylic acid in response to drought stress, which improves root growth, the ratio of roots to shoots, water uptake at the roots, and hydraulic conductance. Root endodermal silicification and suberization are also boosted by Si-mediated alterations in root growth, which help plants better retain water and tolerate the negative effects of drought (Wang *et al.*, 2021). In comparison to plants treated with SS and SOS, SFW and SDW in plants treated with SOES and SOBS may have increased due to a decrease in ROS products and membrane lipid peroxidation. Furthermore, RFW and RDW increased in plants treated with SOBS throughout all stress stages, and these increases were induced by the presence of cytokinins, enzymes (lipase, amylase, protease, and chitinase), *Bacillus subtilis*, and *Pseudomonas putida*. These components of the SOBS treatment improve root area and volume by degrading organic matter and boosting phosphorus availability in the soil (Kim *et al.*, 2016). *Bacillus subtilis* and *Pseudomonas putida* invade plant rhizospheres and produce volatile organic chemicals that can affect plant development and root architecture in a variety of plants (Bavaresco *et al.*, 2020; Ortiz-Castro *et al.*, 2020).

Drought stress, on the other hand, can alter the chemical composition of fruits. Organic acids (malic and citric acids) and soluble sugars are among the primary osmotic components found in ripe fruits (Medyouni *et al.*, 2021). Organic acids are stored by plants in order to reduce their osmotic potential and prevent cell turgor pressure from decreasing (Menezes-Silva *et al.*, 2017; Ma *et al.*, 2022). Vitamin C, which also known as ascorbic acid, is found in all parts of plants, it plays a pivotal role in the development and growth of plants. Ascorbic acid is the plant's primary antioxidant, which neutralizes the active forms of oxygen. Our results showed that the ascorbic acid content of the red fruit of the stressed plant increased due to the water shortage. This increase in ASC may be vital for detoxifying ROS. Antioxidant capability is determined by the phenolic contents of tomato fruits (TPC), and an increase in TPC amount results in a decrease in oxidative alterations in cells due to a lower concentration of free radicals (Zhu, 2001; Wang *et al.*, 2014).

Fructose and glucose levels both increase sharply when tomatoes ripen. The total soluble solids (TSS) concentration is influenced by the carbohydrates, organic acids, proteins, fats, and mineral components. Our results suggest that shifts in the glucose/fructose ratio and organic acid levels may be responsible for the observed reduction in TSS in our investigation (Medyouni *et al.*, 2021). Compared to SS and SOS circumstances, the availability of silane, heptasiloxane, octasiloxane, and 2,4-Di-tert-butylphenol increase SSC during SOES application, which decreases ROS production by triggering antioxidant systems.

The results of the accession effects revealed that tomato accessions responded differentially to SS, SOS, SOES, and SOBS applications under water stress. According to ASC, CAC, SSC, and TPC data, drought stress reduced the quality of tomato tolerant accessions treated with SOS, SOES, and SOBS.

Different reactions were seen in terms of the leaf biochemical responses in plants treated with SS, SOS, SOES, and SOBS under stressful conditions. The highest levels of lipid peroxidation (LP), a metabolic process that results in the oxidative degradation of lipids by ROS, were observed in the untreated and stressed accessions conditions (SS). As a result of this process, the lipids in the cell membrane may break down, which can damage the cell and lead to its death. Low accumulations of biochemical compounds such as TPC, PC, SSC, AC, GPA, and CAT are responsible for this increase in LP. The accessions treated under SOES and SOBS conditions, on the other hand, showed the highest levels of TPC, PC, SSC, AC, GPA, and CAT, which led to the reduction of LP. Furthermore, the SOES application may have induced the antioxidant systems, which may have contributed to the availability of silane, heptasiloxane, octasiloxane, and 2,4-Di-tert-butylphenol in the leaf extract.

Different responses were observed for the tolerant and sensitive accessions during stress stages. Due to the low accumulation of SSC, PC, TPC, AC, GPA, and CAT in sensitive geometries, the findings of leaf biochemical parameters showed the maximum LP. Different response profiles between the tolerant accessions were found. Under the first stages of stress, Raza Pashayi (AC61) demonstrated the highest levels of TPC, GPA, and CAT, whereas Sandra (AC63) accessions had the highest levels of SSC, PC, and AC. Raza Pashyi (AC61) recorded the highest values for GPA, AC, and SSC traits during the second stress stage, while Sandra had the maximum values for TPC, CAT, and AC.

### 4.2.4 Heavy metal (Cd) stress

Heavy metals pollution is currently one of the most important environmental issues. Because of their tenacity, removing heavy metals from the environment is a major challenge (Karnib *et al.*, 2014). Plants respond to heavy metal stressors in a variety of ways, ranging from gene expression to cellular metabolism to growth and production (Farid *et al.*, 2013). Heavy metals including cadmium, copper, lead, chromium, manganese, iron, and mercury are substantial environmental contaminants (Al Khateeb and Al-Qwasemeh, 2014). The deleterious effects of heavy metal accumulation in soils on food safety, marketability, and crop growth due to phytotoxicity are causes for concern in agricultural production (Asati *et al.*, 2016). Cadmium (Cd) is a non-essential element that is considered the most toxic among the heavy metals. It has a negative impact on plant metabolism by causing oxidative stress (Gratão *et al.,* 2015). Cd is widespread in the environment in trace amounts or can be introduced through anthropogenic activities, such as the use of pesticides, fertilizers, and domestic and industrial effluent, where it is absorbed by plants from contaminated soil or water. Secondary metabolites and phytohormones play a precise role in reducing the adverse effects of heavy metals by chelating metal ions of cadmium and other heavy metal forms, reducing the level of ROS, limiting the synthesis of free radicals, and providing an osmotic homeostasis balance of nutrients (Ashfaque *et al.,* 2020).

Under *in vitro* conditions by using different Cd concentrations (0, 150, 300, and 450 µM Cd), indicated that all tomato accessions have different response to all Cd concentrations, according to all morphological and biochemical traits. The *in vitro* test for Cd concentrations revealed a huge variation among the studied accessions, which indicated different growth characteristics in the control and stress conditions of the sixty-four tomato accessions, and significant decreases were obtained as the concentration of Cd increased. These findings are corresponding with previous research (Rehman *et al.,* 2011). The morphological parameters GP, RL, SL, FW and DW were significantly reduced as the Cd concentration increased, GP value was 92.38%, 89.52%, 86.54% and 84.48%, under T0, T1, T2 and T3 respectively. SL values were 7.25, 6.08, 4.71 and 2.57 cm under T0, T1, T2 and T3, respectively. These results are similar with the finding by (Hediji *et al.,* 2010). The biochemical traits PC, SSC, TPC, AC, GPA, CAT, and LP were gradually increased, in both T1 an T2 treatment, while decreased in the highest Cd concentration used 450 µM. PC values were 888.81, 1631.77, 2242.51 and 1750.51 µg/g FW, and SSC values were 124.53, 189.8204.56, 262.27 and 243.73 µg/g FW, under T0, T1, T2 and T3 respectively. Under cadmium stress, the decrease in growth may be due to stronger bonds between pectin molecules in the cell wall, which is associated with a reduction in the size of intercellular space. On the other hand, Rahmatizadeh *et al.,* (2019) stated that the injection of lignin into the cell wall under cadmium stress leads to hardening and decreasing expansion of the wall. This stress also leads to increasing the production of ROS, which is followed by damage to the cell membrane and macromolecules (Rahmatizadeh *et al.,* 2019). It is well documented that heavy metals stress in plants increases the production of ROS, causing oxidative stress and damaging macromolecules such as lipids, proteins, and nucleic acids (Gratão *et al.* 2015).

Under greenhouse conditions, plant morphological characteristics, physiological traits, and chemical characteristics are affected by heavy metals. In this study, four tomato accessions, (two sensitive (AC32, AC56) and two tolerant (AC05, AC07)) to Cd stress were studied with two different treatments compared with control. For the first treatment, Cd only used which mixed with the soil, and oak leaf residue as a biosorbent in another treatment was used to reduce the effect of Cd in the soil. In the treatment of Cd+Soil, morphological traits such as root length, shoot length, shoot fresh weight, shoot dry weight and total fruit weight per plant were significantly reduced compared with control, while these traits under Cd+Soil+Oak treatment decreased slightly. Similar results was found in tomato by (Djebali *et al.,* 2005; Dong *et al.,* 2005; Cherian *et al.,* 2007; Gratão *et al.,* 2008; Rehman *et al.,* 2011; Al Khateeb and Al-Qwasemeh, 2014; Piotto *et al.,* 2018) and in musk melon (Zhang *et al.,* 2020). Root fresh weight, root dry weight and total chlorophyll content were increased in both Cd treatments, which was similar to results from a previous study (Rehman *et al.,* 2011). In the case of accessions, AC05 and AC07 are resistant to cadmium absorption compared to AC32 and AC56. The resistant accessions were slightly affected in biomass traits, leaf biochemical traits, and Cd concentration in fruit by Cd stress in both treatments, especially under the Cd+Soil+Oak treatment, compared to their control. In contrast, the two sensitive accessions (AC32 and AC56) were negatively affected by both Cd stress treatments, particularly in the Cd+Soil treatment, morphological traits and leaf phytochemical contents were reduced. Similar results were found by (Hasan *et al.,* 2009; Lima *et al.,* 2017; Borges *et al.,* 2019) and Cd concentration in the fruits recorded the higher levels, which are more than the threshold value, which is 0.100 ppm (Ishaq *et al.,* 2020; Romero-Estévez *et al.,* 2020). Arduini *et al.,* (2004) indicated that the root surface plasma membranes become hyperpolarized when cadmium levels are low, and increasing the trans-membrane potential, which is an energy source for cation uptake. Otherwise, low levels of Cd induce genes which increase plant growth (Arduini *et al.,* 2004). In the case of high levels of Cd, it causes cell damage through the loss of cellular turgor brought on by the physiological drought caused by Cd. Since the ability of the cell wall to grow affects how much the cell grows, as the cell wall grows, the cell will grow less, and the size of the cell will shrink (Hasan *et al.*, 2009). The effect of Cd toxicity in plants is to reduce growth processes and hence a decrease in photosynthetic apparatus activity. In order to understand the responses of tolerant and sensitive accessions to Cd stress, the biochemical tests of the leaves were evaluated under Cd+Soil and Cd+Soil+Oak treatments. Comparing both treatments to control, a high concentration of biochemical components was observed. When exposed to Cd stress in the presence of oak leaf residue pretreated with NaOH, tomato plants tended to accumulate high levels of PC, SSC, AC, and GPA, indicating that they used these osmoprotectant and antioxidant molecules to tolerate the Cd toxicity. On the other hand, when exposed to Cd stress in the absence of oak leaf residue pretreated with NaOH, tomato plants tended to accumulate high levels of TPC and CAT, indicating that they used these molecules to reduce the negative effects of Cd. Cd-induced oxidative damage and antioxidant defenses in plants. It also has the role of reactive oxygen and nitrogen species in Cd toxicity (Gallego *et al.*, 2012). There are some researches revealed the role of plant growth regulators (plant hormones) such as gibberellins (GAs), cytokinins (CKs), auxins, abscisic acid (ABA), and ethylene, jasmonate (JA), brassinosteroids (BR), and salicylic acid (SA) on plant resistance to Cd absorption, and are well-known for controlling a variety of physiological processes and enhancing resistance to heavy metal stress (Singh *et al.*, 2016). Furthermore, abscisic acid (ABA) enhances plant resistance to Cd stress, so the plant can decrease Cd absorption (Wang *et al.*, 2016) and increased susceptibility to Cd in tomato seedlings due to endogenous jasmonic acid insufficiency and it controls tomato plant response to Cd stress (Zhao *et al.*, 2016). ROS function as signaling molecules involved in the regulation of several important physiological processes, including root hair development, stomatal movement, cell proliferation, and cell differentiation, when precisely regulated and controlled by an antioxidative defense system (Tsukagoshi *et al.*, 2010). The importance of antioxidants is based on the fact that their increased and/or decreased levels are generally related to stress tolerance of stressed plants. An antioxidant system comprises two types of components: enzymatic and non-enzymatic. Water-soluble antioxidants include ascorbate, glutathione, proline, and α-tocopherol. Both can directly quench ROS and regulate the gene expression associated with biotic and abiotic stress responses (Singh *et al.*, 2016).

### 4.2.5 Nematode resistance

Plant Parasitic Nematodes (PPNs) attack the majority of economically important crops (Holbein *et al.*, 2016). The most well-known plant parasitic nematodes are cyst (*Globodera and Heterodera* spp.) and root-knot nematodes (RKN) (*Meloidogyne* spp.), which cause significant damage to crops such as soybean, potato, tomato, and sugar beet (Holterman *et al.*, 2006). *Meloidogyne incognita*, a tropical, is a polyphagous species that has been dubbed the world's most damaging crop pathogen (Coyne *et al.*, 2018). In our study, four tomato accessions (AC14, AC43, AC53, and AC63) were evaluated for root-knot nematode resistance. The eggs of *Meloidogyne incognita* and *Meloidogyne javanica* were used. The results showed that the best resistant accession was AC63, which can survive to the end of growing season. According to the previous studies, *Mi* gene may be present in this accession, each tomato genotype has this gene, can resist the nematode infection at all (Bozbuga *et al.*, 2020). The using oak leaf powder in our study, showed that there is no effect on the lowering the effect of nematode infection, this result is completely different from the study which used plant extracts and the results showed a reduction in the nematode infection while plant extracts were used (Abo-Elyousr *et al.*, 2010).

# Conclusions

The major conclusions from this study include:

- High and significant genetic variation was found among the 64 accessions tested at both the fruit quality traits and molecular levels. This wide range of variance will help germplasm management, classification, and preservation.

- The results confirmed the efficacy of ISSR, SCoT, and CDDP markers as useful methods for assessing tomato diversity. The best two types of markers for the study of genetic diversity in tomato, according to the Mantel test, were ISSR and SCoT, followed by the combination of all three types of markers.

- To the best of our knowledge, this is the first study to use morphological characteristics and molecular markers to assess the diversity of Iraqi Kurdistan region tomato accessions. Our findings pave the way for the selection of parental lines for tomato improvement programs aimed at producing novel abiotic- and biotic-tolerant tomato varieties, which will lead to farmer preferred cultivars with desirable characteristics.

- According to the results from *in vitro* drought tolerance, the accessions of Sandra and Raza Pashayi were highly tolerant to drought stress.

- According to our findings under greenhouse, the accessions responded differently to the application of SS, SOS, SOES, and SOBS at various stress stages. In contrast to untreated and stressed plants, tomato plants treated with SOS, SOES, and SOBS showed a slight decrease in the morpho-physiological and fruit physicochemical attributes in response to drought stress.

- All tomato accessions exposed to SOES and SOBS exhibited significant levels of TPC, ASC, and SSC characteristics along with low amounts of TA in fruit. In fruit TPC, ASC, TSS, CAC, and SSC, the *in vitro* tolerant accessions (Sandra and Raza Pashyi) outperformed the *in vitro* intolerant accessions (Braw and Yadgar).

- In the leaf tolerant accessions treated with SOES and SOBS, the lowest levels of lipid peroxidation and the highest levels of TPC, AC, SSC, PC, GPA, and CAT were found. Based on the findings of this study, Raza Pashyi and Sandra are ideal for growing in places with limited water availability. Furthermore, these accessions are beneficial for breeding projects aimed at developing drought-tolerant tomato cultivars.

- The use of oak leaf powder, oak leaf extract, and biofertilizer reduced the effect of drought stress on tomato plants.

- According to the result of Cd stress under *in vitro* conditions, the accessions of Karazi and Sirin were highly tolerant to Cd stress.

- The results obtained under greenhouse conditions to Cd stress, confirmed that both accessions of Karazi and Sirin have same ability to tolerate Cd stress.
- Our results, confirmed that the majority of absorbed Cd by tomato plant, was accumulated in the root, particularly, in the tolerant accessions.
- In the case of nematode infection, the results indicated that the accession of Sandra was resistant to the nematode infection.

# Recommendations

- In this regard, more investigations are needed to confirm the distinction in the gene pool between our collection and other tomato accessions from other parts of the world. Finally, the presence of genetic diversity in this crop may be advantageous in evolving and adapting to current climate changes, maintaining the agricultural production system's sustainability.

- The usage of the combination of leaf crude extract, oak leaf powder, and arbuscular mycorrhizal fungus should be investigated further under stress conditions. In order to determine the biostimulation effects of oak leaf powder and oak leaf extract, it is important to test their impacts on plant growth and production under normal conditions.

- The oak leaf powder and oak leaf extract may be described as novel agricultural practices because they are low-cost, simple to use, and time consuming, and they can meet the growing demands of the agricultural sector by providing environmentally sustainable techniques for enhancing plant resistance to abiotic stresses.

- The use of oak leaf powder to reduce the heavy metal stress should be more investigated in the future.

- It can be recommended to cultivate the accessions of Karazi (AC5) and Sirin (AC7) that showed the highest tolerance to Cd stress, in the regions which polluted by heavy metals.

- It can be recommended, that using oak leaf extract to reduce the nematode infection instead of oak leaf powder.

- The detection of resistant genes for drought, Cd stress, and nematode in tomato accessions may be helpful for selection of the resistant accessions.

285-292.